\documentclass[12pt,a4paper]{article} 
\pdfoutput=1
\usepackage{style}
\usepackage{amsmath}
\usepackage{amssymb}
\usepackage{amsthm}
\usepackage{psfrag}
\usepackage{graphicx}
\usepackage{hyperref}
\usepackage{feynmp}
\usepackage{color}
\usepackage{bm}
\usepackage{mathrsfs}

\newcommand{\be}{\begin{equation}}
\newcommand{\ee}{\end{equation}}
\newcommand{\bea}{\begin{eqnarray}}
\newcommand{\eea}{\end{eqnarray}}
\newcommand{\beqa}{\begin{eqnarray}}
\newcommand{\eeqa}{\end{eqnarray}}

\newcommand\m{\mu}
\newcommand\p{{\bf p}}

\newcommand\D{\Delta}
\newcommand\G{\Gamma}

\newcommand\s{\sigma}

\renewcommand\a{\alpha}
\renewcommand\b{\beta}

\newcommand{\T}{\Theta}

\newcommand{\HH}{{\cal H}}

\newcommand\x{{\bf x}}
\renewcommand\k{{\bf k}}
\newcommand\q{{\bf q}}
\newcommand{\e}{\eta}

\newcommand{\ve}{\varepsilon}

\def\d{\partial}
\newcommand{\bseq}{\begin{subequations}}
\newcommand{\eseq}{\end{subequations}}

\renewcommand{\ln}{\mathop{\rm ln}\nolimits}

\renewcommand{\H}{\mathcal H}
\renewcommand{\S}{\mathcal{S}}
\renewcommand{\D}{\mathcal{D}}

\newcommand{\F}{\mathcal{F}}

\renewcommand{\L}{\Lambda}

\renewcommand{\k}{{\bf k}}
\newcommand{\z}{{\bf z}}
\renewcommand{\s}{{\bf s}}

\relax

\title{Infrared Resummation 
for Biased Tracers
in Redshift Space}

\author[a,b,c]{Mikhail M. Ivanov\footnote{\texttt{ivanov@ias.edu}}}
\author[b,d,c]{Sergey Sibiryakov\footnote{\texttt{sergey.sibiryakov@cern.ch}}}

\affiliation[a]{School of Natural Sciences, Institute for Advanced Study, \\
1 Einstein Drive, Princeton, NJ 08540, United States}
\affiliation[b]{FSB/ITP/LPPC, \'Ecole Polytechnique F\'ed\'erale de Lausanne, \normalsize\it \\
CH-1015, Lausanne, Switzerland}
\affiliation[c]{Institute for Nuclear Research of the
Russian Academy of Sciences, \\ 
\normalsize \it  60th October Anniversary Prospect, 7a, 117312
Moscow, Russia}
\affiliation[d]{Physics Department, Theory Department, CERN, CH-1211 Gen\`eve 23, Switzerland}

\abstract{
We incorporate the effects 
of redshift space distortions and non-linear bias in
\textit{time-sliced perturbation theory} (TSPT). 
This is done via a new method that allows to map 
cosmological correlation functions from real to redshift space.
This mapping preserves a transparent infrared (IR) structure of the theory
and provides us with an efficient tool to study non-linear infrared 
effects altering the pattern of baryon acoustic oscillations (BAO) in redshift space.
We give an accurate description of the BAO 
by means of a systematic resummation 
of Feynman diagrams guided by well-defined power counting rules.
This establishes IR resummation within TSPT as a 
robust and complete procedure and provides  
a consistent theoretical
model for the BAO feature in the statistics of biased tracers 
in redshift space.
}

\begin{document}

\begin{flushright}
CERN-TH-2018-076\\
INR-TH-2018-006
\end{flushright}

\maketitle

\section{Introduction}

Baryon acoustic oscillations (BAO) are one of the most powerful tools of precision
cosmology. The BAO pattern has been  
observed across various redshifts in the 2-point correlation function of the 
distribution of galaxies 
(see \cite{Eisenstein:2005su,Cole:2005sx} for the first measurements
and \cite{Alam:2016hwk,Abbott:2017wcz} for recent ones),
Ly$\a$ forest absorption \cite{Delubac:2014aqe,Bautista:2017zgn}, 
quasars \cite{Ata:2017dya,Hou:2018yny}, 
and voids \cite{Liang:2015oqc,Zhao:2018hoo}. 
Recently, the BAO signal has also been detected in the 3-point
correlation function
\cite{Slepian:2015hca,Slepian:2016kfz,Pearson:2017wtw}. 
The significance of the BAO measurements for cosmology calls for
improving the analytic understanding 
of the BAO feature in the non-linear regime and 
robustly controlling the theoretical uncertainty.

It is well known that the BAO peak in position space 
(located at $r_{BAO}\sim 110\,h$/Mpc) 
is prone 
to non-linear damping produced by large-scale bulk flows.
Qualitatively, 
bulk flows move the pairs of tracers that 
used to be
at separations $r_{BAO}$ to larger or smaller distances,
which degrades their spatial correlation and thus reduces the observed
BAO signature. 
This effect is more severe in redshift space 
where the apparent separation 
of tracers
along the line-of-sight is additionally altered by peculiar velocities.
Besides, the BAO signal is further deformed by non-linear bias.

Several approaches have been put forward to deal with these effects.
Most of the smearing of the BAO is produced by Lagrangian displacements 
of matter. 
Thus, the process of erasing the BAO signal can be undone
by reversing tracers' trajectories and moving them back to their initial 
Lagrangian positions. This method, known as reconstruction 
\cite{Eisenstein:2006nk,
Seo:2009fp,
Schmittfull:2015mja,
Obuljen:2016urm,
Zhu:2016sjc,Zhu:2016xyy,
Schmittfull:2017uhh}, has become 
a standard tool in the BAO data analysis. 
Typically, reconstruction is used to increase the signal
in measurements 
of the BAO scale obtained upon
marginalizing over the broad-band shape and amplitude of the underlying 
correlation function (or power spectrum), 
see e.g. \cite{Beutler:2016ixs,Ross:2016gvb}. 
On the other hand, 
full shape measurements without reconstruction reveal
the rich cosmological information encoded in the entire power spectrum, see e.g. 
\cite{Peloso:2015jua,Ding:2017gad,Baumann:2017lmt,Baumann:2018qnt}.
In particular, the full shape measurements yield constrains on
structure growth rate through 
redshift-space distortions \cite{Beutler:2016arn,Satpathy:2016tct}.

The interpretation of the full shape BAO measurements relies on
theoretical modeling based  
on perturbation theory. 
Following early works on the subject 
\cite{Crocce:2005xy,Crocce:2007dt,Eisenstein:2006nj,Matsubara:2007wj},
it has been realized how the physical effects of bulk flows 
can be resummed to all orders in 
perturbation theory 
\cite{Senatore:2014via,
Vlah:2015zda,
Blas:2016sfa,
delaBella:2017qjy,
Senatore:2017pbn},
and how this procedure, 
called IR-resummation,
is related to the equivalence 
principle \cite{Baldauf:2015xfa,Mirbabayi:2014gda}.
The analysis was also extended to the power spectrum 
of biased tracers in redshift space~\cite{Matsubara:2008wx,Senatore:2014vja,Vlah:2016bcl}.

An 
efficient and systematic 
framework for
IR resummation has recently been proposed in
Ref.~\cite{Blas:2016sfa}. 
So far the analysis has been performed for
arbitrary $n$-point statistics of
matter in real space.
In this paper we generalize this framework 
in order to capture the non-linear regime of 
BAO for biased tracers in redshift space. 
The basis for our study is \textit{time-sliced 
perturbation theory} (TSPT) \cite{Blas:2015qsi}. 
This description is free from spurious IR-divergences
plaguing other methods
and thus clearly reveals the
physical structures relevant for the BAO physics. 

TSPT describes the evolution of the statistical distribution for
cosmological fields from 
the initial time slice to the final one. 
At a first step one solves for the time dependence of the probability
distribution function (PDF), 
which is governed by the Liouville equation.
In perturbation theory this generates 
an hierarchy 
of equations defining the time evolution of statistical cumulants that can be 
solved recursively. 
At a second step one computes correlators 
of the density and velocity fields using a diagrammatic technique
similar to
Feynman diagrams of a 3-dimensional Euclidean
quantum field theory of a scalar field.

In this paper we show that TSPT provides us with a convenient 
framework to study redshift space distortions and biased tracers.
The key observation is that the coordinate transformation relating real 
and redshift spaces can be seen as a free 1-dimensional fluid flow.
We introduce a fictitious time, over which this flow evolves,
and study the evolution of statistical properties of the flow
along the lines of TSPT. 
This auxiliary time will be loosely referred to as 
``redshift time''.
In this picture the initial redshift time slice corresponds to real space, 
the final one to redshift space.
Using this scheme, the redshift space statistical 
cumulants can be easily obtained from their
real space counterparts.  

Our method gives an alternative way to compute 
equal-time correlation 
functions of cosmological fields in redshift space 
that explicitly retains their IR safety.
This property helps us identify the physical IR-enhanced contributions 
and resum them in a systematic and controllable way,
which provides us with a powerful tool to explore the non-linear BAO
physics in redshift space. 

In the second part of the paper we discuss how to incorporate bias
into our framework. In the case of deterministic bias the 
tracers' density is a function of the 
matter density field and thus it
is not a statistically independent variable. 
Such variables are naturally described in TSPT as 
composite operators. 
We will show how the correlation functions of biased tracers can be 
obtained within TSPT and discuss the effect of IR resummation on them.

The paper is organized as follows.
In Section~\ref{sec:standard rsd} we review the standard approach to
redshift space distortions. 
In Section~\ref{sec:1d} we introduce a new redshift space mapping by means 
of the 1D flow analogy. 
In Section~\ref{sec:tsptrsd} we construct 
the redshift-space probability distribution function and the corresponding TSPT 
generating functional.
In Section~\ref{sec:Irres} we discuss 
the IR resummation of matter correlation functions in redshift space.
In Section~\ref{sec:bias} we include bias in our 
IR resummation procedure.
In Section~\ref{sec:pract} we describe how to practically evaluate 
IR resummed power spectra and bispectra at leading and next-to-leading order.
Section~\ref{sec:numerics} is devoted to a quantitative analysis of our results 
and their comparison to N-body data.
Section~\ref{sec:summary} 
draws conclusions and points future directions. 
We give a brief review of TSPT in Appendix~\ref{app:A}. 
Appendix~\ref{app:B} contains the derivation of the soft limit
of TSPT vertices. 
Appendix~\ref{app:C} is devoted to some useful expressions for the
bias and RSD kernels.  
In Appendix~\ref{app:D} we discuss possible simplifications of the
redshift-space IR-resummed integrands that make them convenient for
numerical evaluation.

\section{Review of standard redshift space mapping}
\label{sec:standard rsd}

Peculiar velocities alter the apparent picture of clustering 
along the line-of-sight and lead to the so-called 
redshift-space distortions (RSD) \cite{Jackson:2008yv,Sargent:1977me,Tully:1978me,Jennings:2010uv,Kaiser:1987qv,Hamilton:1992zz}. 
RSD break the full rotation symmetry of cosmological correlation functions 
down to a little group of azimuthal rotations along the line-of-sight.
Qualitatively, one can distinguish two main effects.
At large scales, galaxies in redshift space
appear to be closer along-the-line of sight due to mutually directed infall velocities,
which is observed as an enhancement of 
the amplitude of fluctuations in this direction.
At small scales, the velocity dispersion in virialized halos
elongate structures along the line-of-sight, 
which is known as the ``fingers of God'' effect \cite{Jackson:2008yv,Tully:1978me,Jennings:2010uv,Peacock:1993xg}. 
This elongation washes out observed structures and
results in a suppression of apparent short-scale power in the line-of-sight direction.

In what follows we will work in the plane-parallel (flat sky) approximation
valid for separations between points  
in redshift space much smaller than 
the distances from these points to the observer.
This approximation is justified for mildly non-linear 
scales  $\sim 100$ Mpc$/h$ typical for perturbation theory considerations.

In the plane-parallel regime the relation between the real space coordinate $\x$ 
and the redshift space coordinate $\s$ is inferred using Hubble's law,
\be
\label{eq:rs}
\s=\x+\hat{\bf z}\frac{v^{(r)}_z(\tau,\x)}{\HH}\,,
\ee
where $\HH$ is the conformal Hubble parameter, 
$\hat{\bf z}$ is the unit vector along the 
line-of-sight,
$v^{(r)}_z(\tau,\x)$ is the projection of the peculiar velocity field on 
the line-of-sight and
$\tau$ is conformal time.
Following the standard convention, we will denote 
real space quantities 
by the superscript $(r)$, whereas 
their redshift space counterparts will be denoted by $(s)$.

The redshift space matter 
density in the Eulerian picture is obtained via 
\be
\label{eq:Jac}
(1+\delta^{(s)}(\tau,\s))d^3s= (1+\delta^{(r)}(\tau,\x))d^3x\,,
\ee
which is dictated by the conservation of mass. 
In Fourier space\footnote{
We use the following conventions: $$\delta_\k\equiv \delta(\k)=\int \frac{d^3x}{(2\pi)^3}\delta(\x) e^{-i\k\cdot \x}~,~~~
\quad\delta(\x)=\int d^3k\,\delta_\k e^{i\k\cdot \x}~,~~~
\langle \delta_\k \delta_{\k'} \rangle= P(k)\delta^{(3)}(\k+\k')\;.$$
} the above equation can be rewritten as 
\be 
\label{eq:dens}
\delta^{(s)}(\tau,\k)=\delta^{(r)}(\tau,\k)+\int
\frac{d^3x}{(2\pi)^3}e^{-i\k\cdot
  \x}\Big(e^{-ik_zv^{(r)}_z(\tau,\x)/\HH}-1\Big)\big(1+\delta^{(r)}(\tau,\x)\big)\,. 
\ee
In the Eulerian standard perturbation theory (SPT)
\cite{Bernardeau:2001qr} the velocity 
field is fully characterized by its suitably normalized divergence $\T^{(r)}$, 
\be
v^{(r)}_i =-f\H\frac{\d_i\T^{(r)}}{\Delta} \,,
\ee
where we introduced the logarithmic growth rate $f$ defined as
\be
f(\tau)=\frac{d\ln D}{d\ln a} \,,
\ee
$D(\tau)$ is the linear theory growth factor and $a(\tau)$ is the scale factor.
In what follows we will also use the rescaled time variable $\e \equiv \ln D$.

Working within SPT,
one Taylor expands the exponent
containing the velocity field in Eq.~\eqref{eq:dens}.
Next, one uses the SPT expansion for the real space 
density 
\be 
\label{eq:sptr}
\begin{split}
\delta^{(r)}(\e,\k)=\sum_{n=1}^\infty D(\e)^n \int [dq]^n\,\delta^{(3)}(\k-\q_{1...n}) F_n(\q_1,...,\q_n)\delta_0(\q_1)...\delta_0(\q_n)\,,
\end{split}
\ee
where we introduced the short-hand notations
\[
[dq]^n=d^3q_1...d^3q_n\,,\quad \quad \quad \q_{1...n}\equiv \q_1+...+\q_n\,,
\]
and an analogous expansion for the velocity divergence $\T^{(r)}$ with
the $G_n$ kernels, instead of $F_n$. This allows one to obtain the formal expression
\be 
\label{eq:deltas}
\begin{split}
\delta^{(s)}(\e,\k)=\sum_{n=1}^\infty D(\e)^n \int [dq]^n\,\delta^{(3)}(\k-\q_{1...n}) Z_n(\q_1,...,\q_n)\delta_0(\q_1)...\delta_0(\q_n)\,,
\end{split}
\ee
where $Z_n$ kernels now contain RSD contributions.
Expressions for a first few of them are given in Appendix \ref{app:C}.
Various correlators of the redshift density field are computed using 
the statistical distribution of the initial density field $\delta_0$,
which is typically assumed to be Gaussian. 

In a matter dominated universe, the linear growth factor
  coincides with the scale factor, $D(\tau)=a(\tau)$, so that
  $f(\tau)=1$ and the kernels $F_n$ in (\ref{eq:sptr}) are
  time-independent. This is no longer true in the presence of
  cosmological constant or dark
  energy. Still, it is known that the use of (\ref{eq:sptr}) with the
  time-independent kernels $F_n$ together with the correct growth
  factor $D(\tau)$ provides an accurate approximation to the exact SPT
  expression for the density in the real space
  \cite{Pietroni:2008jx}. This is known as the Einstein--de Sitter
  (EdS) 
  approximation. Following the common practice, we will adopt it this
  paper; corrections to it can, in principle, be taken into account
  perturbatively. Notice that we do not assume any simplifications in
  the redshift space mapping \eqref{eq:dens}, so that the redshift
  space kernels $Z_n$ explicitly contain the factors $f(\tau)$ with
  the full time dependence.

A notorious drawback of SPT is spurious IR sensitivity
that arises due to homogeneous
translations of small-scale density fluctuations 
by soft modes.
Due to the equivalence principle, 
these large-scale translations must have no effect on equal-time correlation functions 
\cite{Scoccimarro:1995if,Creminelli:2013mca}. 
Indeed, the correlation between two galaxies 
should be insensitive to the acceleration produced by a
long-wavelength fluctuation.  
Technically, the sensitivity of the density field to large-scale translations
is encoded in the poles of the kernels $F_n$, $G_n$ 
at low momenta. 
The presence of these poles translates into an IR enhancement\footnote{
This enhancement is sometimes called ``IR divergence'', 
referring to the fact that 
the loop integrals would be divergent in IR 
for power-law spectra $P(k)\propto k^\nu$ with
$\nu \leq -1$.  
The $\L$CDM power spectrum vanishes quickly at small $k$, so these
integrals are actually convergent.
} 
of SPT loop diagrams composed out of $F_n$ or $G_n$ kernels.
The explicit cancellation of this spurious IR enhancement happens only 
after summing up all diagrams of a given loop order \cite{Vishniac}, 
and becomes intricate for higher-point functions and at 
higher orders in perturbation theory.\footnote{
The cancellation has been formally proven in real space
for leading IR divergences
to all orders in perturbation theory in \cite{Jain:1995kx} and
for subleading IR divergences in 
\cite{Blas:2013bpa, Carrasco:2013sva, Sugiyama:2013gza,
  Kehagias:2013yd, Peloso:2013zw,Blas:2015qsi}. } 

The presence of a feature in the initial power spectrum with 
a characteristic scale $k_{osc}$
makes the IR cancellation incomplete.
The translations produced by modes with momenta $q\gtrsim k_{osc}$ 
give large contributions that have to be resummed.
However, the identification of these physical IR contributions in SPT 
is obscured by 
the
spurious IR enhancement.

The situation becomes worse in redshift space. The exponent of the velocity
field in Eq.~\eqref{eq:dens} produces new poles compared to those
already present in real space, 
which brings in new spurious IR enhanced terms and further 
complicates the calculations.
The way to avoid these difficulties is to work directly in terms of
equal-time correlation functions
which are protected from IR divergences by the equivalence principle. 
This is precisely 
the core idea of TSPT. In order to realize this program and 
explicitly retain IR-safety in redshift space one has to perform
a
mapping from real to redshift space at the level of equal-time 
correlation functions. 
We introduce such a mapping in the next section.

Before moving on, we briefly comment on another problem, namely
an unphysical sensitivity of SPT to short-scale (UV) modes. 
This sensitivity arises because the 
perfect fluid hydrodynamical description breaks down at short scales.
This problem has been addressed within the effective field theory of large scale structure 
\cite{Baumann:2010tm,Carrasco:2012cv}, which 
captures the departures from Eulerian hydrodynamics 
by introducing so-called UV counterterms.
Since TSPT deals directly with $n$-point functions as a traditional quantum field theory,
it provides a natural framework to implement the ideas of UV renormalization.
However, we will refrain from doing it in this paper in order to focus
on IR resummation and its impact  
on the BAO physics. 
UV-renormalization will be addressed elsewhere.

\section{Redshift space transformation as a 1D fluid flow}
\label{sec:1d}

In this section we present a new mapping procedure that allows us to obtain redshift
space correlators directly from real space ones.
The core observation is that Eq.~\eqref{eq:rs} can be equivalently rewritten in the form
\be
\label{eq:xrs}
\s=\x+\hat{\bf z} \,v^{(r)}_z(\x)\mathcal{T}\,,
\ee
where $\mathcal{T}\equiv 1/\HH$. 
Now assume that the parameter $\mathcal{T}$ ranges from 0 to $1/\HH$.
Then Eq.~\eqref{eq:xrs} turns into an equation describing a flow of particles 
with Lagrangian coordinates $\x$ along the $z$-axis with initial velocity $v^{(r)}_z(\x)$.
The parameter $\mathcal{T}$ plays a role of time in the fictitious
dynamics described by Eq.~\eqref{eq:xrs}.  
This fictitious dynamics can be described 
in the Eulerian picture upon introducing 
the density $\delta^{(s)}$ and velocity ${\bm v}^{(s)}$
of this flow.
If we set the initial conditions 
\be
\label{eq:ics}
\begin{split}
& {\bm v}^{(s)}\Big|_{\mathcal{T}=0}={\bm v}^{(r)}(\eta,\x)\,,\\
& \delta^{(s)}\Big|_{\mathcal{T}=0}=\delta^{(r)}(\eta,\x)\,,
\end{split} 
\ee
then the value of $\delta^{(s)}$ at $\mathcal{T}=1/\HH$ will give us
the redshift space density, 
while ${\bm v}^{(s)}(\mathcal{T}=1/\HH,\s)$ will have the meaning of the
fluid velocity at a given position in redshift space. Note that only
orthogonal to the line-of-sight components of this velocity can, in
principle, be observed, so the physical relevance of this quantity is
not clear. 
However, it appears convenient to use this variable 
in intermediate steps when computing the density correlators.

There are no external forces in our fictitious evolution, thus the velocity is conserved
along the flow:
\be
\label{eq:eomvs}
\frac{D v^{(s),i}}{D \mathcal{T}}=\d_\mathcal{T} v^{(s),i} +
v^{(s)}_z\d_z v^{(s),i} =0 \,. 
\ee
This equation conserves vorticity. In real-space 
Eulerian perturbation theory the velocity field is longitudinal.
Then the initial conditions \eqref{eq:ics} imply that $v^{(s)}$ is
longitudinal as well, i.e. 
\be 
v^{(s)}_i  =-f\H \frac{\d_i \T^{(s)}}{\Delta} \,.
\ee
It is convenient to rescale our auxiliary time as
\be 
{\cal T}\to {\cal F}= f{\cal T} \H \quad \text{with}\quad {\cal F}\in [0,f]\,.
\ee
In this case 
the equation of motion for the 
velocity divergence obtained from \eqref{eq:eomvs}
takes a very simple form independent of cosmology,
\be 
\label{eq:Teq}
\d_\F \T^{(s)}(\F,\s;\e) = \d_i\bigg(\frac{\d_i \d_z\T^{(s)}}{\Delta} 
\frac{\d_z \T^{(s)}}{\Delta}\bigg)\,,
\ee
where we have emphasized that in this equation $\T^{(s)}$ 
depends parametrically on the cosmic time $\e$.

Since Eq.~\eqref{eq:xrs} 
describes a simple Lagrangian flow of particles, its density current
\[
{\bm j}^{(s)}=(1+\delta^{(s)}){\bm v}^{(s)}
\]
is conserved, which implies the continuity equation:
\be
\label{eq:long2}
\d_{\cal T}\delta^{(s)}+\d_z[v^{(s)}_z(1+\delta^{(s)})]=0\,.
\ee
Collecting together Eqs.~\eqref{eq:Teq} and~\eqref{eq:long2}
and switching to Fourier space we obtain the final system
\be
\label{eq:finaleoms}
\begin{split}
& \d_\F\delta^{(s)}_\k -\frac{k_z^2}{k^2}\T^{(s)}_\k=\int [dq]^2\delta^{(3)}(\k-\q_{12})
\alpha^{(s)}(\q_1,\q_2)\T^{(s)}_{\q_1}\delta^{(s)}_{\q_2}\,,\\
& \d_\F \Theta^{(s)}_\k  =  \int [dq]^2\delta^{(3)}(\k-\q_{12})\beta^{(s)}(\q_1,\q_2)\T^{(s)}_{\q_1}\T^{(s)}_{\q_2}\,,
\end{split}
\ee
where 
\be 
\label{eq:kernab}
\a^{(s)}(\q_1,\q_2)\equiv
\frac{q_{1,z}(q_{1,z}+q_{2,z})}{q_1^2}\,,\quad\quad \quad  
\b^{(s)}(\q_1,\q_2)\equiv \frac{(\q_1+\q_2)^2q_{1z}q_{2z}}{2q^2_1q_2^2}\,.
\ee
Note that the system of equations \eqref{eq:finaleoms}
contains a closed equation for the velocity divergence field 
and in this respect is quite similar to the Zel'dovich approximation in the Eulerian picture.

\section{TSPT partition function and vertices}
\label{sec:tsptrsd}

Our next step is to build a generating functional which produces the correlation functions of the $\T^{(s)}$ field\footnote{We will discuss the density field $\delta^{(s)}$ in the next subsection.}. 
This can be done by applying the ideas of TSPT to the system
\eqref{eq:finaleoms}.  
A detailed description of the TSPT framework
is given in Ref.~\cite{Blas:2015qsi} for a generic system, 
here we only outline the main steps. 
Some details on the TSPT in real space are summarized in Appendix \ref{app:A}.

The PDF
of the velocity divergence field undergoes certain evolution in
the auxiliary time $\F$. The initial  
distribution is given by the PDF in real space and the final one 
corresponds to the PDF in redshift space that we are looking for. 
In order to describe this evolution, consider 
the TSPT generating functional at a finite slice of the redshift time $\F$:
\be
\label{eq:genfuncTs}
\begin{split}
Z[J;\F]
=\int \mathcal{D}\T^{(s)}\; {\mathcal P}[\T^{(s)};\F]
\exp\left\{\int [dk] \T^{(s)}_\k J(-\k)\right\}\,,
\end{split}
\ee
where the PDF ${\mathcal P}$ is perturbatively expanded as
\be
\label{eq:PDFs}
{\mathcal P}[\T^{(s)};\F]=
\exp\left\{-\sum_{n=1}^\infty
\frac{1}{n!}\int [dq]^n \Gamma^{(s)\,tot}_n(\F;\k_1,...,\k_n)\prod_{j=1}^n 
\T^{(s)}_{\q_j}\right\}
\ee
Conservation of probability under the change of redshift time implies
\be
\D[\T^{(s)}+\delta\T^{(s)}]\;
{\mathcal P}[\T^{(s)}+\delta \T^{(s)};\F+\delta \F]=
\D[\T^{(s)}]\;
{\mathcal P}[\T^{(s)};\F]\,.
\ee
This leads to the following evolution equations for the TSPT 
vertices $\G^{tot\,(s)}_n$:
\be
\label{eq:TSPTeq1}
\begin{split}
& \d_\F \Gamma^{(s)\,tot}_n(\F;\k_1,...,\k_n)
\\
&+\sum_{m=1}^n \frac{1}{(n-m)!m!}\sum_{\sigma}
I^{(s)}_m(\k_{\sigma(1)},...,\k_{\sigma(m)}) \, 
\Gamma^{(s)\,tot}_{n-m+1}\Big(\F;\sum_{l=1}^m\k_{\sigma(l)},\k_{\sigma(m+1)},...,\k_{\sigma(n)}\Big)\\
&=\delta^{(3)}\left(\sum_{i=1}^n \k_i \right)\int [dp]
I^{(s)}_{n+1}(\F;\p,\k_1,...,\k_n)\,, 
\end{split} 
\ee
where in the second line the sum runs over all permutations $\sigma$
of $n$ indices and 
$I^{(s)}_m$ are the kernels 
determining the 
dynamical evolution of field  
$\T^{(s)}$:
\be 
\d_\F \T^{(s)}_\k=\sum_{n=1}^\infty \frac{1}{n!}\int [dq]^n\delta^{(3)}(\k-\q_{1...n}) I^{(s)}_n(\q_1,...,\q_n)\T^{(s)}_{\q_1}...\T^{(s)}_{\q_n}\,.
\ee
In the case of the system~\eqref{eq:finaleoms} we simply have $I^{(s)}_2=2\beta_z$
with all other kernels vanishing. 
In particular, $I_1^{(s)}=0$, in contrast to $I_1^{(r)}$=1 in real space, which
makes the structure of solution to (\ref{eq:TSPTeq1}) somewhat
different from the case of real-space TSPT 
\cite{Blas:2015qsi} (see also Appendix~\ref{app:A}). 


It is convenient to split the solution of Eq.~\eqref{eq:TSPTeq1}
into the solution of the 
homogeneous equation $\G^{(s)}_n$ and `counterterms' $C^{(s)}_n$
sourced by the singular r.h.s.
The corresponding initial conditions are
\be 
\label{eq:incond}
\begin{split}
  \G^{(s)}_n\Big|_{\F=0}= \G_n^{(r)}\,, \quad  C^{(s)}_n\Big|_{\F=0}= C_n^{(r)}\,,
\end{split}
\ee
where $\G_n^{(r)}$ and $C_n^{(r)}$ are TSPT vertices in real space.
Their structure is discussed in Appendix~\ref{app:A}. In particular,
for Gaussian initial conditions and in Einstein-de Sitter
approximation (which we adopt in this paper), the counterterms $C_n^{(r)}$ are time independent,
whereas the time dependence of $\G_n^{(r)}$ factorizes (here we are
talking about the dependence on the physical time $\e$),
\be
\label{eq:timedep}
\G^{(r)}_{n}=\frac{\bar{\G}^{(r)}_{n}}{g^2(\e)}\,.
\ee
In this expression we have denoted by $g(\e)$ the linear growth factor
\be\label{coupling}
  g(\eta) \equiv e^\eta = D(z)\,.
\ee
This notation agrees with Ref.~\cite{Blas:2015qsi} and emphasizes that
$g$ plays the role of the expansion parameter in TSPT, similar to a
coupling constant in quantum field theory. We will interchangeably use
$g$ and $D$ to denote the linear growth factor in what follows.

The equations for the $\G^{(s)}_n$
vertices take the following form:
\be 
\label{eq:Gammaseq}
\d_\F \Gamma^{(s)}_n(\F;\k_1,...,\k_n)
+\sum_{i<j}^n I^{(s)}_2(\k_{i},\k_{j}) 
\Gamma^{(s)}_{n-1}(\F;\k_1,...,\check{\k_i},...,\check{\k_j},...,\k_i+\k_j)=0\,,
\ee
where $\check{\k_j}$ means that $\k_j$ is {\em not} included 
in the arguments of $\Gamma^{(s)}_{n-1}$.
Let us start with the first non-trivial vertex $\G_2$. We have:
\be 
\label{G2s}
\d_\F \G^{(s)}_2=0\,, \quad \Rightarrow \quad \G^{(s)}_2= \G_2^{(r)}=\frac{\delta^{(3)}(\k'+\k)}{g^2\bar P(k)}\,,
\ee
where $\bar P(k)$ is the linear power spectrum at $\e=0$ and we have
used Eq.~(\ref{recGED1}) in the last equality.
Note that the inverse of $\G_2^{(s)}$ gives the linear power spectrum
of $\Theta^{(s)}$. From (\ref{G2s}) we conclude that the latter
coincides with the linear power spectrum of matter overdensities
$g^2(\e) \bar P(k)$. 
For $n\geq 3$ we consider the Ansatz,
\be
\label{eq:Gammals}
\Gamma^{(s)}_n=\sum_{l=0}^{n-2} \Gamma^{(s)}_{n,\,l}\;\F^l\,.
\ee
Plugging it into Eq.~\eqref{eq:Gammaseq} leads to the following recursion relation
\be
\label{eq:RecRSD}
 \Gamma^{(s)}_{n,\,l} =-\frac{1}{l}\sum_{i<j}^n I^{(s)}_2(\k_{i},\k_{j}) 
\Gamma^{(s)}_{n-1,\,l-1}(\k_1,...,\check{\k_i},...,\check{\k_j},...,\k_i+\k_j)\,,
\ee
with the initial condition for $l=0$,
\be
\G^{(s)}_{n,\,0}= \G^{(r)}_n\,.
\ee
The recursion relation \eqref{eq:RecRSD} allows us 
to obtain all redshift space velocity vertices from 
the real space ones (given in Appendix \ref{app:A}). 
Since the redshift space vertices are sourced by the
real space ones through the linear recursion relation
Eq.~\eqref{eq:RecRSD}, 
in the case of Gaussian initial conditions
they inherit factorized dependence on the coupling constant $g(\e)$
given by \eqref{eq:timedep}.
Another important property of the RSD vertices is that they are IR safe. 
The proof essentially repeats the proof of IR safety of the standard
TSPT vertices given in Ref.~\cite{Blas:2015qsi} and we do not present
it here.

The singular counterterms $C^{(s)}_n$ satisfy the following equations,
\be
\begin{split}
& \d_\F C^{(s)}_1(\F;\k)
=\delta^{(3)}\left( \k \right)\int [dp] I^{(s)}_2(\p,\k)\,, \\
& \d_\F C^{(s)}_n(\F;\k_1,...,\k_n)
\!+\!\! \sum_{i<j} I^{(s)}_2(\k_{i},\k_{j}) 
C^{(s)}_{n-1}(\F; \k_{1},...,\check{\k_i},...,\check{\k_j},...
\k_i\!+\!\k_j)=0\,, \quad 
n>1.
\end{split} 
\ee
Using the Ansatz $C^{(s)}_n=\sum_{l=0}^{n}C^{(s)}_{n,\,l}\; \F^{l}$, we find
the recursion relations similar to Eq.~\eqref{eq:RecRSD},
 \be 
 \begin{split}
 & C_{1,1}^{(s)}=\delta^{(3)}\left( \k \right)\int [dp] I^{(s)}_{2}(\p,\k)\,,\\
 & C^{(s)}_{n,\,l}=-\frac{1}{l}\sum_{i<j} I^{(s)}_2(\k_{i},\k_{j}) 
C^{(s)}_{n-1,\,l-1}(\k_{1},...,\check{\k_i},...,\check{\k_j},
...,\k_i+\k_j)\,, \quad 
n>1\,,
 \end{split}
\ee
with $C^{(s)}_{n,\,0}=C_n^{(r)}$.
Note that the $C_n$ counterterms 
appear already in the perfect fluid description. Their structure 
is totally fixed by the relevant equations of motion.

\subsection{Density field as a composite operator}
\label{sec:denscomp}

In cosmological perturbation theory with
adiabatic initial conditions there is only one statistically independent field
which can appear as the integration variable in the generating functional.
For studies of the IR structure it appears convenient to choose 
the velocity field, as we did above. 
In this subsection we express the redshift density
field in terms of $\T^{(s)}$
as a composite operator.
We focus on the matter density for the time being. Biased tracers
will be studied in 
Sec.~\ref{sec:bias}.
We introduce the Ansatz, 
\be
\label{eq:Kn}
\delta^{(s)}_\k=\sum_{n=1}^\infty \frac{1}{n!}\int[dq]^n \delta^{(3)}(\k-\q_{1...n})
K^{(s)}_n(\F;\q_1,...,\q_n)
\T^{(s)}_{\q_1}...\T^{(s)}_{\q_n}\,.
\ee
Plugging \eqref{eq:Kn} into the equations of motion \eqref{eq:finaleoms}, 
we obtain
\be 
\label{eq:Meqs}
\begin{split}
& \d_\F K^{(s)}_1(\k_1)=\frac{k^2_z}{k^2}\,,\\
& \d_\F K^{(s)}_n (\k_1,...,\k_n)= \sum_{i=1}^n K^{(s)}_{n-1}(\k_1,...,\check{\k}_i,...,\k_{n})
\a^{(s)}\left(\k_i,\sum_{j\neq i}\k_j\right)\\
&~~~~~~~~~~~~~~~~~~~~~-2\sum_{i< j}^n 
K^{(s)}_{n-1}(\k_1,...,\check{\k}_i,...,\check{\k}_j,...,\k_{n},\k_i+\k_j)
\beta^{(s)}\left(\k_i,\k_j\right)\,,\quad n>1\,.
\end{split} 
\ee
The kernels $K^{(s)}_n$ satisfy the following initial conditions:
\be
\label{eq:Mnini}
K^{(s)}_n\Big|_{\F=0} = K_n^{(r)}\,, 
\ee
where $K_n^{(r)}$ are TSPT kernels relating the density 
and velocity field in real space
(see Appendix~\ref{app:A}).
The first two kernels read,
\be
\begin{split}
 K^{(s)}_1(\k_1)&=1+\frac{k_z^2}{k^2} f \,,\\
  K^{(s)}_2(\k_1,\k_2) &= K_2^{(r)}(\k_1,\k_2)+
\left\{\frac{k_{1z}^2}{k_1^2}+\frac{k_{2z}^2}{k_2^2}
-2\frac{(\k_1\cdot \k_2) k_{1z}k_{2z}}{k_1^2k_2^2}
\right\} f\,,
\end{split} 
\ee
where we have made the substitution $\F\to f$
in the final expressions.
Proceeding along the
lines of Ref.~\cite{Blas:2015qsi}
one can easily prove that the kernels $K^{(s)}_n$ are IR safe.

\subsection{Feynman rules}

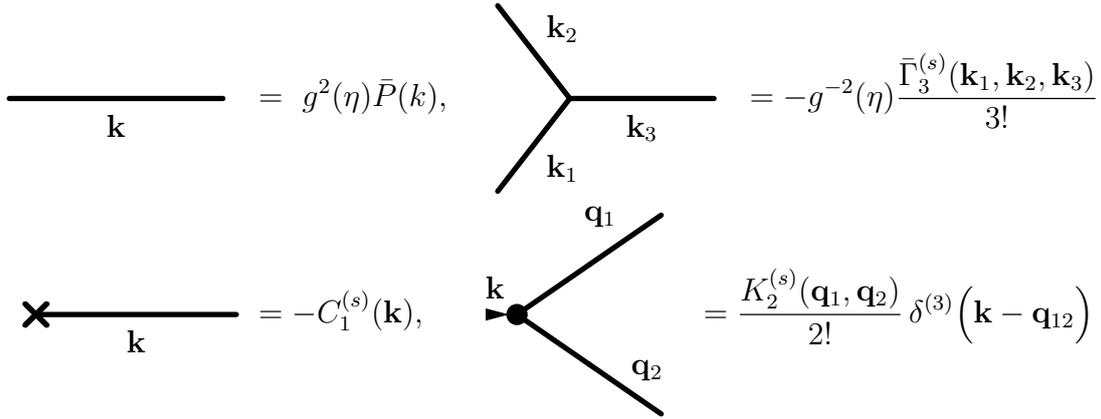
\begin{figure}
\begin{align*}
\begin{fmffile}{example-ps}
\parbox{90pt}{
\begin{fmfgraph*}(80,80)
\fmfpen{thick}
\fmfleft{l1}
\fmfright{r1}
\fmf{plain,label=${\bf k}$}{l1,r1}
\end{fmfgraph*}}
\end{fmffile}&=~g^2(\e)\bar P(k), ~~
\begin{fmffile}{example-bisp_f}
\parbox{100pt}{
\begin{fmfgraph*}(90,70)
\fmfpen{thick}
\fmfleft{l1,l2}
\fmfright{r1}
\fmf{plain,label=${\bf k}_1$,label.side=right}{l1,b1}
\fmf{plain,label=${\bf k}_2$,label.side=left}{l2,b1}
\fmf{plain,label=${\bf k}_3$}{b1,r1}     
\end{fmfgraph*}}
\end{fmffile}=-g^{-2}(\e)\frac{\bar\G^{(s)}_3(\k_1,\k_2,\k_3)}{3!}\\
\begin{fmffile}{example-C1}
\parbox{80pt}{
\begin{fmfgraph*}(75,75)
\fmfpen{thick}
\fmfleft{l1}
\fmfright{r1}
\fmf{plain,label=${\bf k}$}{l1,r1}
\fmfv{d.sh=cross,d.si=.15w}{l1}
\end{fmfgraph*}}
\end{fmffile}&=-C^{(s)}_1(\k),~~~~~
\begin{fmffile}{composite}
\parbox{80pt}{
\begin{fmfgraph*}(75,75)
\fmfpen{thick}
\fmfleft{l1}
\fmfright{r5,r6}
\fmfv{d.sh=circle,d.filled=full,d.si=3thick}{b1}
\fmf{phantom_arrow,tension=4,label=$\k$,l.s=left}{l1,b1}
\fmf{plain}{b1,v6}
\fmf{plain,label=$\q_1$,l.s=left}{v6,r6}
\fmf{plain}{b1,v5}
\fmf{plain,label=$\q_2$,l.s=left,l.d=4}{v5,r5}
\end{fmfgraph*}}
\end{fmffile}=\frac{K^{(s)}_2(\q_1,\q_2)}{2!}\,\delta^{(3)}\Big(\k-\q_{12}\Big)
\end{align*}
\caption{Examples of TSPT Feynman rules in redshift space.\label{fig:feynmanrules}}
\end{figure}

The TSPT perturbative expansion is produced by expanding the PDF
${\mathcal P}$ in the
generating functional \eqref{eq:genfuncTs} over its Gaussian part, 
which is equivalent to an expansion
in the coupling constant $g(\e)$. 
This calculation can be represented
as a sum of Feynman diagrams.
Our redshift space mapping 
does not produce new diagrammatic elements,
thus we can use the same Feynman rules as in real space, see Ref.~\cite{Blas:2015qsi}.
The first elements of the perturbative expansion in redshift space are shown in 
Fig.~\ref{fig:feynmanrules}: the linear power spectrum (inverse of $\G^{(s)}_2$) 
is represented by a line
(propagator), the different elements  $\G^{(s)}_n$ (with $n>2$) and $C^{(s)}_n$
correspond to vertices, and $K^{(s)}_n$ are depicted as vertices with an
extra arrow. To compute an $n$-point correlation
function of the velocity divergence $\T^{(s)}$
one needs to draw all diagrams with $n$ external legs.
For the correlators of the density field $\delta^{(s)}$ one has to add diagrams with external arrows (composite operators) and multiply each external line 
with momentum $\k$
by a factor $K^{(s)}_1(\k)$. 
For instance, at linear order we have the following 
expression for the correlator of the $\delta^{(s)}$ field,
\be 
\begin{split}
P^{(s)}_{mm}(\e;k)=~~~\begin{fmffile}{ps-delta}
\begin{fmfgraph*}(80,10)
\fmfpen{thick}
\fmfleft{ll1}
\fmfright{rr1}
\fmf{phantom_arrow,tension=2,label=$K^{(s)}_1$,l.s=right}{ll1,l1}
\fmf{phantom_arrow,tension=2,label=$K^{(s)}_1$,l.s=left}{rr1,r1}
\fmf{plain,label=$ $}{l1,v1}
\fmf{plain,label=$ $}{v1,r1}
\fmfposition
\end{fmfgraph*}
\end{fmffile}~~~&=(K^{(s)}_1(\k))^2 g^2\bar P(k)=
\left(1+f(\e)\frac{k_z^2}{k^2}\right)^2g^2(\e)\bar P(k)\,,
\end{split}
\ee
which reproduces the famous Kaiser formula \cite{Kaiser:1987qv}.

\section{IR resummation}
\label{sec:Irres}

The absence of spurious IR enhancement of loop integrals in TSPT
allows one to easily extract the physical IR effects responsible for
deforming the BAO pattern in redshift space.  
In this section we work out the ingredients necessary for systematic
IR resummation 
along the lines of \cite{Blas:2016sfa}:
perform the decomposition of the redshift space vertices into `wiggly'
and `smooth' parts, 
introduce power counting rules,
identify the leading IR contributions, and resum them.
In this section we will be discussing only the redshift space
quantities and omit the superscript $(s)$
on TSPT vertices to simplify notations. In all vertices and kernels we set
$\F\to f$. 
We also introduce primed notations for quantities stripped of the
momentum delta functions, e.g.,
\be 
\G^{(s)}_n(\k_1,...,\k_n)=\delta^{(3)}\left(\k_{1...n} \right)
\G'^{(s)}_n(\k_1,...,\k_n)\,.
\ee

\subsection{Wiggly-smooth decomposition}

One starts from the observation that the linear power spectrum can be
decomposed into an oscillating (wiggly) component corresponding to BAO
and a smooth (non-wiggly) part\footnote{In \cite{Blas:2016sfa} 
  the subscript ``$s$'' was used to denote the smooth part. In this paper
  we change this notation to ``$nw$'' (non-wiggly) in order 
not to be  
confused with the superscript $(s)$ referring to the redshift-space
quantities.}, 
\be
\label{eq:wsmooth}
\bar P(k)=\bar P_{nw}(k) + \bar P_w(k)\,.
\ee
The period of oscillations of $\bar P_w$ is set by
$k_{osc}=r_{BAO}^{-1}\sim 9\cdot 10^{-3} h/{\rm Mpc}$. Interaction
with long-wavelength modes differently affects these two components
leading to exponential damping of the wiggly part in the non-linear
power spectrum. In principle, the decomposition (\ref{eq:wsmooth}) is
not unique; two possible algorithms are described in
\cite{Blas:2016sfa}. In practice, the two algorithms lead to
essentially identical results in real space and we expect this to be
true also with inclusion of RSD. 

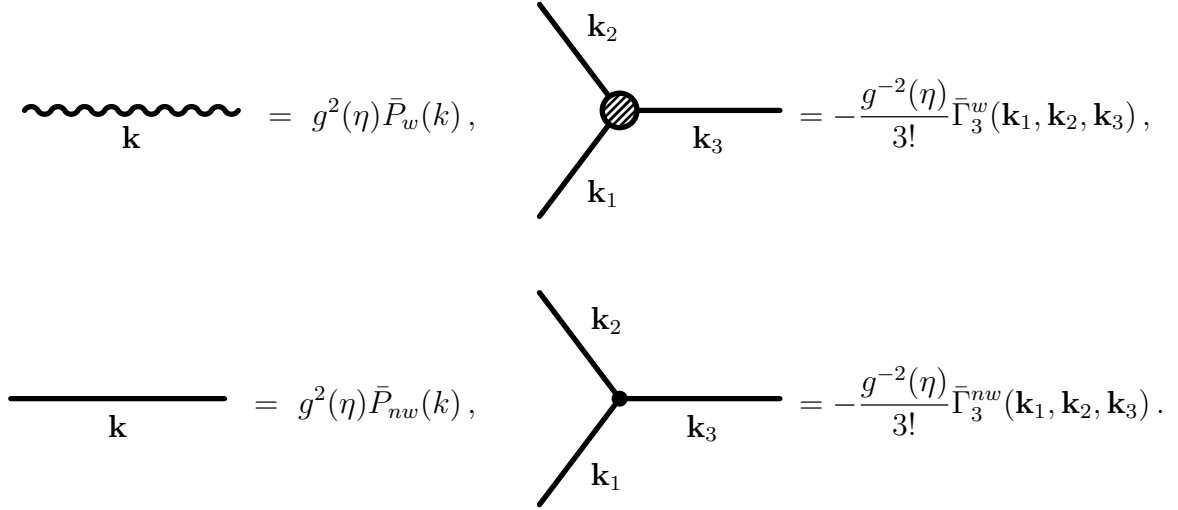
\begin{figure}[h]
\begin{align*}
\begin{fmffile}{w-ps}
	    \begin{gathered}
        \begin{fmfgraph*}(80,80)
        \fmfpen{thick}
        \fmfleft{l1}
        \fmfright{r1}
		\fmf{wiggly,label=$\textbf{k}$}{l1,r1}
	    \end{fmfgraph*}
	     \end{gathered}
	    \end{fmffile}~~=~g^2(\eta) \bar P_w(k) \,,
	    &
	     ~~~ \begin{fmffile}{w-bs}
	    \begin{gathered}
        \begin{fmfgraph*}(100,80)
        \fmfpen{thick}
        \fmfleft{l1,l2}
        \fmfv{d.sh=circle,d.filled=shaded,d.si=.13w,label=$ $,l.a=-90,l.d=.00w}{b1}
        \fmfright{r1}
		\fmf{plain,label=${\bf k}_1$}{l1,b1}
		\fmf{plain,label=${\bf k}_2$}{l2,b1}
		\fmf{plain,label=${\bf k}_3$}{b1,r1}   
	    \end{fmfgraph*}
	    \end{gathered}
	    \end{fmffile}~=-\frac{g^{-2}(\eta)}{3!}\bar\Gamma^w_3(\k_1,\k_2,\k_3)\,,\\
	     \\
	    	  \begin{fmffile}{s-ps}
	    \begin{gathered}
        \begin{fmfgraph*}(80,80)
        \fmfpen{thick}
        \fmfleft{l1}
        \fmfright{r1}
		\fmf{plain,label=$\textbf{k}$}{l1,r1}
	    \end{fmfgraph*}
	     \end{gathered}
	    \end{fmffile}~~=~g^2(\eta) \bar P_{nw}(k) \,,
	    &
	     ~~~ \begin{fmffile}{s-bs}
	    \begin{gathered}
        \begin{fmfgraph*}(100,80)
        \fmfpen{thick}
        \fmfleft{l1,l2}
        \fmfv{d.sh=circle,d.filled=full,d.si=.04w,label=$ $,l.a=-90,l.d=.00w}{b1}
        \fmfright{r1}
		\fmf{plain,label=${\bf k}_1$,label.side=right}{l1,b1}
		\fmf{plain,label=${\bf k}_2$,label.side=left}{l2,b1}
		\fmf{plain,label=${\bf k}_3$}{b1,r1}   
	    \end{fmfgraph*}
	    \end{gathered}
	    \end{fmffile}~=-\frac{g^{-2}(\eta)}{3!}\bar\Gamma^{nw}_3(\k_1,\k_2,\k_3)\,.
\end{align*}
	    \caption{Examples of Feynman rules for wiggly and smooth elements in redshift space.}    \label{feynmanruleswiggly}
\end{figure}

Since the TSPT vertices $\bar\G_n$ depend on the
linear power spectrum,
the decomposition \eqref{eq:wsmooth} produces a similar decomposition
of vertices,
\be
\bar\G_n=\bar\G_n^{nw}+\bar\G_n^w\,.
\ee
Here $\bar\G_n^w$ is of order $O(\bar P_w/\bar P_{nw})$ and one 
can neglect terms 
$O(\bar P_w^2/\bar P_{nw}^2)$ as they produce sub-percent
corrections.   
The counterterms $C_n$ and kernels $K_n$ are not subject to
wiggly-smooth decomposition as they are not functionals of the initial
power spectrum. Their momentum dependence is purely smooth.
Throughout the paper we will use the same graphic representation for
the 
redshift-space propagators and vertices 
as in \cite{Blas:2016sfa},
see Fig.~\ref{feynmanruleswiggly}.

\subsection{IR-enhanced diagrams and power counting}
\label{sec:pc}

Consider a TSPT $n$-point vertex $\bar\G_n(\k_1,...,\k_n)$
whose arguments $\k_i$ may belong to two different domains: 
either the soft one, denoted by $q$, or the hard one, denoted by $ k$, with 
\be 
\label{eq:qk}
q\ll k\,.
\ee
Let us first take a look at the wiggly three-point vertex
$\bar\G'^w_3(\k_1,\k_2,\k_3)$. 
From Eq.~(\ref{eq:RecRSD})
it is found to be
\be
\label{eq:G3w}
 \bar\G'^w_3(\k,\q,\!-\k-\q)\!=\!
 J_2(\k,\q)\frac{\bar P_w(|\k\!+\!\q|)}{\bar P^2_{nw}(|\k\!+\!\q|)}
 +J_2(-\k-\q,\q)\frac{\bar P_w(k)}{\bar P^2_{nw}(k)}
 +J_2(\k,\!-\k-\q)\frac{\bar P_w(q)}{\bar P^2_{nw}(q)}\,,
\ee
where we have defined
\be
J_2(\k_1,\k_2)\equiv \frac{(\k_1+\k_2)^2}{k_1^2k_2^2}\big((\k_1\cdot \k_2)+f
k_{1,z}k_{2,z}\big)
\,.
\ee
In the limit \eqref{eq:qk} the rightmost term in Eq.~\eqref{eq:G3w} is negligibly small, while the other two terms yield the expression 
\be 
\label{eq:G3w2}
\bar\G'^w_3(\k,\q,-\k-\q)=
 \frac{(\k\cdot \q)+f k_zq_z}{q^2}
 \frac{\bar P_w(|\k+\q|)-\bar P_w(k)}{\bar P^2_{nw}(k)}+O(1)\,.
\ee
Here we have Taylor expanded the smooth power spectrum  $\bar
P_{nw}(|\k+\q|)=\bar P_{nw}(k)+O(q/k)$.
We observe that at 
\be
\label{qinterm}
k_{osc}<q\ll k
\ee 
the first term is enhanced by
$O(k/q)$. For yet softer $q\ll k_{osc}$, one can write,
\be
\label{eq:taylor}
\bar P_w(|\k+\q|)-\bar P_w(\k)\approx \frac{(\k\cdot\q)}{k}\frac{d
  \bar P_w}{dk}\;.
\ee   
Taking into account that $d\bar P_w/dk\sim \bar P_w/k_{osc}$, we see
that the enhancement of (\ref{eq:G3w2}) becomes
$O(k/k_{osc})$. Note that despite the enhancement, 
the vertex (\ref{eq:G3w2}) remains finite in the limit $q\to 0$, 
in line with the IR safety of the TSPT expansion discussed in
Sec.~\ref{sec:tsptrsd}. 

For general values of $q$
in the range (\ref{qinterm}) the expansion (\ref{eq:taylor}) does not
provide a good approximation to the finite difference on the
l.h.s. Indeed, the latter oscillates with the period $q\sim 2\pi
k_{osc}$, whereas the r.h.s. of (\ref{eq:taylor}) is linear in $q$.  
As we want to include the range (\ref{qinterm}) in our
analysis, we work in what follows with the representation
(\ref{eq:G3w2}), where
the finite difference of
the wiggly power 
spectra is kept explicitly.

The enhanced contribution (\ref{eq:G3w2}) can be written in a compact form
by introducing a linear operator $\D^{(s)}_\q$ acting on the wiggly
power spectrum,
\be
\begin{split}
\D^{(s)}_\q [\bar P_w(k)]&=
\frac{(\k\cdot \q)+f k_zq_z}{q^2}
\big(\bar P_w(|\k+\q|)-\bar P_w(k)\big)\\
&=\frac{\mathcal{P}_{ab}k^a q^b}{q^2}(e^{\q\cdot \nabla_{\k'}}-1)\bar P_w(k')\Big|_{k'=k}\,,
\end{split}
\ee 
where 
\be 
\label{eq:proj}
\mathcal{P}_{ab}\equiv \delta_{ab}+f \hat{z}_a\hat{z}_b\,.
\ee
This operator has the following properties.
First, it scales as 
\be  
\D^{(s)}_\q[\bar P_w]=O(1/\varepsilon)\bar P_w\,,
\ee
where $\varepsilon \sim q/k$. 
We will come back to the precise definition of this parameter shortly.
Second, this operator commutes with itself and 
in any expression acts only on occurrences of $\bar P_w$, leaving the
smooth components intact.
It is a simple generalization of the operator 
$\D^{(r)}_\q$
which controls the IR enhancement in 
real space and is obtained from $\D^{(s)}_\q$ by replacing
$\mathcal{P}_{ab}$ with the Kronecker symbol $\delta_{ab}$
(see Appendix \ref{app:A}). 

In Appendix \ref{app:B} we prove that the expression (\ref{eq:G3w2})
generalizes
to an arbitrary $n$-point vertex with 
$m$ hard momenta $\k_i$ and $n-m$ soft momenta $\q_j$ uniformly going to zero,
\be
\label{eq:GnmindMain}
\begin{split}
\bar \Gamma'^{w}_n\Big(\k_1,...,&\k_m -\sum_{i=1}^{n-m} \q_i, \q_1,...,\q_{n-m}\Big)\\
&=(-1)^{n-m}\left(\prod_{j=1}^{n-m} \D^{(s)}_{\q_j}\right)\big[\bar
\G'^{w}_{m}(\k_1,...,\k_m)\big]\,(1+O(\varepsilon))\,.
\end{split}
\ee
The leading IR enhancement of this vertex is $O(\varepsilon^{-n+m})$, 
and the maximum is achieved for $n-2$ soft wavenumbers, 
\be 
\label{eq:Gnn-2}
\begin{split}
\bar \Gamma'^{w}_n &\Big(\k_,-\k-\sum_{i=1}^{n-2} \q_i, \q_1,...,\q_{n-2}\Big)
=(-1)^{n-2}\left(\prod_{j=1}^{n-2} \D^{(s)}_{\q_j}\right)\bar \G'^{w}_{2}(\k,-\k)(1+O(\varepsilon))\,\\
& =(-1)^{n-1}\left(\prod_{j=1}^{n-2}\frac{\mathcal{P}_{ab}k^aq^b_j}{q_j^2}(e^{\q_j\cdot \nabla_\k}-1)\right)\frac{\bar P_w(k')}{\bar P^2_{nw}(k)}\Bigg|_{k'=k}(1+O(\varepsilon))\,.
\end{split}
\ee
On the other hand, it is straightforward to verify that 
the smooth vertices do not receive IR enhancements
as their arguments go to zero, 
in line with the fact that bulk flows have significant effect only 
on wiggly correlation functions.

We now discuss the power counting rules that will help us identify 
IR enhanced diagrams. 
These rules are completely similar to those discussed in Ref.~\cite{Blas:2016sfa},
where the reader can find further details.
The first relevant parameter is the separation scale $k_S$ that cuts loops into 
hard and soft parts. 
This scale should be lower than the characteristic momentum $k$ of
interest, but at the same time high enough for the soft part of the
loop to capture as many relevant IR modes as possible. 
Apart from that, the precise choice of $k_S$ is arbitrary. 
As discussed below, the
sensitivity of the final result to $k_S$ can be used to estimate the
theoretical uncertainty.
The second parameter is the characteristic IR scale $q\equiv \varepsilon k$ at
which the IR loop integrals are saturated. 
This is the scale where the loop integrands peak and roughly 
it is of the order of the maximum point of the linear power spectrum
corresponding to the modes entering the horizon at the
radiation--matter equality,
$k_{eq}\sim 0.02\,h$/Mpc. 
The two other parameters are the variances of the linear power
spectrum in the IR and UV, 
which control the loop corrections coming from 
the corresponding domains,
\bseq
\begin{align}
  \sigma_S^2 &\simeq  g^2\int_{q<k_S} d^3q\, \bar P_{nw}(q) \,, \label{sigmas}\\
  \sigma_h^2 &\simeq g^2\int_{q>k_S} d^3q\, \bar P_{nw}(q) \,. \label{eq:sigmah}
\end{align}
\eseq
Although $\sigma_S^2$ is small, 
the IR loops involving wiggly vertices receive an 
enhancement by inverse powers of $\ve$, resulting in $O(1)$
corrections at low redshift. 
These are the corrections we intend to resum. 
Owing to Eq.~\eqref{eq:GnmindMain}, 
the resummation procedure is totally analogous to the 
one discussed in \cite{Blas:2016sfa}, with the only difference that we have to 
substitute the real-space operator $\D^{(r)}_\q$, kernels and vertices 
with their redshift-space counterparts.

It is instructive to consider the leading IR correction to the matter
power spectrum 
at one loop. It is given by the following graph:

\vspace{0.5cm}
\be
\label{1lpw}
\begin{gathered}
	   \begin{fmffile}{w-vertex4}
          \begin{fmfgraph*}(120,20)
        \fmfpen{thick}
        \fmfkeep{1loop}
        \fmfleft{ll1}
        \fmfright{rr1}
        \fmfv{d.sh=circle,d.filled=shaded,d.si=.1w,label=
$\bar\G^w_4$,l.a=-90,l.d=.15w}{b1} 
	\fmf{plain,tension=1.0,label=$ $}{l1,b1}
			\fmf{plain,tension=1.0,label=$ $}{r1,b1}
				\fmf{phantom_arrow,tension=3.0,label=$
                                  $}{ll1,l1} 
			\fmf{phantom_arrow,tension=3.0,label=$ $}{rr1,r1}
	     \fmf{plain,right=0.5,tension=0.8,label=$ $}{b1,b1}
	    \end{fmfgraph*}
  \end{fmffile}
  	     \end{gathered}
  	    \!=\frac{g^4}{2}  
K_1^2(\k) \!\!\int_{q< k_S} [dq]\bar P_{nw}(q){\cal D}^{(s)}_\q {\cal
  D}^{(s)}_{-\q} \bar P_w(k) \equiv  - g^4 K_1^2 \S^{(s)}[\bar P_w]\,,
\ee 
where in the last equality we defined a new linear operator $\S^{(s)}$
acting on the wiggly power spectrum,
\be 
\label{Slead}
\begin{split}
\S^{(s)}[\bar P_w]&= \mathcal{P}_{ab}\mathcal{P}_{cd}k^a k^c   
\int_{q<k_S}[dq]\bar P_{nw}(q)\frac{q^b q^d}{q^4}\big(1-
\cosh\left( \q\cdot\nabla_{k'} \right)\big)\bar P_w(k')\Big|_{k'=k}\,.
\end{split} 
\ee
Within our power counting rules,
\be
 g^2\S^{(s)}[\bar P_w] \sim   O(1/\varepsilon^2\times \sigma_S^2)\; 
\bar P_w\;. 
\ee
As discussed previously, the product $1/\ve^2\times \sigma_S^2$ is
$O(1)$ at low redshifts and 
therefore this one-loop contribution is of the same order as the
linear wiggly power spectrum, 
which points to the need for IR resummation.

To determine the order of
an arbitrary TSPT diagram with $L$ loops (\emph{i.e.} scaling as
$g^{2L}$), 
one must 

\begin{enumerate}

\item choose for each propagator and each vertex whether it is smooth
  or wiggly. Since we are interested in diagrams that contain one
  power of $\bar P_w$,
at most one element (either propagator or vertex) can be wiggly. 

\item assign each loop to be either hard ($q> k_S$) or soft ($q<
  k_S$). 
The number of hard loops is denoted by $L_h$ and the number of soft
loops by $L_s$.  
As $L=L_h+L_s$, the diagram contributes at order
$(\sigma_S^2)^{L_s}\times (\sigma_h^2)^{L_h}$. 

\item count the number of soft lines that are attached to the wiggly
  vertex. We call this number $l$.  
According to \eqref{eq:GnmindMain}, this vertex brings an IR enhancement of order $1/\varepsilon^l$.
\end{enumerate}
Then the order of a contribution  characterized by the numbers $(L_h,
L_s, l)$ is given by
\be\label{eq:powerc}
O\left( (\sigma_S^2)^{L_s}\times (\sigma_h^2)^{L_h} \times 1/\varepsilon^l \right)\,.
\ee
For a diagram without wiggly vertices $l=0$ and there is
no IR-enhancement.
The most IR-enhanced contributions have
the largest value of $l$. 
As a single loop cannot contain more than two lines attached to the same vertex, we have the inequality 
$l \leq 2L_s$. 

\subsection{IR resummation at leading order}

Let us first consider the density power spectrum.
The most IR-enhanced contributions correspond to 
$l=2L$ and $L_h=0$, i.e. all loops are soft and attached to wiggly vertices. They scale as 
\be
(\sigma^2_S \times 1/\ve^2)^{L} \,.
\ee
Resummation of these \textit{daisy} diagrams is graphically
represented as follows,
\begin{align}
\label{fig:daisy}
	&P^{(s)\, \text{IR res,LO}}_{mm,w}(\e;\k)	~=~   
	\begin{fmffile}{w-ps_a}
	    \begin{gathered}
        \begin{fmfgraph*}(80,80)
        \fmfpen{thick}
        \fmfleft{ll1}
        \fmfright{rr1}
		\fmf{wiggly,label=$ $}{l1,r1}
                \fmf{phantom_arrow,tension=3.0,label=$ $}{ll1,l1} 
		\fmf{phantom_arrow,tension=3.0,label=$ $}{rr1,r1}
	    \end{fmfgraph*}
	     \end{gathered}
	    \end{fmffile} 
~+~
	\begin{fmffile}{w-vertex4_3_a}
	    \begin{gathered}
          \begin{fmfgraph*}(90,60)
        \fmfpen{thick}
        \fmfkeep{1loop}
        \fmfleft{l1,ll2,l3}
        \fmfright{r1,rr2,r3}
        \fmfv{d.sh=circle,d.filled=shaded,d.si=.15w,label=$ $,l.a=-90,l.d=.0w}{b1}
		\fmf{plain,label=$\bar \G^w_4$}{b1,l2}
		\fmf{plain,label=$ $}{b1,r2}
\fmf{phantom}{l1,u1}
		\fmf{phantom}{u2,r1}
		\fmf{phantom}{l3,v1}
		\fmf{phantom}{v2,r3}
	    \fmf{phantom,right=0.5,tension=0.01,l.side=left}{b1,u1}
	     \fmf{phantom,right=0.5,tension=2,label=$ $}{u1,u2}
	   \fmf{phantom,left=0.5,tension=0.01}{b1,u2}
	   	    \fmf{plain,left=0.5,tension=0.01,l.side=right}{b1,v1}
	     \fmf{plain,left=0.5,tension=2,label=$ $}{v1,v2}
	   \fmf{plain,right=0.5,tension=0.01}{b1,v2}
\fmf{phantom_arrow,tension=3.0,label=$ $}{ll2,l2} 
		\fmf{phantom_arrow,tension=3.0,label=$ $}{rr2,r2}
	    \end{fmfgraph*}
 	     \end{gathered}   
  \end{fmffile}\\
  \nonumber
  &~+~
	    \begin{fmffile}{w-vertex6}
     \begin{gathered}
        \begin{fmfgraph*}(90,60)
        \fmfpen{thick}
        \fmfkeep{2loop}
        \fmfleft{l1,ll2,l3}
        \fmfright{r1,rr2,r3}
        \fmfv{d.sh=circle,d.filled=shaded,d.si=.15w,label=$ $,l.a=155,l.d=.0w}{b1}
		\fmf{plain,label=$\bar \G^w_6$}{b1,l2}
		\fmf{plain,label=$ $}{b1,r2}
		\fmf{phantom}{l1,u1}
		\fmf{phantom}{u2,r1}
		\fmf{phantom}{l3,v1}
		\fmf{phantom}{v2,r3}
	    \fmf{plain,right=0.5,tension=0.01,l.side=left}{b1,u1}
	     \fmf{plain,right=0.5,tension=2,label=$ $}{u1,u2}
	   \fmf{plain,left=0.5,tension=0.01}{b1,u2}
	   	    \fmf{plain,left=0.5,tension=0.01,l.side=right}{b1,v1}
	     \fmf{plain,left=0.5,tension=2,label=$ $}{v1,v2}
	   \fmf{plain,right=0.5,tension=0.01}{b1,v2}  
\fmf{phantom_arrow,tension=3.0,label=$ $}{ll2,l2} 
		\fmf{phantom_arrow,tension=3.0,label=$ $}{rr2,r2}
	    \end{fmfgraph*}
	    	    	    \end{gathered}
	    	  \quad  	    
\!\!\!+ 
	    \end{fmffile}
	    \begin{fmffile}{w-vertex8}
	    \begin{gathered}
        \begin{fmfgraph*}(105,80)
        \fmfpen{thick}
        \fmfkeep{3loop}
        \fmfleft{l1,ll2,l3}
        \fmfright{r1,rr2,r3}
        \fmfv{d.sh=circle,d.filled=shaded,d.si=.12w,label=$ $,l.a=90,l.d=.00w}{b1}
		\fmf{plain,l.side=left,label=$\bar \G^w_8$}{l2,b1}
		\fmf{plain,l.side=left,label=$ $}{b1,r2}
		\fmf{phantom}{l1,u1}
		\fmf{phantom}{u2,r1}
		\fmf{phantom,tension=2}{l3,v1}
		\fmf{phantom,tension=2}{v1,r3}
	    \fmf{plain,right=0.55,tension=0.01,l.side=left}{b1,u1}
	   \fmf{plain,left=0.55,tension=0.01,l.side=right,label=$ $}{b1,u1}
	   	    \fmf{plain,right=0.55,tension=0.01,l.side=left,label=$ $}{b1,u2}
	   \fmf{plain,left=0.55,tension=0.01}{b1,u2}
	     \fmf{plain,right=0.7,tension=0.00001,label=$ $}{b1,v1}
	   \fmf{plain,left=0.7,tension=0.00001}{b1,v1}
\fmf{phantom_arrow,tension=3.0,label=$ $}{ll2,l2} 
		\fmf{phantom_arrow,tension=3.0,label=$ $}{rr2,r2}
	    \end{fmfgraph*}
	    	    	    \end{gathered}
	    \end{fmffile}
	  ~+~ \begin{fmffile}{w-vertex10}
	    \begin{gathered}
        \begin{fmfgraph*}(105,80)
        \fmfpen{thick}
        \fmfkeep{3loop}
        \fmfleft{l1,ll2,l3}
        \fmfright{r1,rr2,r3}
        \fmfv{d.sh=circle,d.filled=shaded,d.si=.14w,label=$ $,l.a=90,l.d=.0w}{b1}
		\fmf{plain,l.side=right,label=$ $,l.d=.15w}{b1,l2}
		\fmf{plain,l.side=left,label=$ $}{b1,r2}
		\fmf{phantom}{l1,u1}
		\fmf{phantom}{u2,r1}
		\fmf{phantom}{l3,v1}
		\fmf{phantom}{v2,r3}
	    \fmf{plain,right=0.55,tension=0.01,l.side=left}{b1,u1}
	   \fmf{plain,left=0.55,tension=0.01,l.side=right,label=$ $}{b1,u1}
	   	    \fmf{plain,right=0.55,tension=0.01,l.side=left,label=$ $}{b1,u2}
	   \fmf{plain,left=0.55,tension=0.01}{b1,u2}   
	      \fmf{plain,right=0.55,tension=0.01,l.side=left}{b1,v1}
	     \fmf{plain,left=0.55,tension=0.01,l.side=right,label=$ $}{b1,v1}
	   	 \fmf{plain,right=0.55,tension=0.01,l.side=left,label=$ $}{b1,v2}
	   	   \fmf{plain,left=0.55,tension=0.01}{b1,v2}
\fmf{phantom_arrow,tension=3.0,label=$ $}{ll2,l2} 
		\fmf{phantom_arrow,tension=3.0,label=$ $}{rr2,r2}
	    \end{fmfgraph*}
	    	    	    \end{gathered}
	    \end{fmffile}
	    +...
	    \end{align}
Using Eq.~(\ref{eq:GnmindMain}) we find that the $L$-th contribution
here has the form,
\be
\begin{split}
&g^2 K_1^2(\k)\bar P_{nw}^2(k) \frac{1}{L!}\bigg[\frac{g^2}{2}\int_{q<k_S}[dq]
\bar P_{nw}(q)\D_\q^{(s)}\D_{-\q}^{(s)}\bigg]^L \bar\G_2'^w(\k,-k)\\
&=g^2 K_1^2(\k)\cdot\frac{1}{L!}\big(-g^2\S^{(s)}\big)^L\bar P_w(k)\;,
\end{split}
\ee
up to $\ve$--suppressed corrections. 
Summing the series
(\ref{fig:daisy}) 
leads to
the exponentiation of the operator \eqref{Slead}, which 
we have already encountered at one loop order, i.e.
\be 
\label{PdLO}
P^{(s)\, \text{IR res,LO}}_{mm}(\e;\k)=g^2
K^2_1\big(\bar P_{nw}+e^{-g^2\S^{(s)}}\bar P_w\big)
\,,
\ee
where we have also added the smooth part which is unaffected by IR
resummation. The time dependence of the resummed power spectrum comes
from its explicit dependence on $g(\e)$, as well as implicitly through
the dependence of the kernel $K_1$ and the operator $\S^{(s)}$ on
$f(\e)$. The practical method to evaluate the exponential operator
appearing in (\ref{PdLO}) will be discussed in Sec.~\ref{sec:pract}.  

Similarly, one can show by following the arguments of
\cite{Blas:2016sfa} (see also Appendix~\ref{app:A}) that IR
resummation of an arbitrary $n$-point function at the 
leading order (LO) 
amounts to simply substituting the wiggly part of the linear spectrum,
$\bar P_w$ by its resummed version $e^{-g^2\S^{(s)}}\bar P_w$ 
in all tree-level diagrams. 
This can be summarized in the following compact form,
\be
\label{Cncomp}
\mathfrak{C}_n^{(s)\,\text{IR res,LO}}(\k_1,...,\k_n)=
\mathfrak{C}_n^{(s)\,tree}\big[
\bar P_{nw}+e^{-g^2\S^{(s)}}\bar P_w\big](\k_1,...,\k_n)\;,
\ee
where $\mathfrak{C}_n^{tree}$ should be understood as a functional of
the linear power 
spectrum.  Note that the leading IR-enhanced contributions are
essentially the
same for 
velocity and density correlators.

\subsection{Next-to-leading order corrections and hard loops}
\label{sec:nlo}

There are two different types of next-to-leading order corrections to
the above results:
\begin{enumerate}
\item[(1)] 
Soft diagrams
with non-maximal IR enhancement, characterized by $l=2L_s-1$ (see
Eq.~(\ref{eq:powerc})), as well as subleading terms in the
daisy diagrams considered above.
Formally, these contributions
are suppressed by one power of $\varepsilon$ relative to the leading order. 
\item[(2)] 
Diagrams with
one hard loop, $L_h=1$, and otherwise maximal IR enhancement
$l=2L_s$. These diagrams are  
suppressed by one factor of $\sigma_h^2$ relative to the leading order.
\end{enumerate}

Naively, one expects the corrections of the first type to scale as
\be
\label{eq:NLOscalenaive}
\ve \times (\sigma^2_S \times 1/\ve^2)^{L_s} \,,
\ee
in which case they should have $O(\ve)\sim k_{eq}/k\sim  10\%$ effect.
However, in contradiction to this expectation,
in Ref.~\cite{Blas:2016sfa} these corrections were found to 
have a sub-percent effect at the BAO scales.
We now argue that the smallness of the NLO soft corrections
is a consequence of the specific shape of the linear power spectrum
in the $\Lambda$CDM cosmology.
As was shown in \cite{Blas:2016sfa}, the integrands of the LO and NLO
soft contributions  
are different, so that the estimate (\ref{eq:NLOscalenaive}) should be
properly written as 
\be 
\label{eq:NLOscaletrue}
(\sigma^2_{S,\,{\rm NLO}}\times 1/\ve)\times (\sigma^2_{S,\,{\rm LO}}
\times 1/\ve^2)^{L_s-1}\,.
\ee
Here $\sigma_{S,\,{\rm LO}}^2$ receives contributions from momenta
$k_{osc}\lesssim q\lesssim k_S$ and is saturated in the vicinity of
the maximum of the power spectrum at $q\sim k_{eq}>k_{osc}$. It is
indeed of the order\footnote{Essentially, $\sigma_{S,\,{\rm LO}}^2$ coincides
  with $k_{eq}^2\Sigma^2$, where $\Sigma^2$ is the BAO damping factor given in
Eq.~(\ref{Sigmanorm}) below. Its numerical value is plotted in
Fig.~\ref{fig:SigmaL}.} (\ref{sigmas}). 
On the other hand, the integrand in the subleading soft loop
corrections schematically has the form,
\be
\label{eq:estNLO}
\begin{split}
(\sigma^2_{S,\,{\rm NLO}} 
\times 1/\ve )
& \propto g^2 \int_{q<k_S} d^3q \frac{(\q\cdot \k)}{q^2}\, \bar P_{nw}(q)\,
\big(1-e^{i\frac{\q\cdot \hat{\k}}{k_{osc}}}\big) \\
 & \sim g^2 k\int_0^{k_S} dq\,q\bar P_{nw}(q) \left[\,j_1\left(\frac{q}{k_{osc}}\right)\right]\,,
  \end{split}
\ee
where $j_1$ is the spherical Bessel function. The integral is
effectively cut at $q\sim k_{osc}$, before the linear power spectrum reaches its maximum. 
Recalling that in this region the
$\L$CDM power spectrum behaves as $\bar P_{nw}(q)\propto q$, we find
that
\be 
\label{eq:sigmasNLO}
\sigma^2_{S,\,{\rm NLO}}/\sigma^2_{S,\,{\rm LO}}\sim (k_{osc}/k_{eq})^3\sim 0.1\;. 
\ee
This leads to additional numerical suppression of the NLO soft corrections.
The same should be true for redshift space, as our argument only appeals to 
the shape of the $\Lambda$CDM power spectrum and the structure of mode
coupling which is similar in real and redshift
space\footnote{Qualitatively, the result \eqref{eq:sigmasNLO} can be
  understood as follows.  
The NLO soft corrections are responsible for the shift of the BAO peak. 
This shift can be seen as shrinking of the BAO scale in 
an overdense region that locally behaves as 
a universe with positive spatial curvature \cite{Sherwin:2012nh}. 
Hence, the shift is sensitive only to the curvature of this ``universe'', 
which is generated by modes with wavelengths 
bigger than $r_{BAO}$. 
Thus, the NLO soft contributions should be saturated at $k_{osc}$. 
On the other hand, the damping of the BAO feature (which is produced
by LO soft corrections) 
is affected by modes with wavelengths down to 
the width of the BAO peak \cite{Baldauf:2015xfa}.
Thus, the LO soft corrections should include contributions 
from wavenumbers $q \gg k_{osc}$.
}.

Note that in a hypothetical universe with $k_{eq}~\ll ~k_{osc}$ 
the situation would be different, 
with $\sigma^2_{S,\,{\rm NLO}}$ being of the same order as
$\sigma^2_{S,\,{\rm LO}}$. 
The power-counting rules of Sec.~\ref{sec:pc} are formulated in full generality 
and do not rely on the precise shape of the 
linear power spectrum.

The upshot of our discussion is that in the $\Lambda$CDM cosmology
the  
soft NLO corrections are numerically suppressed and
can be neglected for the purposes of this paper. 
Their resummation for the matter power spectrum in real space was performed in
\cite{Blas:2016sfa}. 
We leave the analysis of the modifications due to redshift space and bias
for future work. 
We point out that, albeit small, these corrections are necessary for
a robust estimation of the shift of the BAO peak

We now focus on contributions with one hard loop and maximal IR
enhancement. These contributions scale as
\be
\label{eq:NLOh}
\sigma^2_h\times (\sigma^2_S\times 1/\ve^2)^{L_s}
\ee 
and their resummation proceeds in a straightforward manner along the
lines of Ref.~\cite{Blas:2016sfa}. 
The key observation is that due to Eq.~\eqref{eq:GnmindMain}
the redshift space vertices have the same 
factorization property as the real space vertices, see Eq.~\eqref{vertdress0}.
Thus, dressing hard-loop diagrams with soft loops results in 
the simple replacement of the wiggly power spectrum appearing in
propagators and vertices 
with its resummed version.
For instance, the IR-resummed matter power spectrum at NLO reads
\be
\label{eq:PhNLO}
P^{(s)\,\text{IR res,LO+NLO}}_{mm}\! =g^2K_1^2
\big[\bar P_{nw}+(1+g^2\S^{(s)})e^{-g^2\S^{(s)}}\bar P_w\big]
+P^{(s)\,1-loop}_{mm}\big[\bar P_{nw}+e^{-g^2\S^{(s)}}\bar P_w\big]\,,
\ee
where $P^{(s)\,1-loop}_{mm}$ is the one-loop contribution
understood as a functional of the linear power
spectrum\footnote{Formally, $P_{mm}^{(s)\,1-loop}$ should contain
  only the hard part of the loop. However, it is convenient to extend
  it to include soft momenta. This introduces a difference of order of
soft NLO corrections which, as we argued, are numerically small.}.
The above formula has a simple meaning: one has to use the leading
order IR-resummed linear power spectrum as an input in the 1-loop
calculation and correct the tree-level result 
in order to avoid double-counting. 
We emphasize that Eq.~\eqref{eq:PhNLO} is not a phenomenological model
but an outcome of the rigorous resummation of IR-enhanced corrections
at order \eqref{eq:NLOh}. 
The result \eqref{eq:PhNLO} can be easily generalized to higher-order statistics,
i.e.
for an arbitrary $n-$point function one obtains
\be
\label{finalNLOhCn}
\mathfrak{C}_{n}^{(s)\,\text{IR res,LO+NLO}}
\!=\mathfrak{C}_n^{(s)\,tree}\big[\bar P_{nw}+(1+g^2\S^{(s)})e^{-g^2\S^{(s)}}
\bar P_w\big]
+\mathfrak{C}_{n}^{(s)\,1-loop}\big[\bar P_{nw}+e^{-g^2\S^{(s)}}\bar P_w\big]\,.
\ee
Further, 
it is possible to include higher order hard loop corrections, i.e. to resum the 
graphs that scale as $(\sigma^2_h)^2\times (\sigma^2_S\times 1/\ve^2)^{L_s}$. 
For the power spectrum the net result reads
\be
\begin{split}
\label{eq:PhNNLO}
& P^{(s),\text{IR res, LO+NLO}+\text{NNLO}}_{mm} =
g^2K_1^2\bigg[\bar
P_{nw}+\bigg(1+g^2\S^{(s)}+\frac{1}{2}\big(g^2\S^{(s)}\big)^2\bigg)
e^{-g^2\S^{(s)}}\bar P_w\bigg]\\
&+P_{mm}^{(s)\,1-loop}\big[\bar
P_{nw}+(1+g^2\S^{(s)})e^{-g^2\S^{(s)}}\bar P_w\big]
+P_{mm}^{(s)\,2-loop}[\bar P_{nw}+e^{-g^2\S^{(s)}}\bar P_w]
\,.
\end{split}
\ee
Generalization to other correlation functions and higher hard-loop
orders is straightforward.

\section{Bias}
\label{sec:bias}

Bias is the relation between the density of observed tracers (e.g. galaxies, halos, etc.)
and the density of the underlying matter field 
\cite{Kaiser:1984sw,McDonald:2009dh,Baldauf:2011bh,Senatore:2014eva,
Mirbabayi:2014zca,Angulo:2015eqa}, see
\cite{Desjacques:2016bnm} for a recent comprehensive review.
This relation can be written involving 
the matter density field at the initial (Lagrangian biasing) 
or final (Eulerian biasing) time slice. 
As TSPT is formulated in terms of Eulerian fields at a finite time slice,
in what follows we adopt the Eulerian biasing scheme.
As long as perturbative treatment is valid, it is possible to 
describe \textit{deterministic} bias
as a local in time and space operator expansion 
\cite{Mirbabayi:2014zca,Desjacques:2016bnm},
\be 
\begin{split}
\delta^{(r)}_h(\tau,\x)=\sum_{n} \sum_{\mathcal{O}^{(n)}} 
b_{\mathcal{O}^{(n)}}(\tau)\mathcal{O}^{(n)}(\tau,\x)
\end{split}
\ee
where $\delta^{(r)}_{h}$ stands for the density contrast of biased
tracers in real space and
$\mathcal{O}^{(n)}$ are operators constructed out of the density field
to the $n$'th power, i.e. $\mathcal{O}^{(n)}\sim O(\delta^n)$. The
coefficients 
$b_{\mathcal{O}^{(n)}}$ are called bias parameters; in general, they
are functions of time.
The first sum runs over orders in perturbation theory,
and the second sum 
runs over all independent operators at a given order.
Note that in general the bias expansion should also include stochastic
(noise) contributions,  
generated by small-scale fluctuations that are uncorrelated with the
long-wavelength  
density field. 
The formal inclusion of stochastic terms into TSPT is straightforward. 
However, we defer the detailed treatment of these contributions for two reasons.
First, the effect of stochastic bias is expected  
to be negligibly 
small 
at the BAO scales.
Second, noise terms clearly have a UV origin and thus should be
treated on the same footing  
as the UV counterterms, which are left beyond the scope of this paper.

Due to the equivalence principle, the density of tracers cannot depend
on the value of the Newtonian potential and its first
derivatives. Thus, the operators $\mathcal{O}^{(n)}$ must be
constructed using the tidal tensor 
\be
\label{eq:pi1}
\Pi^{[1]}_{ij}= \d_i\d_j\Phi\,
\ee
and its derivatives. Here 
$\Phi$ is the suitably normalized gravitational potential
related to the standard Newtonian potential $\phi$
via $\Phi \equiv {2\phi}/{(3\Omega_m\HH^2)} $.
A convenient basis is constructed as follows. One introduces a
sequence of tensors,
\bseq 
\label{eq:biasbasis}
\begin{align}
\label{eq:biasbasis1}
& \Pi^{[1]}_{ij}=\d_i\d_j\Phi\,,\\
& \Pi^{[n]}_{ij}=\frac{1}{(n-1)!}\left[\frac{1}{f\HH}\frac{D}{D\tau}
\Pi^{[n-1]}_{ij}-(n-1)\Pi^{[n-1]}_{ij}
\right]\,,
\label{eq:biasbasis2}
\end{align}
\eseq
where ${D}/{D\tau}$ is the convective derivative,
\be 
\label{eq:partder}
\frac{D}{D\tau}=\frac{\d}{\d \tau}+v^i\d_i=\frac{\d}{\d \tau}-f\HH \frac{\d_i\T^{(r)}}{\Delta}\d_i\,,
\ee
and in passing to the last equality we used that 
only the longitudinal component of the peculiar velocity is present in
perturbation theory. The use of convective derivative accounts for the
fact that the evolution of tracers is determined by the physical
conditions along the fluid flow \cite{Desjacques:2016bnm}. The second
term in (\ref{eq:biasbasis2}) is adjusted to subtract
$O(\delta^{n-1})$ contributions, so that $\Pi_{ij}^{[n]}$ has
homogeneous dependence on $\delta$ of order $O(\delta^{n})$.
Despite the fact that the tensors \eqref{eq:biasbasis2} 
contain partial time derivatives,
it is always possible to eliminate them by using the equations 
of motion for matter. 

The bias operators at $n$'th order are given by all possible contractions
of the tensors \eqref{eq:biasbasis} with total order $n$, e.g.
\be
\label{eq:contr}
\begin{split}
& \text{1st} \quad \text{Tr}[\Pi^{[1]}]\,,\\
& \text{2nd} \quad \text{Tr}[(\Pi^{[1]})^2]\,,\quad (\text{Tr}[\Pi^{[1]}])^2\,,\\
& \text{3rd} \quad \text{Tr}[(\Pi^{[1]})^3]\,,\quad \text{Tr}[(\Pi^{[1]})^2]\text{Tr}[\Pi^{[1]}]\,,\quad (\text{Tr}[\Pi^{[1]}])^3\,,\quad
\text{Tr}[\Pi^{[1]}]\text{Tr}[\Pi^{[2]}]\,,
\\
& ...
\end{split}
\ee
Note that 
the terms $\text{Tr}[\Pi^{[n]}]$ are excluded at the $n$'th order (except $n=1$) 
as they are degenerate with other operators in the basis.
The basis \eqref{eq:biasbasis} does not contain
higher-derivative terms.  
In principle, they can always 
be added by applying derivatives to the tidal tensor and making all
possible contractions   
analogous to \eqref{eq:contr}.
As the bias expansion preserves the equivalence principle 
\cite{Mirbabayi:2014zca}, it 
contains no IR poles. We give explicit expressions for a few first
bias operators relevant for one-loop computations in
Appendix \ref{app:C}.

All in all, the bias expansion takes the following form:
\be 
\label{eq:biasdens}
\delta^{(r)}_{h}(\tau,\k)=
\sum_{n=1}^\infty\frac{1}{n!} \int[dq]^n\delta^{(3)}(\k-\q_{1...n})\,
\tilde{M}_n(\tau;\q_1,...,\q_n)\prod_{i=1}^n \delta^{(r)}(\tau,\q_i)\,.
\ee
In order to incorporate bias into TSPT, it is convenient to 
rewrite \eqref{eq:biasdens} in terms of the velocity divergence 
field. 
Using that the matter density field can be expressed in perturbation 
theory through the velocity divergence via \eqref{eq:psi1},
the relation \eqref{eq:biasdens} can
be rearranged in the desired form:
\be 
\label{eq:biasveloc}
\delta^{(r)}_{h}(\tau,\k)=\sum_{n=1}^\infty\frac{1}{n!}
 \int[dq]^n\delta^{(3)}(\k-\q_{1...n})\,
M^{(r)}_n(\tau;\q_1,...,\q_n)\prod_{i=1}^n\T^{(r)}(\tau,\q_i)\,.
\ee
The kernels $M_n^{(r)}$ relevant for the 1-loop calculation are given
in Appendix \ref{app:C}. 
In principle, the $M_n^{(r)}$ kernels may have arbitrary 
time-dependence, that is why we will not treat them as functions of the coupling 
constant $g$
in the TSPT perturbative expansion.
The bias parameters are expected to
evolve slowly, with the rate comparable to that of the growth of matter.
Note that the bias parameters are subject to UV 
renormalization \cite{Mirbabayi:2014zca,Assassi:2014fva}. 
This issue will be addressed elsewhere.

The tracers' velocity field can, in principle, also be biased, so
within the validity of perturbation theory
it will be expressed as a power series in $\T^{(r)}$,
\be 
\label{eq:bvv}
\T^{(r)}_{h}(\tau,\k)=\sum_{n=1}^\infty\frac{1}{n!}
\int[dq]^n\delta^{(3)}(\k-\q_{1...n})\, 
V_n(\eta;\q_1,...,\q_n)\prod_{i=1}^n\T^{(r)}(\tau,\q_i)\,.
\ee
However, as long as the effect of relative velocities between
different matter components can be neglected, the velocity bias will
be absent at the lowest order in spatial derivatives. The difference
of $\T_h^{(r)}$ from $\T^{(r)}$ will appear only at higher
derivatives. For example,
\be
 V_1(\tau;\k)=1+b_{\nabla^2\textbf{v}}(\tau)k^2\,.
\ee
We conclude that the velocity bias has the same order in the derivative expansion 
as the UV counterterms, and thus its treatment goes 
beyond the scope of this paper.
In what follows we will neglect all the effects related to velocity bias. 

\subsection{IR resummation for biased tracers in real and redshift space}

The goal of this section is to incorporate bias into TSPT 
and perform IR resummation for correlation functions of biased tracers
in real and redshift space. 

We start by discussing real space. 
In this case Eq.~\eqref{eq:biasveloc}
describes the density of biased tracers 
as a composite operator analogous to the 
density of matter. 
Thus, we can use the technique developed in 
\cite{Blas:2015qsi} by simply using the kernels $M_n$ instead of $K_n$
in the relevant Feynman diagrams. 
Since the bias vertices are IR safe, they do not produce additional contributions to be resummed.
IR~resummation thus goes in full analogy with the IR resummation 
of the density correlators in real space \cite{Blas:2016sfa}.
The result of this procedure in real space has been anticipated in
Ref.~\cite{Desjacques:2016bnm}: 
one simply has to substitute the linear power spectrum by its
IR-resummed version in all expressions for the correlation functions
of biased tracers.

Generalization to the case of redshift space is straightforward. 
The redshift coordinate of the tracer is related to the real-space one by 
means of the tracer's velocity ${\bm v}^{(r)}_{h}(\x)$,
\be
\label{eq:rsdb}
\s_h=\x+\hat{\textbf{z}}\frac{v^{(r)}_{h,z}(\x)}{\HH}\,.
\ee
As pointed out before, we do not consider velocity bias in this paper, thus 
in the rest of it we will assume that 
tracers are comoving with matter and simply replace
\be 
{\bm v}^{(r)}_{h}\to{\bm v}^{(r)}
\ee
in Eq.~\eqref{eq:rsdb}.
In order to transform the bias kernels $M^{(r)}_n$ into redshift space
we use the same trick of introducing a fictitious 
1D flow described in Sec.~\ref{sec:1d}.
In this way we obtain the same equations of motion as \eqref{eq:finaleoms}, but with 
$\delta^{(s)}$ replaced by $\delta^{(s)}_h$.
At the next step we use these equations to derive the kernels 
relating the tracer density field with the redshift space velocity~$\T^{(s)}$. We obtain
\be
\label{eq:dhbnkern}
\delta^{(s)}_{h}(\F;\tau,\k)=\sum_{n=1}^\infty\frac{1}{n!} \int [dq]^n \, M_n^{(s)}(\F;\q_1,...,\q_n)\T^{(s)}_{\q_1}...\T^{(s)}_{\q_n}\,,
\ee
with $M^{(s)}_n$'s satisfying the same equations of motion
as \eqref{eq:Meqs} with an obvious change in the initial conditions,
\be
M^{(s)}_n\Big|_{\F=0}=M^{(r)}_n\,. 
\ee
This procedure allows us to unambiguously map the real space bias 
parameters to redshift space ones. 
It should be noted that some tracers 
(e.g. Ly$\a$ forest, 21 cm intensity) may have additional biases 
in redshift space \cite{Seljak:2012tp,Desjacques:2016bnm,Desjacques:2018pfv}.
In this case one has to supplement \eqref{eq:dhbnkern} with relevant
extra bias operators.  

An immediate consequence of the above construction
is that the kernels $M^{(s)}_n$ are IR-safe and 
are not functionals of the initial power spectra. Thus, they do not
receive any IR enhancement which, as before, affects solely the
vertices 
$\bar \G^{w\,(s)}_n$.
The diagrams involving these vertices are resummed,
as has been shown in the 
previous sections.
The net result at leading order is that one has to use the ``dressed''
power spectrum
\be
\label{eq:dressedLO}
\bar P_{nw}+e^{-g^2\S^{(s)}}\bar P_w\;,
\ee
instead of the linear one in all tree-level calculations. 
At first order in hard loops one has to use the power spectrum 
\eqref{eq:dressedLO} in the loop diagrams and correct the tree-level
result for double-counting. 
At higher loop order this procedure iterates, as illustrated by
Eq.~\eqref{eq:PhNNLO}.

\section{Practical implementation and comparison with other me\-thods}
\label{sec:pract}

In this section we formulate the practical prescription to evaluate 
the IR-resummed power spectra and bispectra. We then compare our results 
with other analytic approaches.
While our results have been derived within the TSPT 
framework, they can be easily reformulated in the 
language of the standard perturbation theory \cite{Bernardeau:2001qr}, 
which may be convenient for implementation within existing numerical
codes, e.g. \texttt{FAST-PT} \cite{McEwen:2016fjn,Fang:2016wcf} or \texttt{FnFast} \cite{Bertolini:2015fya}. 

\subsection{The power spectrum and bispectrum at leading order}

In order to simplify notations in this section we drop 
the explicit time dependence of the power spectra and use the shorthand
\be
\label{eq:notcomp}
P(k)\equiv D^2(z)\bar{P}(k)\,.
\ee
Having decomposed the linear power spectrum into wiggly and smooth parts,
e.g. using one of the methods described in \cite{Blas:2016sfa}, 
we have to evaluate the derivative operator acting on the wiggly part.
Since $P_w$ is a function oscillating with the period $k_{osc}=h/(110\,$Mpc), we have 
\be
\label{eq:diffoper}
\nabla_{\a_1}\cdots\nabla_{\a_{2n}} P_w(k) = (-1)^n \frac{\hat
  k_{\a_1}\cdots \hat k_{\a_{2n}}}{k_{osc}^{2n}}  
P_w(k)\big(1+O(\varepsilon)\big)\;, 
\ee
where $\varepsilon\simeq k_{eq}/k$ is the small expansion parameter
controlling the IR enhancement and
$\hat \k=\k/k$.  
Then the action of the operator $\S^{(s)}$ (Eq.~(\ref{Slead})) at
leading order in $\ve$ reads,
\be 
g^2\S^{(s)}[P_w(k)]=\mathcal{P}_{ab}\mathcal{P}_{cd}k^ak^c
\int_{q< k_S}[dq]\, P_{nw}(q)\,\frac{q^b q^d}{q^4}\bigg[1-
\cos\bigg( \frac{(\q\cdot\hat\k)}{k_{osc}}\bigg)\bigg]P_w(k)\;,
\ee
which reduces to a $\k$-dependent multiplicative factor.
Evaluating the integral we obtain,
\be
\label{Sleading}
g^2 \S^{(s)}[P_w(k)] = k^2\big[\big(1+f\mu^2(2+f)\big)\Sigma^2+f^2\m^2
(\m^2-1)\delta\Sigma^2\big] P_w(k) \times \big(1+O(\varepsilon)\big)\,,
\ee
with $\mu \equiv k_z/k$ and 
\bseq
\label{Sigmadefk}
\begin{align}
\label{Sigmanorm}
  &\Sigma^2 \equiv \frac{4\pi}{3} \int_{0}^{k_S}dq  \, P_{nw}(q)\bigg[1-j_0\left(\frac{q}{k_{osc}}\right)+2j_2\left(\frac{q}{k_{osc}}\right) \bigg]\,,  \\
  \label{deltaSigma}
   & \delta\Sigma^2 \equiv 4\pi\int_{0}^{k_S}dq  \, P_{nw}(q)\, j_2\left(\frac{q}{k_{osc}}\right)  \,.
\end{align}
\eseq
Here $j_n$ are spherical Bessel functions and $k_S$ is the separation scale 
of long and short modes in the loop integrals. 
Thus, the LO IR-resummed power spectrum of biased tracers (say, halos) 
in redshift space is given by
\be 
\label{eq:finaltree}
\begin{split}
P_{hh}^{(s)\,\text{IR res,LO}}(k,\mu)=& (b_1+f\mu^2 )^2\Big(
P_{nw}(k) +e^{-k^2\Sigma^2_{\text{tot}}(\mu;k_S)} P_w(k)\Big)\,,
\end{split}
\ee
where 
\be 
\label{eq:sigmaeval}
\begin{split}
\Sigma^2_{\text{tot}}(\mu;k_S)~\equiv ~ &\big(1+f\mu^2(2+f)\big)\Sigma^2
+f^2\m^2 (\m^2-1)\delta\Sigma^2\,.
\end{split}
\ee
Note that the damping factor $\Sigma^2$ has already appeared as a
result of IR resummation in real space 
\cite{Baldauf:2015xfa,Blas:2016sfa}, 
while $\delta\Sigma^2$ is a new contribution.
The form of the first term in \eqref{eq:sigmaeval} 
has a simple physical meaning: one has to 
keep the perpendicular (real space) rms displacement 
$\Sigma$ 
of soft modes intact
while multiplying the rms displacements along the 
line-of-sight by a factor $(1+f)$,
as prescribed by the Kaiser formula, i.e.
\be
\label{dampKaiser}
k^2_{\parallel}(1+f)^2\Sigma^2 + k^2_{\perp}\Sigma^2 =\Sigma^2(1+f\mu^2(2+f))\,,
\ee 
where 
$k_{\parallel}=k_z$, $k_{\perp}=\sqrt{k^2-k^2_z}$.
Note that the contribution proportional to $\delta\Sigma^2$ is
negative and thus it somewhat reduces the BAO damping compared to the
simple formula (\ref{dampKaiser}).
The form of exponential damping \eqref{eq:sigmaeval} 
does not depend on 
bias parameters.
This is consistent with physical intuition, 
as the degradation of the BAO feature
in the statistics of biased tracers
is caused by displacements of underlying matter,
in agreement with the equivalence principle~\cite{Desjacques:2016bnm}.

\begin{figure}
\begin{center}
\includegraphics[width=.49\textwidth]{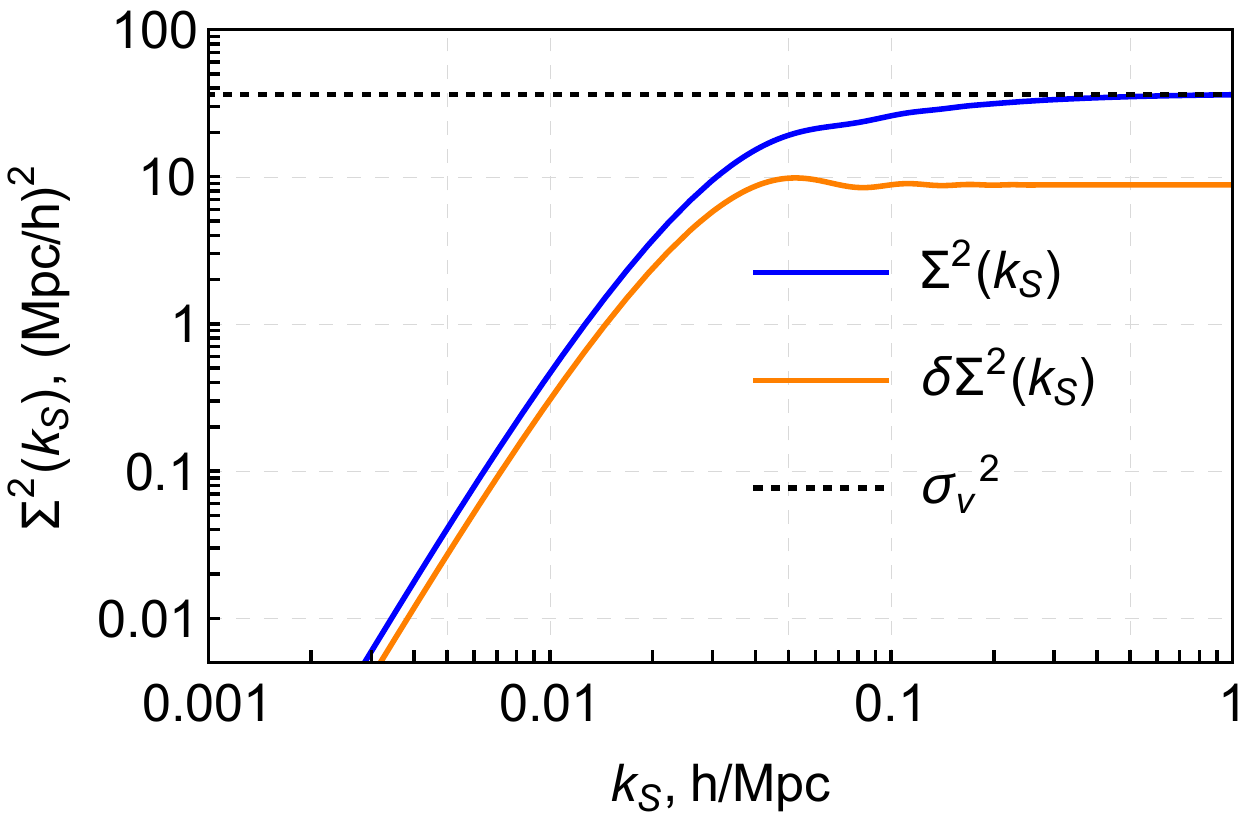}
\includegraphics[width=.49\textwidth]{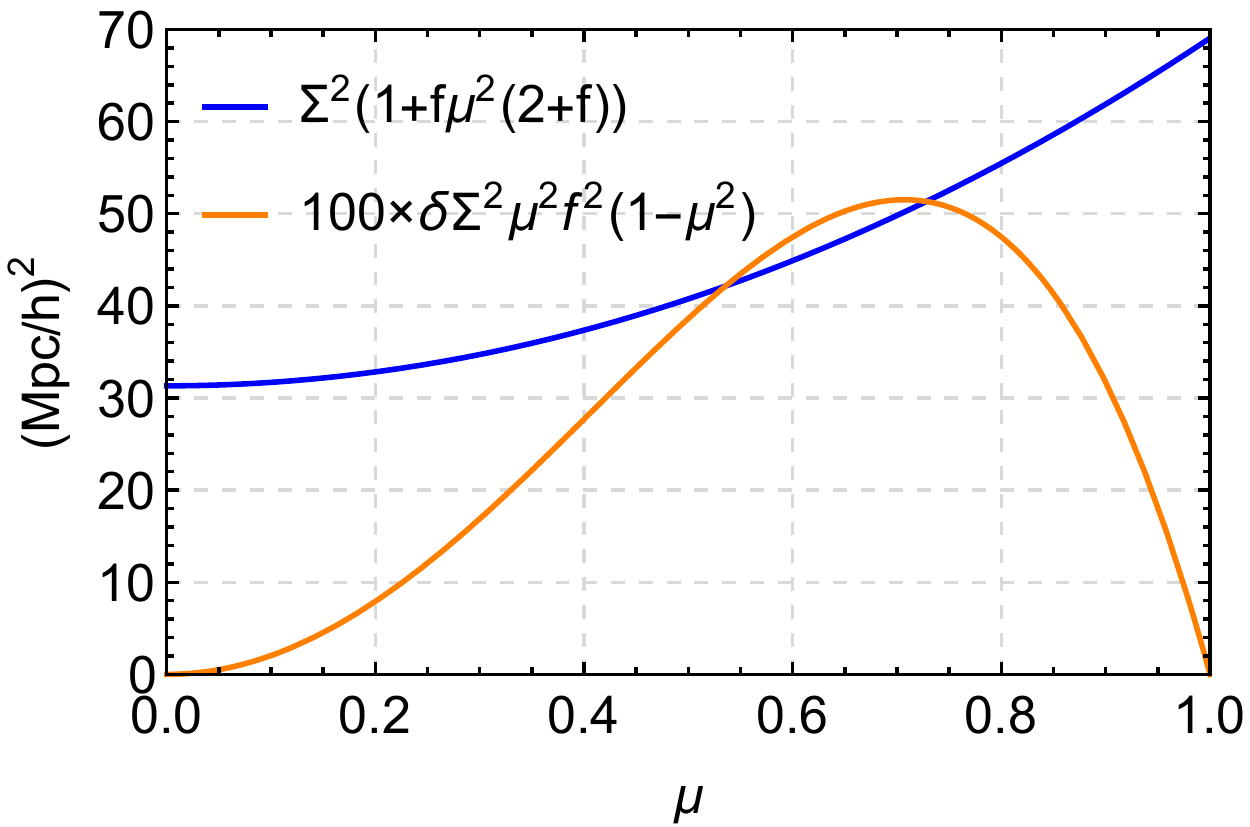}
\caption{\label{fig:SigmaL} 
Left panel: the dependence of the BAO damping factors 
$\Sigma^2$ and $\delta\Sigma^2$ on the separation scale $k_S$ at redshift zero 
(in the cosmological model of \cite{GilMarin:2012nb}, $f=0.483$). 
Right panel: the dependence of two contributions to the damping factor 
on the angle $\mu$ between the Fourier wavevector and the
line-of-sight; $k_S$ is fixed to $0.2\,h/$Mpc.
}
\end{center}
\end{figure}

We plot the dependence of the damping factors $\Sigma^2$
and $\delta\Sigma^2$ on the choice of $k_S$ in the left panel of
Fig. \ref{fig:SigmaL}.
At small $k_S\ll k_{osc}$ the damping functions 
have the following asymptotic behavior:
\be 
\begin{split}
\Sigma^2(k_S)\to \frac{2\pi}{5}\int_0^{k_S}dq\,\frac{q^2}{k_{osc}^2}P_{nw}(q)\,,\quad
\delta\Sigma^2(k_S)\to \frac{4\pi}{15}\int_0^{k_S}dq\,\frac{q^2}{k_{osc}^2}P_{nw}(q)\,.
\end{split}
\ee
The integral in $\delta\Sigma^2(k_S)$ is cut off at $k_{osc}$ by the
Bessel function  
and becomes a constant equal to
\be 
\delta\Sigma^2(k_{osc})\simeq \frac{4\pi}{15 k_{osc}^2}\int_{q\lesssim
  k_{osc}} dq\,q^2 P_{nw}(q)\,, 
\ee
whereas $\Sigma^2(k_S)$ keeps
growing up to $q\sim 0.2\,h/$Mpc where
it approaches its asymptotic value $\sigma_v^2\equiv 4\pi\int dq\,
P_{nw}(q)/3$. 
The integral in $\Sigma^2(k_S)$ receives the dominant contribution
from $q\gtrsim k_{eq}$ 
and is significantly bigger than $\delta\Sigma^2(k_S)$.
This numerical hierarchy is due to the specific shape of the $\L$CDM
power spectrum which is strongly suppressed at $q\lesssim k_{osc}$; if
the power spectrum peaked at momenta smaller than $k_{osc}$,  
the damping factors 
$\Sigma^2(k_S)$  and $\delta\Sigma^2(k_S)$
would have comparable magnitudes.

As discussed in Sec.~\ref{sec:pc}, the scale $k_S$ should be chosen
high enough to include the contributions of all relevant soft
modes. At the same time, it should be smaller than the momentum $k$ of
interest. In the numerical calculations below we will vary $k_S$ in
the range $(0.05\div 0.2)\,h/$Mpc. The dependence of the final result on
the precise choice of $k_S$ should be considered as a measure of 
theoretical uncertainty.   

In order to understand the effect of damping in redshift space
let us rewrite the expression \eqref{eq:sigmaeval} 
as $\Sigma^2_{\text{tot}}=\Sigma_1^2+\Sigma_2^2$
with
\be
\begin{split}
&\Sigma_1^2(\mu;k_S)\equiv (1+f\mu^2(2+f))\Sigma^2\,,\\
&\Sigma_2^2(\mu;k_S)\equiv f^2\m^2(\m^2-1)\delta\Sigma^2.
\end{split}
\ee
$\Sigma_1^2$ and $\Sigma^2_2$ as functions
of $\m$ are plotted in the right panel of Fig.~\ref{fig:SigmaL}. 
We fix $k_S=0.2 \,h$/Mpc.
For visualization purposes 
we multiply $\Sigma_2^2$ by factor $100$ and flip its sign.
We observe that $\Sigma_2^2$ is much smaller than 
$\Sigma^2_1$ for all wavevector directions. 
Its relative effect somewhat 
increases at high redshifts where the suppression by factor $f^2$
is mitigated.
As for the $\Sigma^2_1$ contribution, we see that it grows monotonically
with $\m$ and thus, as expected, the BAO signal is more suppressed
for the wavevectors aligned with the line-of-sight.

The IR-resummed bispectrum at leading order is easily obtained from
the general formula (\ref{Cncomp}). Making use of the well-known SPT
result we get,
\be
\label{BLO}
\begin{split}
&  B_{hhh}^{(s)\,\text{IR res,LO}}(\k_1,\k_2,\k_3)=
2\!\!\sum_{1\leq i<j\leq 3}\!\!
(b_1+f\mu^2_i)(b_1+f\mu^2_j)
Z_2(\k_i,\k_j)\\
&\times
\left(P_{nw}(k_j)P_{nw}(k_i)+e^{-k_j^2\Sigma^2_{\text{tot}}(\m_j)} P_w(k_j)P_{nw}(k_i) +
e^{-k_i^2\Sigma^2_{\text{tot}}(\m_i)} P_w(k_i)P_{nw}(k_j)  \right)\,,
\end{split} 
\ee
where 
$\m_j\equiv \hat{\k}_j\cdot\hat{\bf z}$.
As everywhere else in the paper, we have retained only linear terms in
$P_w$. The expression for the SPT kernel $Z_2$ is given in 
Appendix \ref{app:C}. 

\subsection{The power spectrum and bispectrum at next-to-leading order}
\label{sec:NLOeval}

At NLO the IR-resummed power spectrum can be written as
\be
\label{eq:rsd1loopps}
\begin{split} 
P_{hh}^{(s)\,\text{IR res,LO+NLO}}(k,\mu)=& (b_1+f\mu^2)^2
\left(P_{nw}(k)+(1+k^2\Sigma^2_{\text{tot}}(\mu))
e^{-k^2\Sigma^2_{\text{tot}}(\mu)}P_{w}(k)\right)\\
&+P_{hh}^{(s)\,1-loop}\big[P_{nw}+e^{-k^2\Sigma^2_{\text{tot}}(\mu)}P_{w}\big]\,.
\end{split}
\ee
Note that the power spectrum that must be used as an input in 
the loop 
contribution is anisotropic due to the angular dependence of the
damping factor. This complicates evaluation of the loop integral, as it
prevents from using the standard procedure of integrating over the
azimuthal angle 
and reducing
$P^{(s)\,1-loop}_{hh}$ to a finite series in $\mu^2$.
To cast \eqref{eq:rsd1loopps} 
in a more convenient
form, we isolate the wiggly terms,
\be 
\label{eq:deltaP}
\begin{split}
P_{hh,\,w}^{(s)\,1-loop}=~& 6Z_1(\k) e^{-k^2\Sigma^2_{\text{tot}}(\mu)}P_{w}(k)
\int d^3p\,Z_3(\k,\p,-\p)\,P_{nw}(p)\\\,
&+6Z_1(\k)P_{nw}(k)
\int d^3p\,Z_3(\k,\p,-\p)\,P_w(p) e^{-p^2\Sigma^2_{\text{tot}}(\mu_{\p})}\\
&+4\int d^3p\,\big(Z_2(\p,\k-\p)\big)^2P_{nw}(|\k-\p|)P_w(p)
e^{-p^2\Sigma^2_{\text{tot}}(\mu_\p)}\;,
\end{split}
\ee
where $\m_\p\equiv (\hat{\p}\cdot \hat{\z})$. 
The first term contains an integral of the isotropic smooth power spectrum and its evaluation does not pose any problem.
The second term is an integral of a quickly oscillating function and
is exponentially suppressed within our power counting. Indeed,
approximating $P_w$ with the sine we have\footnote{The kernel
  $Z_3(\k,\p,-\p)$ is not regular at $\p\to 0$. However, this
  singularity does not contribute because $P_w(p)$ vanishes at the origin.},
\be 
\int[dp]\sin(p/k_{osc})f_{\text{smooth}}(p/k)\sim 
e^{-k/k_{osc}}\sim e^{-1/\ve}\,.
\ee
Similarly, the hard part of the third integral, i.e. the contribution
from $|\p-\k|>k_S$, is also exponentially suppressed. On the other
hand, in the vicinity $|\p-\k|<k_S$ the damping factor can be
approximated as
\be
e^{-p^2\Sigma^2_{\rm tot}(\mu_\p)}=e^{-k^2\Sigma^2_{\rm tot}(\mu)}
\big(1+O(\ve)\big)\;.
\ee 
The difference pertains to NLO soft corrections which we neglect in
this paper. We conclude that the damping factor can be pulled out of
the loop integrals without changing the order of approximation in our
power counting.  
This allows us to rewrite the IR-resummed power spectrum
in the form involving only integration over the isotropic
initial power 
spectrum,
\be
\label{eq:rsd1loopps2}
\begin{split} 
P_{hh}^{(s)\,\text{IR res,LO+NLO}}(k,\mu)=& (b_1+f\mu^2)^2
\left(P_{nw}(k)+(1+k^2\Sigma^2_{\text{tot}}(\m))
e^{-k^2\Sigma^2_{\text{tot}}(\m)}P_{w}(k)\right)\\
&+P_{hh}^{(s)\,1-loop}[P_{nw}]
+e^{-k^2\Sigma^2_{\text{tot}}(\m)}P^{(s)\,1-loop}_{hh,\,w}\,,
\end{split}
\ee
where $P_{hh}^{(s)\,1-loop}[P_{nw}]$ is evaluated on 
the smooth power spectrum only and
\be
\label{eq:rsdfinal1loop}
\begin{split} 
P^{(s)\,1-loop}_{hh,\,w}=\, 
&6P_w(k)Z_1(\k)\int d^3p\,Z_3(\p,-\p,\k) P_{nw}(p)\\
&+4\int d^3p\,\big(Z_2(\p,\k-\p)\big)^2P_{nw}(|\k-\p|)
P_w(p)\;.
\end{split}
\ee
An expression similar to \eqref{eq:rsd1loopps2}
was obtained for the 1-loop IR resummed real-space power spectrum in
Ref.~\cite{Baldauf:2015xfa}.

For higher-point correlation functions 
the ``isotropisation'' of the 
IR resummed loop integrands is in general impossible. However, some
partial contributions to the total result may still be
simplified. Thus, in Appendix~\ref{app:D} we show that the 1-loop
bispectrum in redshift space can be written in the following form,   
\be
\label{Bs1loop}
\begin{split}
B^{(s)\,\text{IR res,LO+NLO}}\!=&B^{(s)\,
  tree}\big[P_{nw}\!+\!(1\!+\!k^2\Sigma_{\rm
  tot}^2)e^{-k^2\Sigma_{tot}^2}P_w\big]
\!+\!B^{(s)\,1-loop}\big[P_{nw}\!+\!e^{-k^2\Sigma_{tot}^2}P_w\big]\\
\approx&\, B^{(s)\,
  tree}\big[P_{nw}\!+\!(1\!+\!k^2\Sigma_{\rm
  tot}^2)e^{-k^2\Sigma_{tot}^2}P_w\big] + B^{(s)\,1-loop}\big[P_{nw}\big]\\
&+\tilde B^{(s)}_{411,w}+\tilde B^{(s)}_{321-I,w}+\tilde B^{(s)}_{321-II,w}+
\tilde B^{(s)}_{222,w}\;,
\end{split}
\ee
where all terms except $\tilde B^{(s)}_{222,w}$ involve isotropic
power spectra inside the momentum integrals. The `approximately equal'
sign between the first and second lines means that the two expressions
are equal up to NLO soft correction. The formulae for 
$\tilde B^{(s)}_{411,w}$,
$\tilde B^{(s)}_{321-I,w}$,
$\tilde B^{(s)}_{321-II,w}$ and 
$\tilde B^{(s)}_{222,w}$ are given in Eqs.~(\ref{tildeBsw}) of
Appendix~\ref{app:D}.

\subsection{Comparison with other approaches}

Let us compare our results to other methods.
At the phenomenological level the suppression of the BAO feature
in redshift space 
is well described by a $\mu$-dependent exponential damping acting on the 
wiggly part of the linear power spectrum. 
The aim of analytic approaches is to derive this 
result from first principles and consistently generalize it
to higher orders in perturbation theory, 
where the effects beyond this simple damping are relevant.
In this section we will focus on a few methods 
to describe the BAO peak which are
most common in the literature.

The simplest model describing the suppression of the BAO peak 
in redshift space is given by
\be
\label{eq:dampmodel}
\begin{split}
 P^{(s)}(k,\mu)=
(b_1+f\mu^2)^2\big(P_{nw}(k)+e^{-k^2A^2(1+f\mu^2(2+f))}P_{w}(k)\big)\,,
\end{split}
\ee
with two possible choices of the damping factor $A$:
\begin{subequations}
\label{eq:dampex}
\begin{align}
\label{ex1}
& A^2=\sigma^2_v\equiv \frac{4\pi}{3}\int_0^\infty dq P(q)\,,&&\text{following \cite{Crocce:2005xy,Matsubara:2007wj}}\,,\\
\label{ex2}
& A^2= 
\Sigma^2_{\infty} \equiv \frac{4\pi}{3} \int_{0}^{\infty}dq  P(q)\bigg[1-j_0\left(\frac{q}{k_{osc}}\right)+2j_2\left(\frac{q}{k_{osc}}\right) \bigg]
\,, 
&& \text{following \cite{Eisenstein:2006nj,Seo:2007ns}}\,.
\end{align}
\end{subequations}
The model \eqref{eq:dampmodel} 
explicitly takes into account the fact that bulk flows significantly 
affect only the wiggly part of the power spectrum.
On the other hand, the damping factors in \eqref{eq:dampex} 
are different from ours, given in Eq.~\eqref{eq:sigmaeval}.
The rms velocity displacement in \eqref{ex1} 
is enhanced at very low $q$, whereas the modes with $q\ll k_{osc}$
should not affect the BAO feature 
as dictated by the equivalence principle.
On the other hand, 
the integrand \eqref{ex2}, as well as in our expressions
(\ref{Sigmadefk}),  
tends to zero in the IR, and thus is consistent with the physical
expectations. 
The damping factor \eqref{ex2} has the structure similar to our
\eqref{Sigmanorm} 
but is evaluated at $k_S=\infty$.
This choice of $k_S$ contradicts the logic 
that only long-wavelength modes 
should be resummed and should be contrasted with our expressions
which explicitly reflect this argument.
Indeed, only for the soft modes with $ q \ll k$
the mode coupling affecting the BAO is enhanced.

As shown in Fig.~\ref{fig:SigmaL}, the numerical value 
of our damping factor appears to be quite close 
to both \eqref{ex1} and \eqref{ex2} for the $\L$CDM cosmology.
It was already pointed out in \cite{Baldauf:2015xfa,Blas:2016sfa} that if 
our universe had more power 
at large scales, $q\lesssim k_{osc}$, the damping factor \eqref{eq:sigmaeval}
would be notably different\footnote{
It would be interesting to understand if this 
can account for the discrepancy
between $\sigma^2_v$ and the actual damping factor found in simulations of 
a toy cosmological model with a bigger $k_{osc}$ 
in~\cite{Padmanabhan:2009yr}.
} from $\sigma^2_v$.
On the other hand, if there were more power at short scales,
using \eqref{ex2} one would significantly overdamp 
the BAO signal.
Last but not least, the models \eqref{eq:dampex} do not take into account 
the $\delta \Sigma^2$ contribution, which, albeit small in the $\L$CDM cosmology,
could be sizable if $k_{eq}$ were smaller than $k_{osc}$. 

The leading order IR-resummed power spectrum \eqref{eq:finaltree} 
coincides with the expression found in
Ref.~\cite{Nishimichi:2017gdq}. 
The approach used in \cite{Nishimichi:2017gdq} 
is related to the framework developed in
Refs.~\cite{Peloso:2016qdr,Noda:2017tfh}.
We point out that the accurate description of the BAO feature
requires including loop corrections and 
thus goes beyond the simple exponential damping prescribed by \eqref{eq:dampmodel}.

Our results for the power spectrum are consistent with those obtained 
within the effective field theory
of large scale structure 
\cite{Senatore:2014via,
Senatore:2014vja,
Lewandowski:2015ziq,
Perko:2016puo,
Vlah:2015zda,
delaBella:2017qjy,
Senatore:2017pbn}. 
Our expressions (\ref{eq:PhNLO}) and (\ref{eq:PhNNLO})
agree, up to higher order corrections\footnote{Note
that \cite{Lewandowski:2015ziq} essentially
applies the operator $e^{-g^2\S^{(s)}}$
to the whole power spectrum, including its smooth part.
This is equivalent to a
partial resummation of IR corrections to the smooth power spectrum.
These corrections are not enhanced and therefore their resummation is 
not legitimate within our power-counting
rules.
}, 
with those obtained in Ref.~\cite{Lewandowski:2015ziq}.
We emphasize that TSPT gives a simple diagrammatic 
description of IR~resummation and provides a 
tool to examine and extend the results found in 
Ref.~\cite{Lewandowski:2015ziq}. 
IR resummation in TSPT readily generalizes 
beyond the power spectrum and applies to any $n$-point correlation functions
with an arbitrary number of hard loops.
The power counting, outlined in this paper, allows one to go beyond next-to-leading order in a systematic way. 
In particular, TSPT allows one to systematically compute the subleading soft corrections relevant for the shift of the BAO peak.
We leave their detailed study for future work.

\section{Numerical results and comparison with N-body data}
\label{sec:numerics}

In this section we show the results for the 2-point correlation function and the 
power spectrum of matter in redshift space,  
although our analysis can be easily extended to biased tracers.
We will first discuss the 2-point correlation function, 
which allows us
to clearly illustrate the effect of IR resummation on the BAO feature
due to a better separation between the BAO peak and short scales.
Then we compare our predictions for the IR-resummed power 
spectrum at one loop against N-body data. 
To the best of our knowledge,
there are no publicly available data on the 2-point correlation function in 
redshift space.
That is why in this paper we limit the comparison to the power
spectrum, even though 
it is not optimal for the visualization of the BAO.

As common in redshift space analysis,
we will study Legendre multipoles of the power spectrum 
and the 2-point correlation function, defined via\footnote{
Recall that in our Fourier transform convention
$\xi(\x)~=~\int d^3k e^{i\k\x}P(\k)$.
}
\be
\begin{split}
& P_\ell(k)=\frac{2\ell +1}{2}\int_{-1}^1 L_\ell(\mu) P^{(s)}(k,\mu)d\mu \,,\\
 & \xi_\ell(r)= 4\pi  \,i^\ell\int P_\ell(k)j_\ell(kr) k^2 dk\,,
\end{split}
\ee
where $L_\ell$ is the Legendre polynomial of order $\ell$. We will focus on the monopole, quadrupole and hexadecapole moments ($\ell=0,2,4$), 
which fully
characterize the linear 
correlation function in redshift space.
 
We consider the cosmological model corresponding to the N-body 
simulations performed in \cite{GilMarin:2012nb}. The linear power
spectrum is produced  
with the Boltzmann code \texttt{CLASS} 
\cite{Blas:2011rf}
and then decomposed into 
the wiggly and non-wiggly components 
using the spline approximation of the broadband power spectrum \cite{Blas:2016sfa}.
The redshift space one-loop integrals
are evaluated using the FFTLog algorithm \cite{Hamilton:1999uv,Simonovic:2017mhp}. 
A similar technique is used to compute the correlation function multipoles
from those of the power spectrum.

\subsection{2-point correlation function: quantitative study}

\begin{figure}
\begin{center}
\includegraphics[width=.49\textwidth]{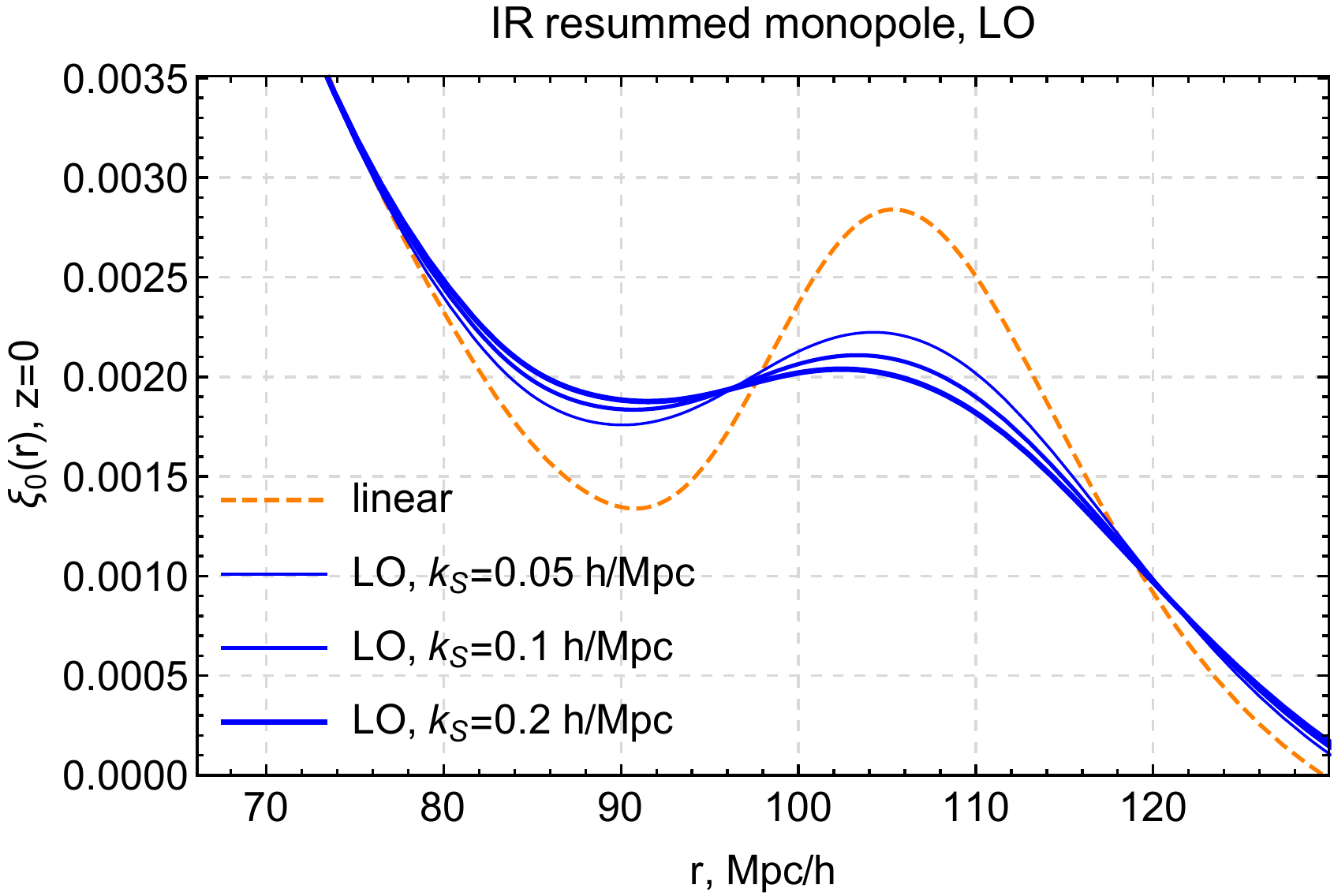}
\includegraphics[width=.49\textwidth]{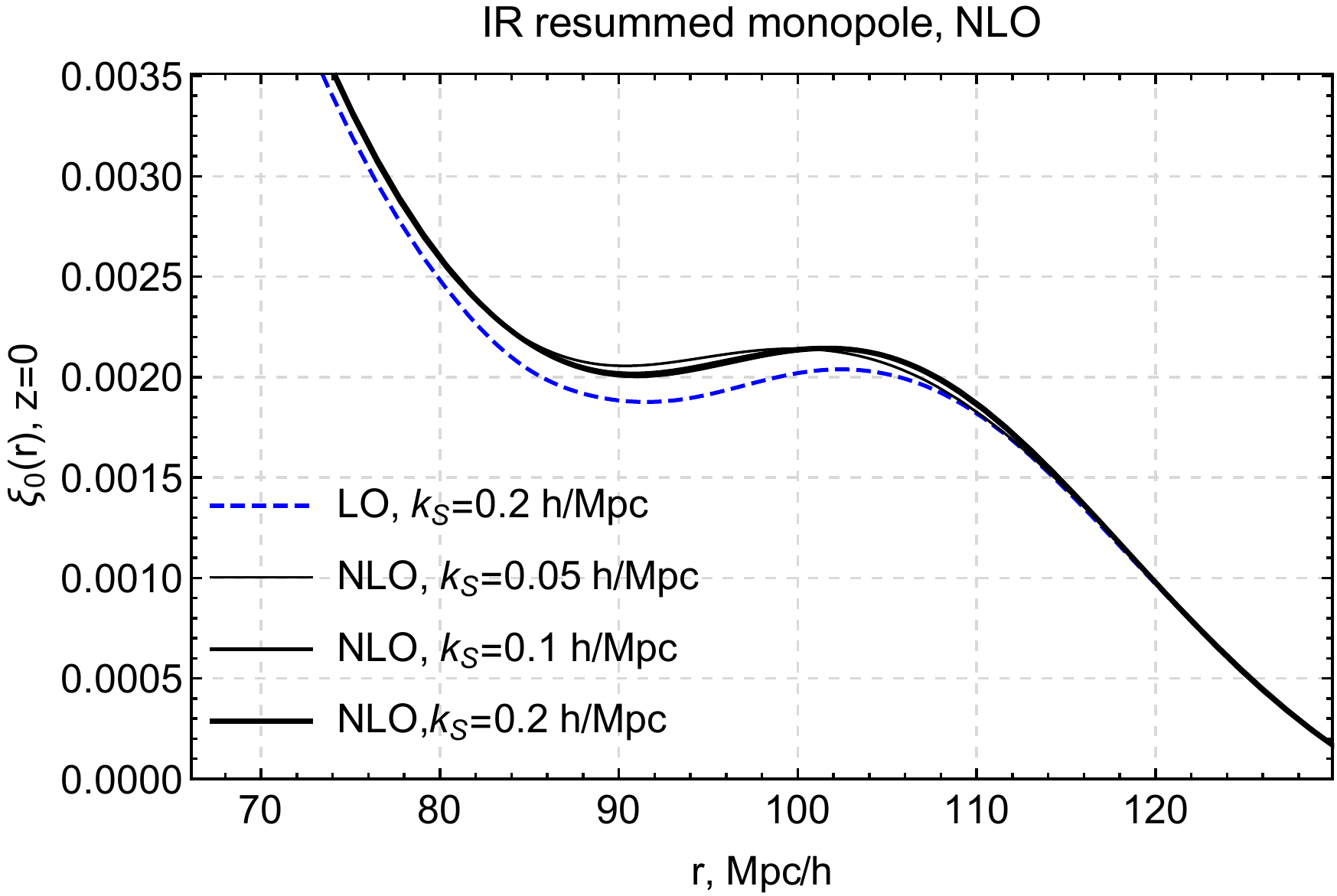}
\caption{\label{fig:ximonopole} 
The monopole ($\ell=0$) moment of the 
2-point correlation function of matter in redshift space at $z=0$. 
\textit{Left panel:} linear theory (orange, dashed) 
vs leading order (LO) IR resummed results for several 
choices of $k_S$ (blue).
\textit{Right panel:} LO for $k_S=0.2 h$/Mpc (blue, dashed)
vs next-to-leading order (NLO) IR resummed results (black).
}
\end{center}
\end{figure}

\begin{figure}
\begin{center}
\includegraphics[width=.49\textwidth]{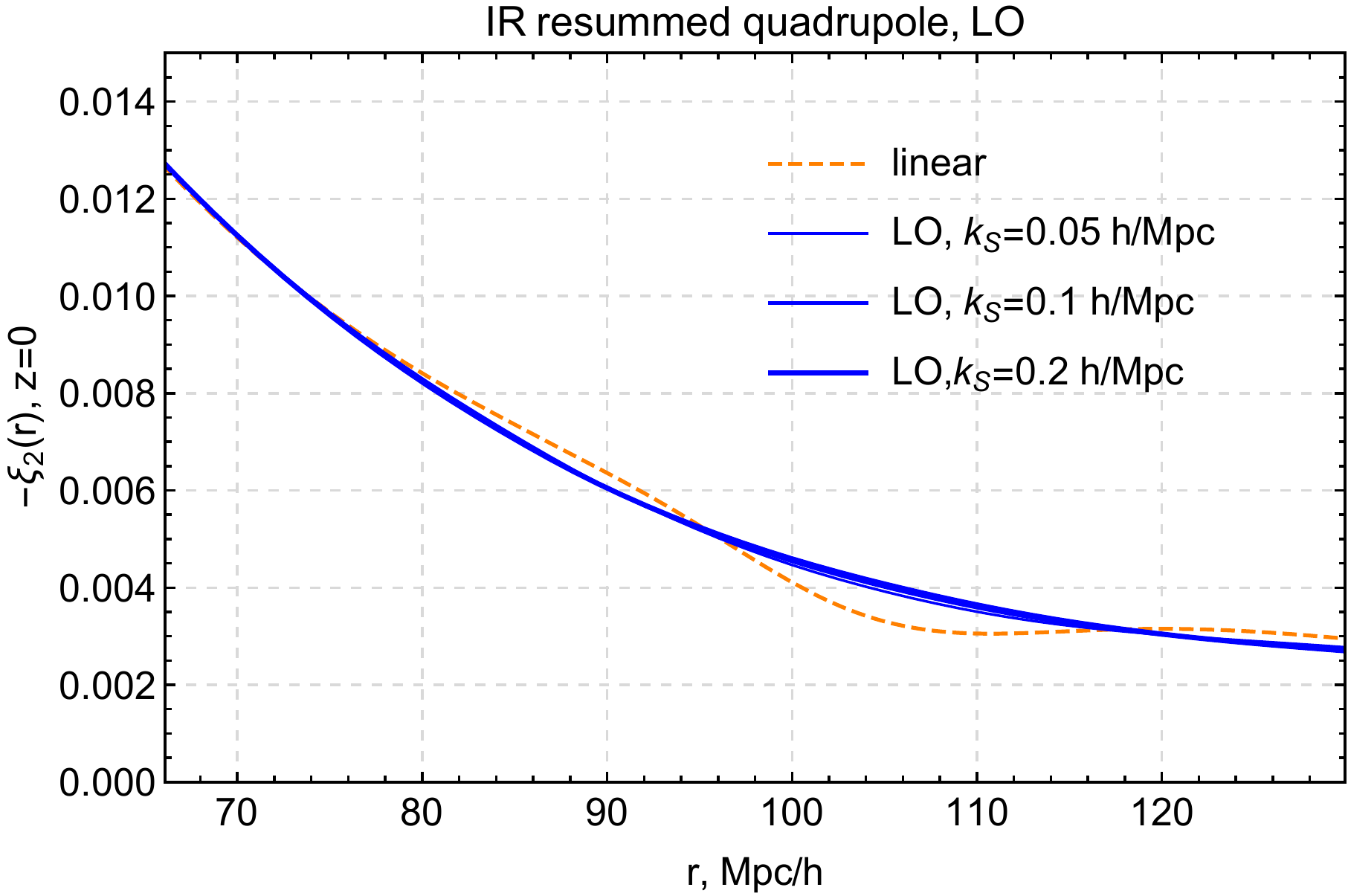}
\includegraphics[width=.49\textwidth]{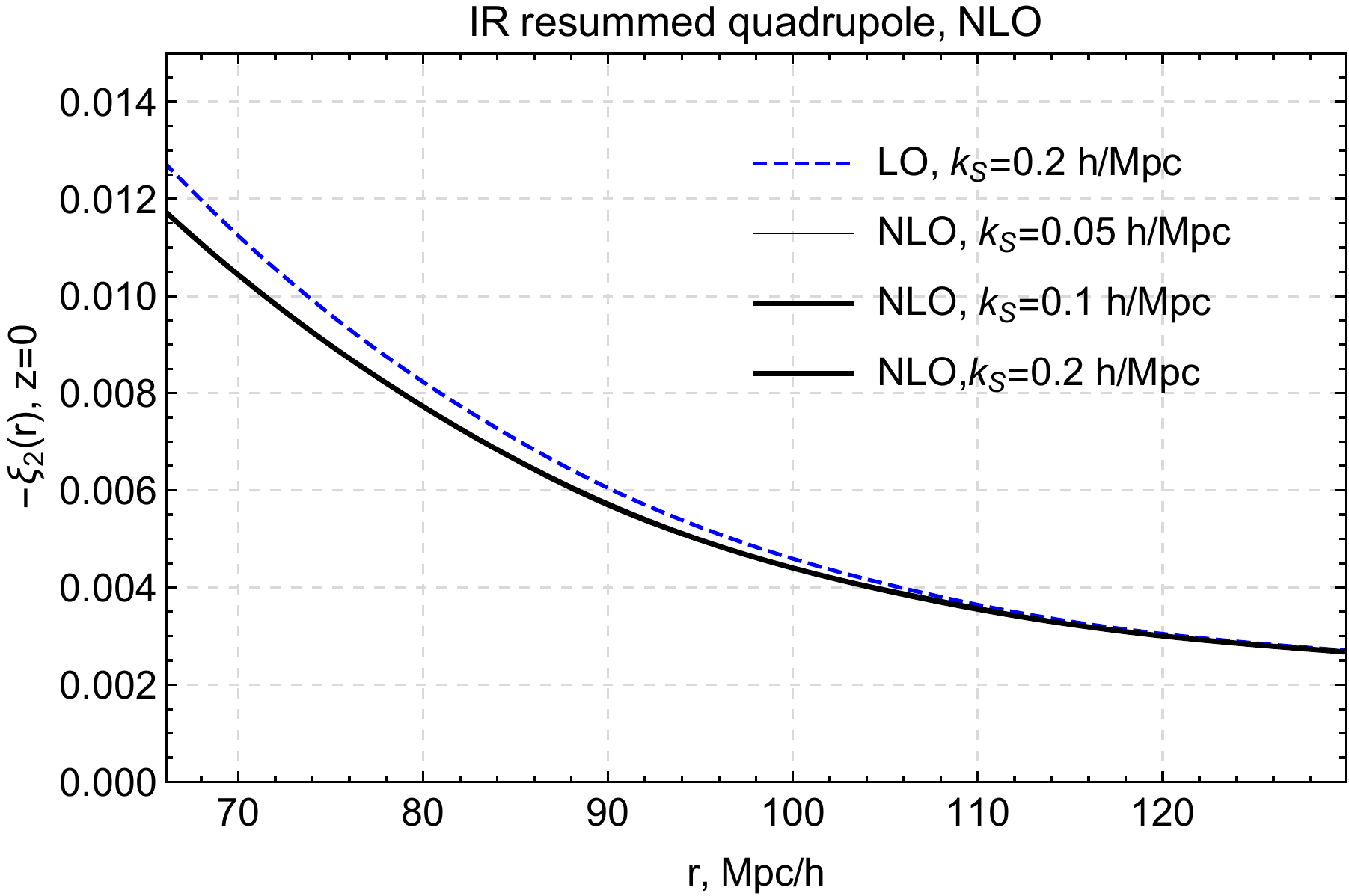}
\caption{\label{fig:xiquad} 
The quadrupole ($\ell=2$) moment of the 
2-point correlation function of matter in redshift space at $z=0$. 
\textit{Left panel:} linear theory (orange, dashed) 
vs LO IR resummed results for several choices of $k_S$ (blue).
\textit{Right panel:} LO for $k_S=0.2 h$/Mpc (blue, dashed) 
vs NLO IR resummed results (black).
The three NLO curves are virtually indistinguishable.
}
\end{center}
\end{figure}

\begin{figure}
\begin{center}
\includegraphics[width=.49\textwidth]{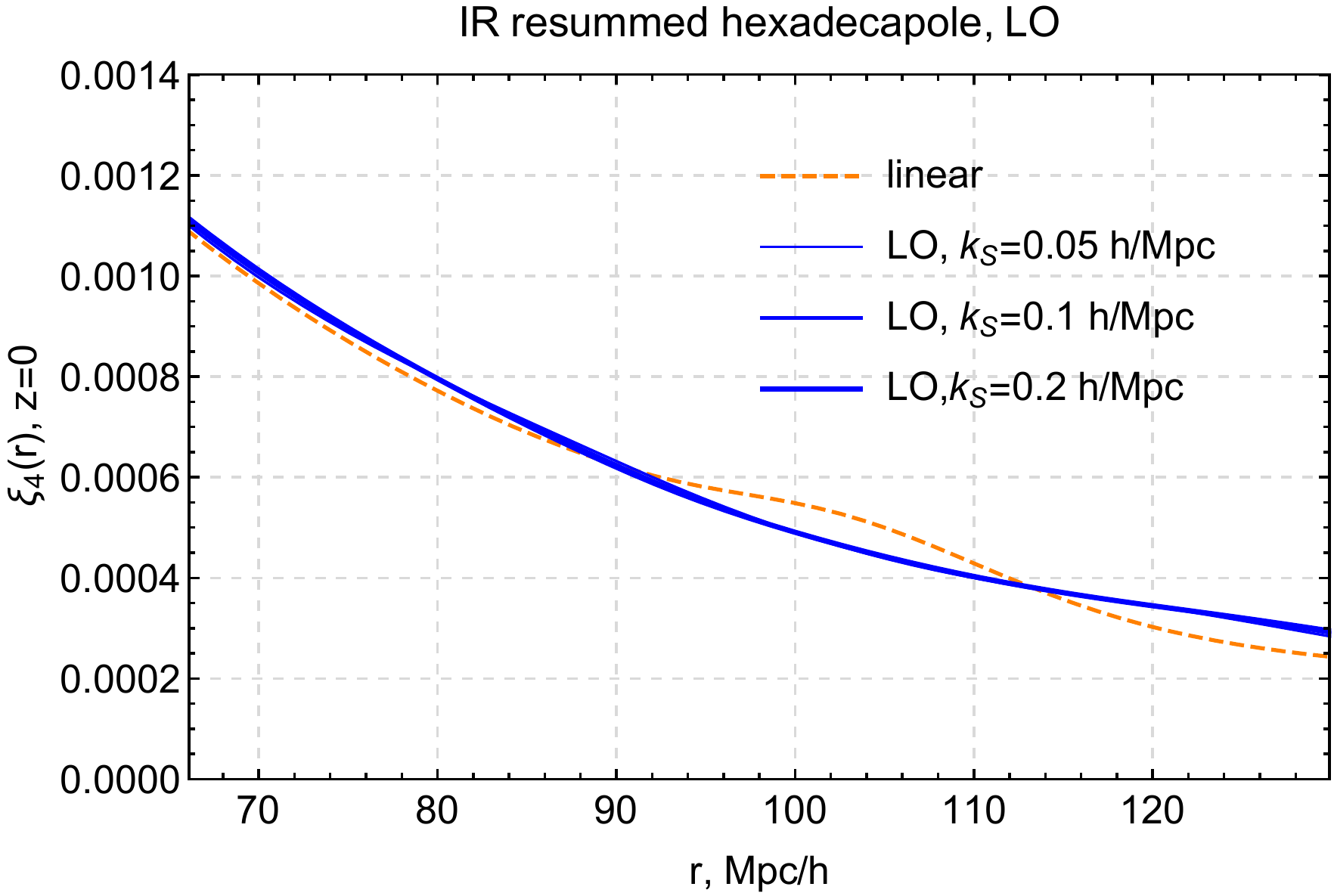}
\includegraphics[width=.49\textwidth]{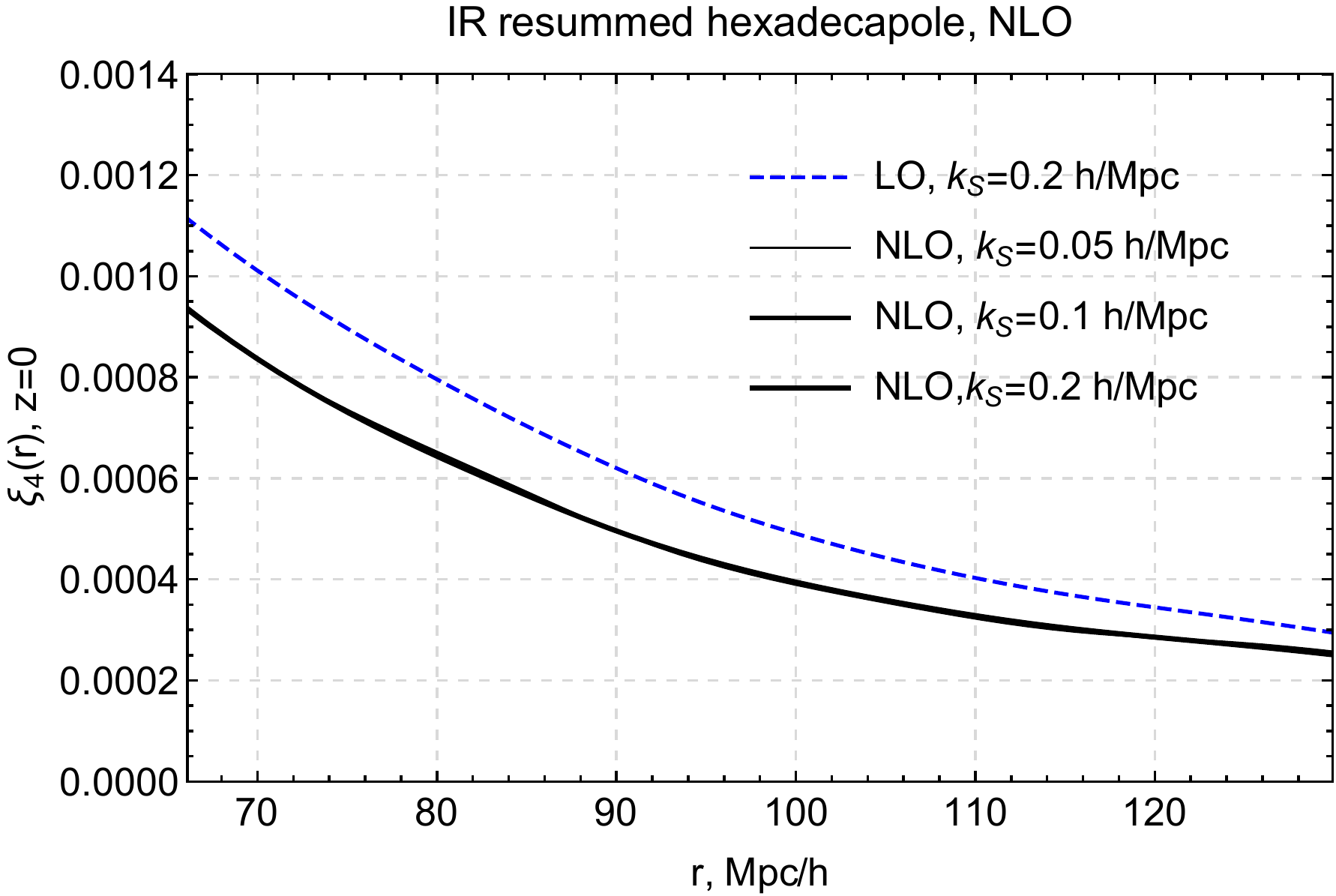}
\caption{\label{fig:xihexa} 
The hexadecapole ($\ell=4$) moment of the 
2-point correlation function of matter in redshift space at $z=0$. 
\textit{Left panel:} linear theory (orange, dashed) 
vs LO IR resummed results for several choices of $k_S$ (blue).
\textit{Right panel:} LO for $k_S=0.2 h$/Mpc (blue, dashed) 
vs NLO IR resummed results (black).
Note that different choices of $k_S$ lead to virtually identical curves.
}
\end{center}
\end{figure}

\begin{figure}
\begin{center}
\includegraphics[width=.49\textwidth]{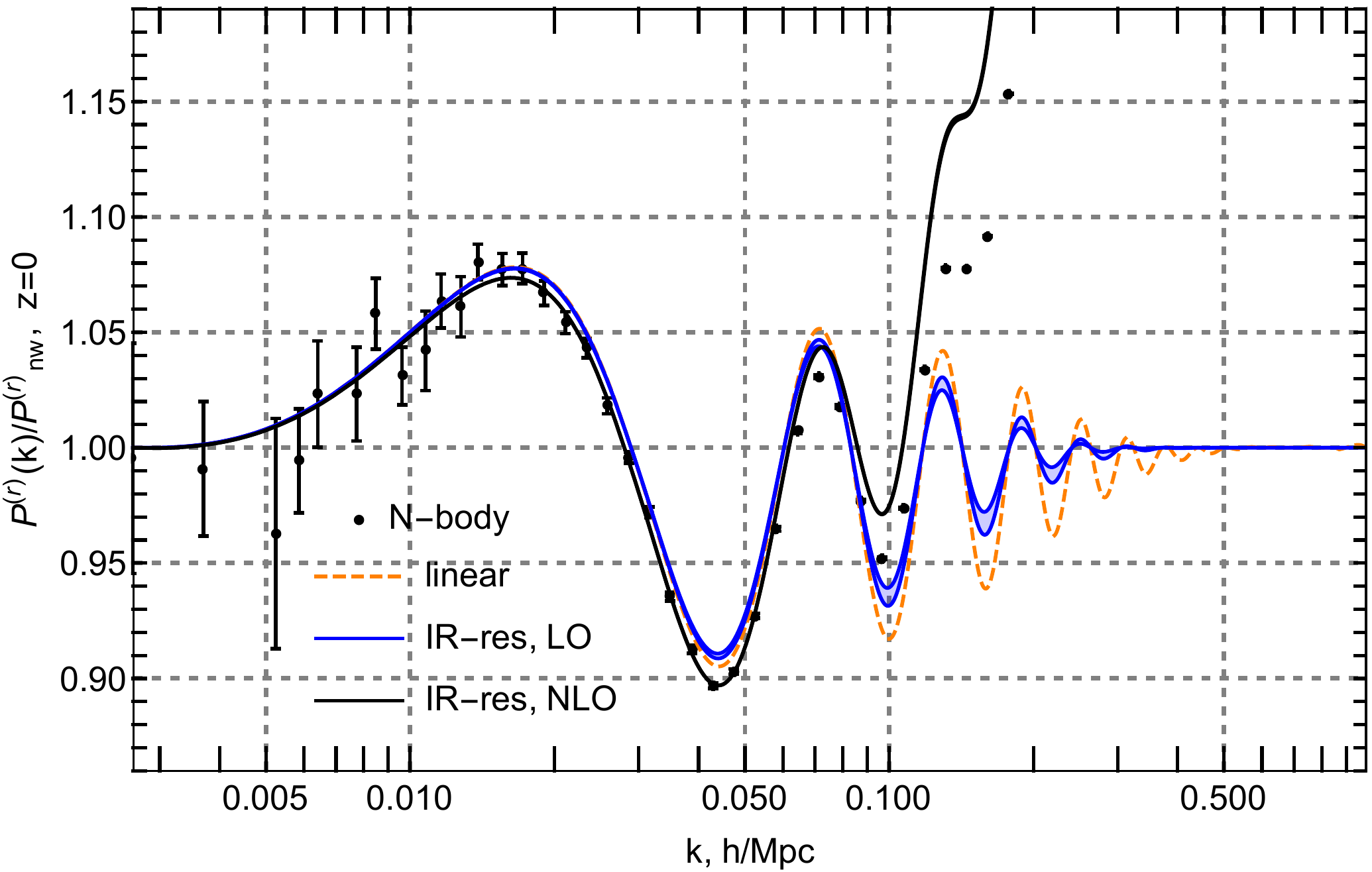}
\includegraphics[width=.49\textwidth]{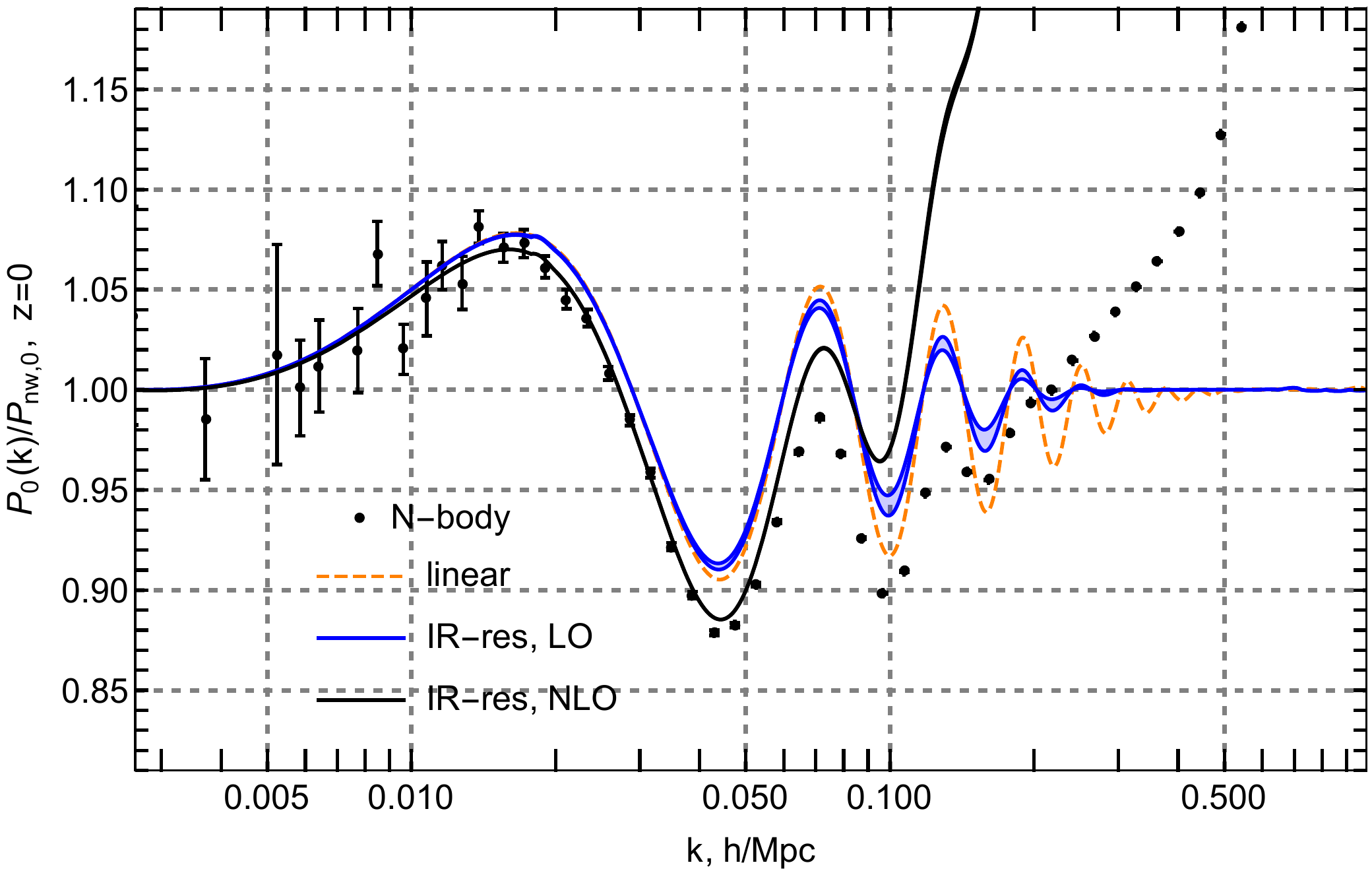}
\includegraphics[width=.49\textwidth]{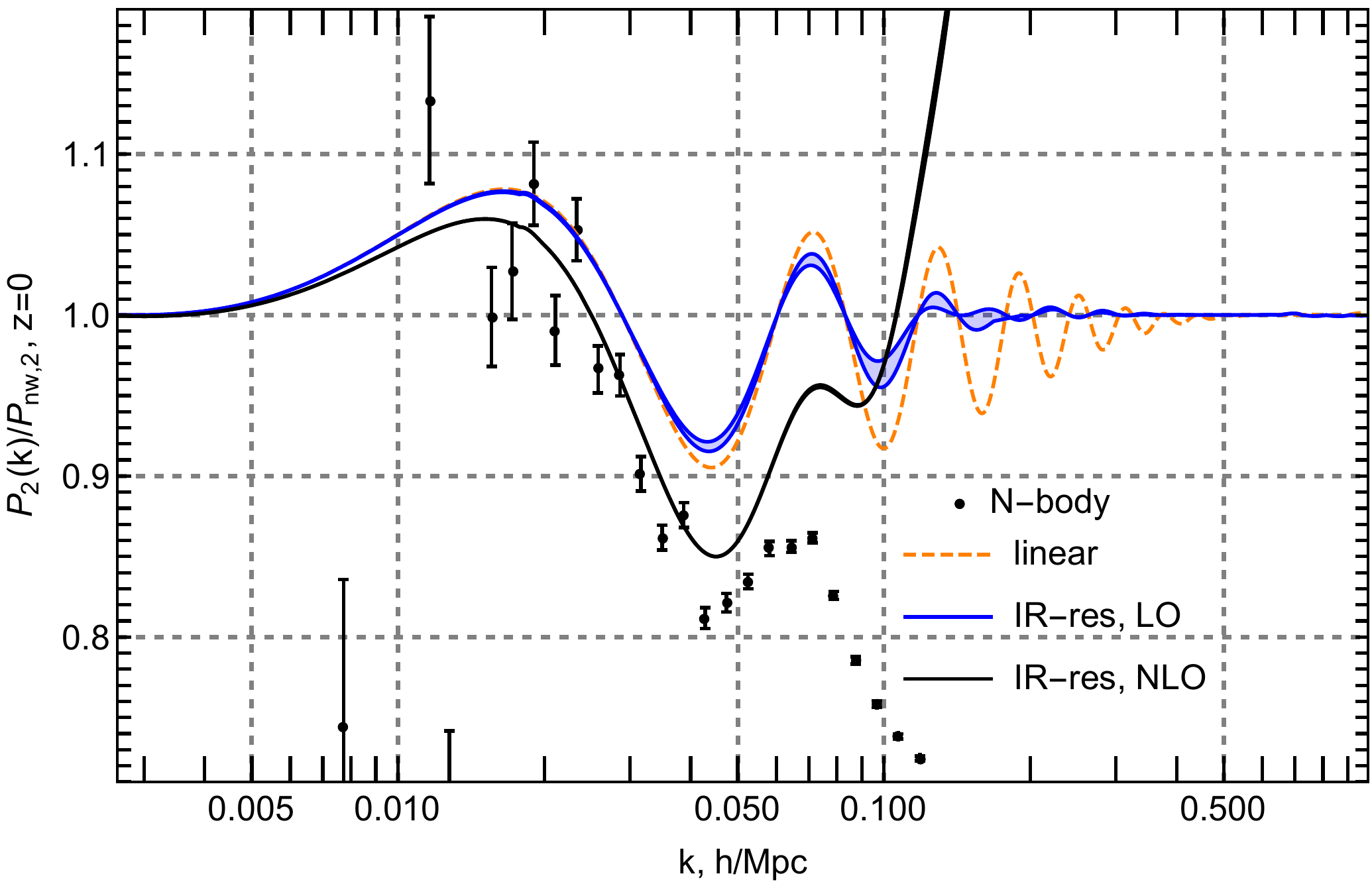}
\includegraphics[width=.49\textwidth]{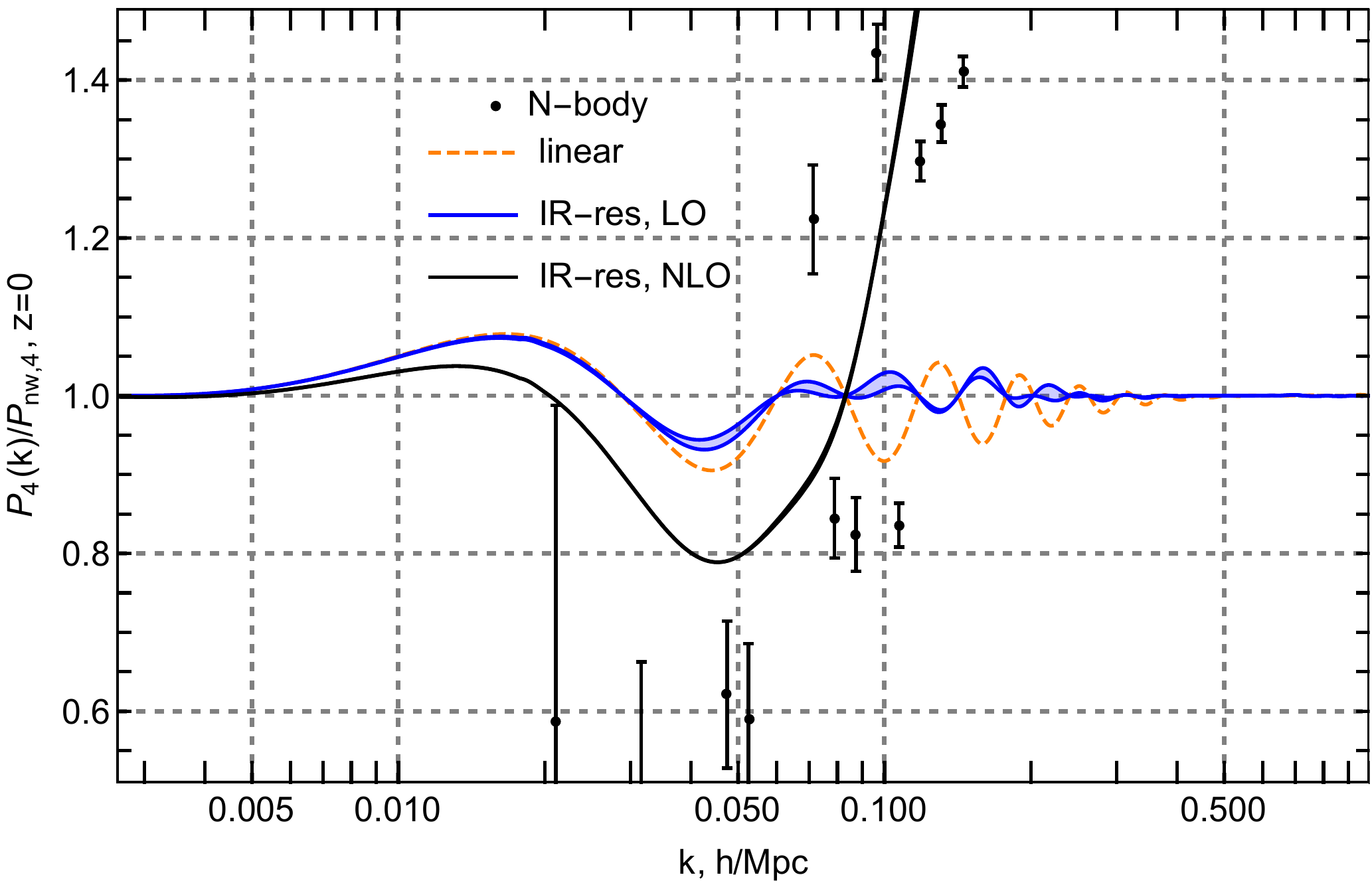}
\caption{\label{fig:pm0z0} 
Matter power spectrum in real space (upper left panel) 
and power spectrum multipoles in redshift space:
monopole (upper right panel), quadrupole (lower left panel), and hexadecapole 
(lower right panel), normalized to the corresponding 
linear non-wiggly power spectra. 
Bands show variation of the IR-resummed results when $k_S$
changes between $0.05$ and $0.2\,h$/Mpc. For NLO results the bands are
barely visible. 
All results are shown for $z=0$, $f=0.483$.
}
\end{center}
\end{figure}

\begin{figure}
\begin{center}
\includegraphics[width=.49\textwidth]{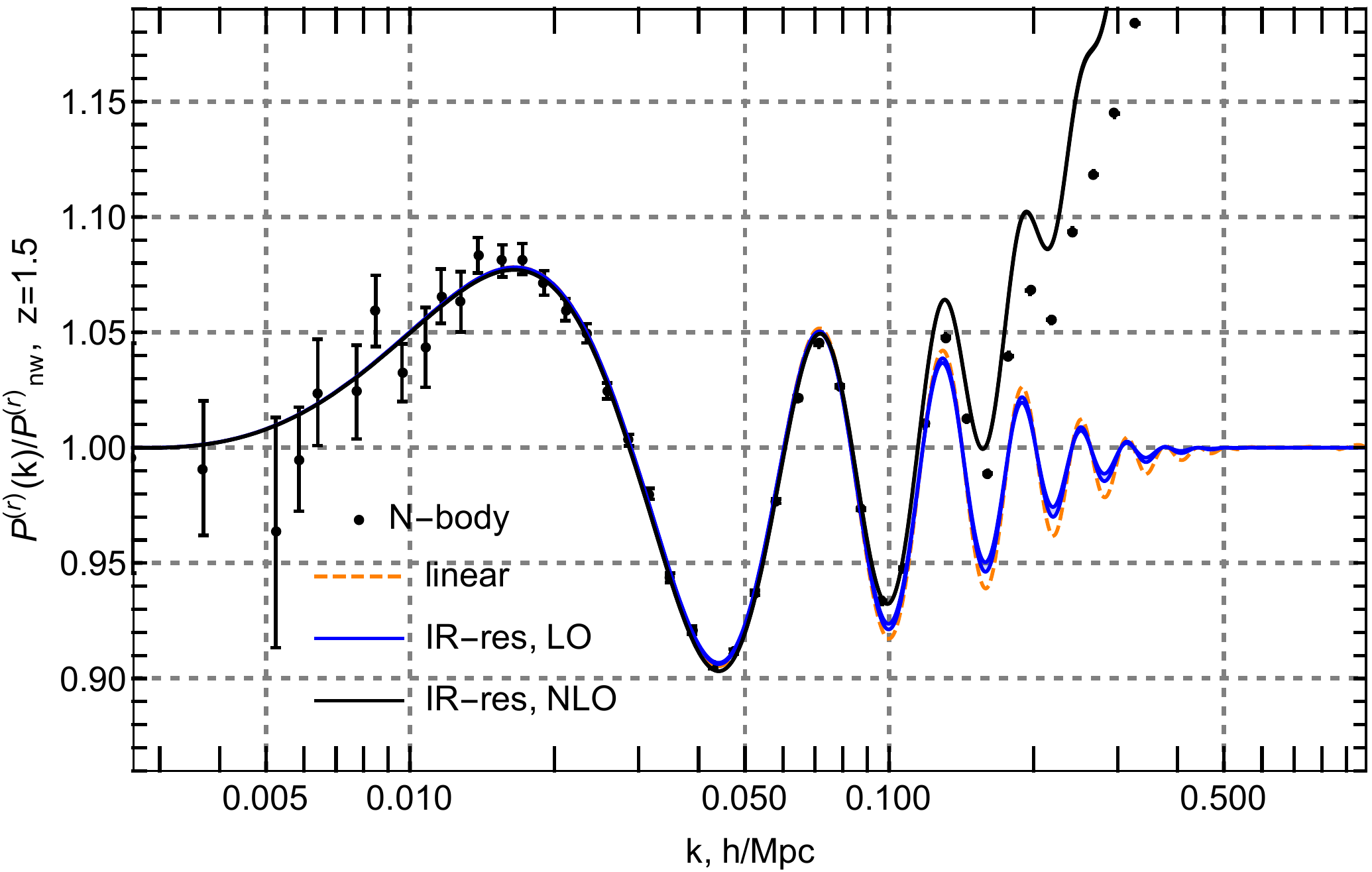}
\includegraphics[width=.49\textwidth]{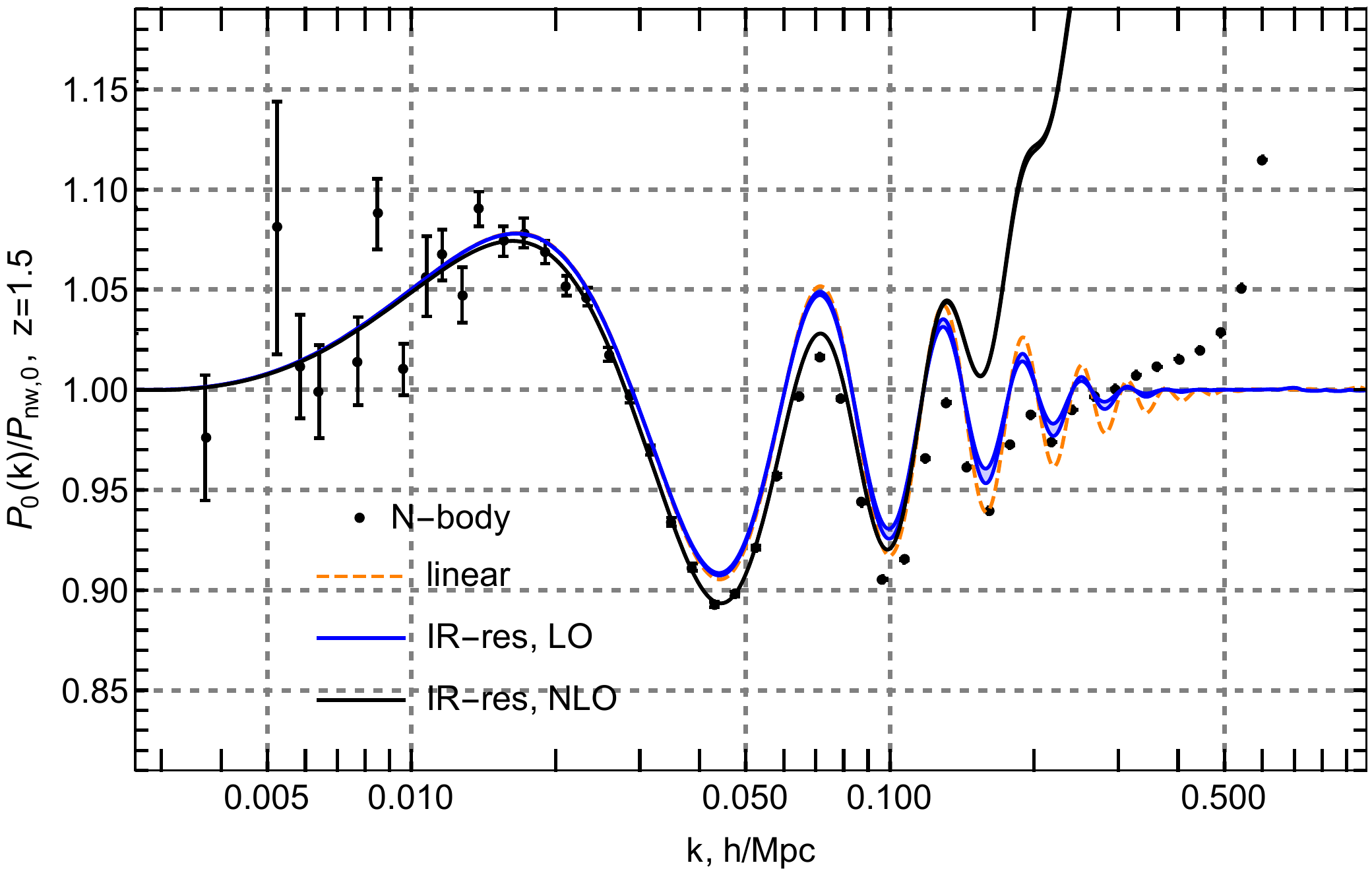}
\includegraphics[width=.49\textwidth]{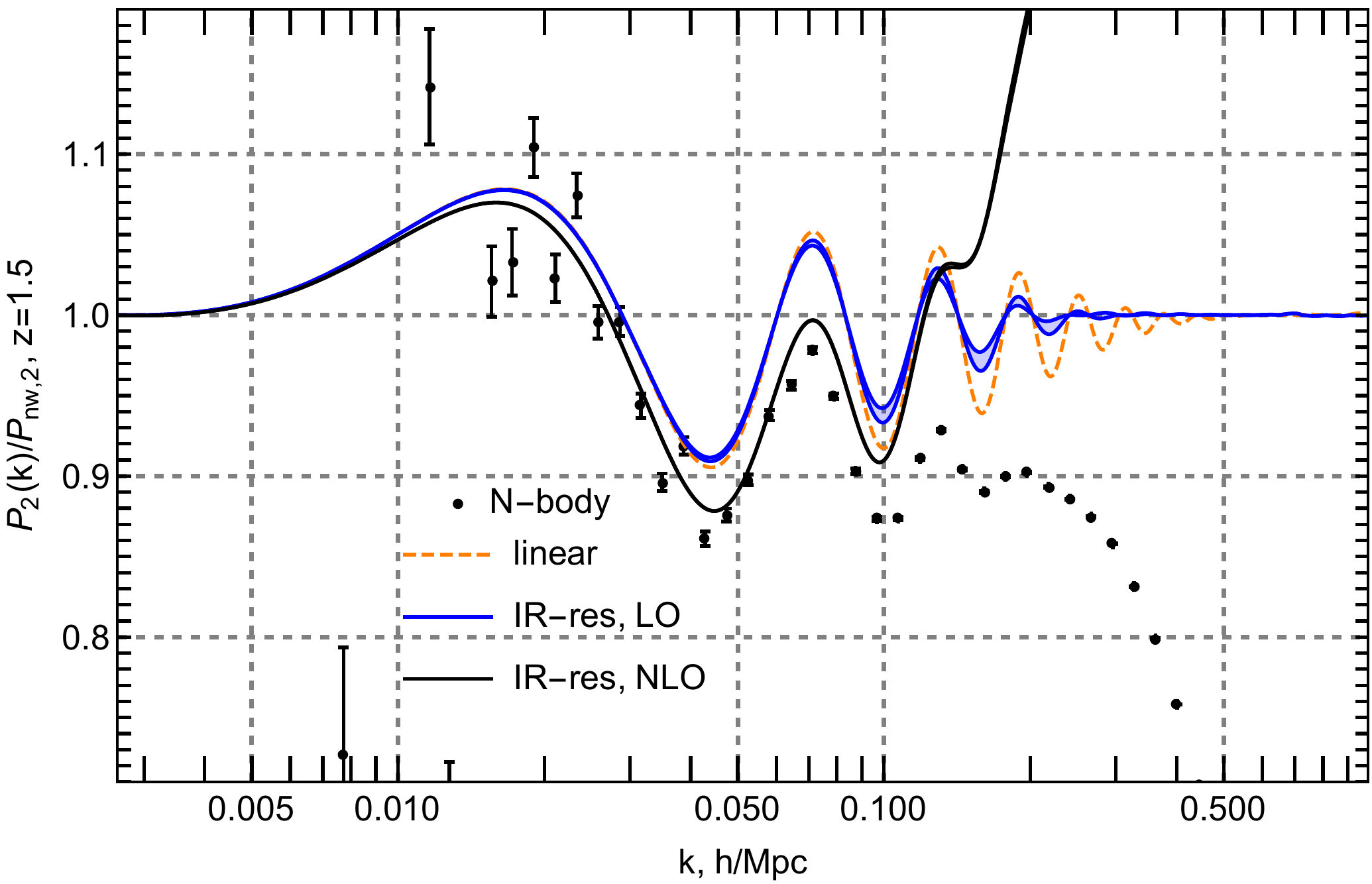}
\includegraphics[width=.49\textwidth]{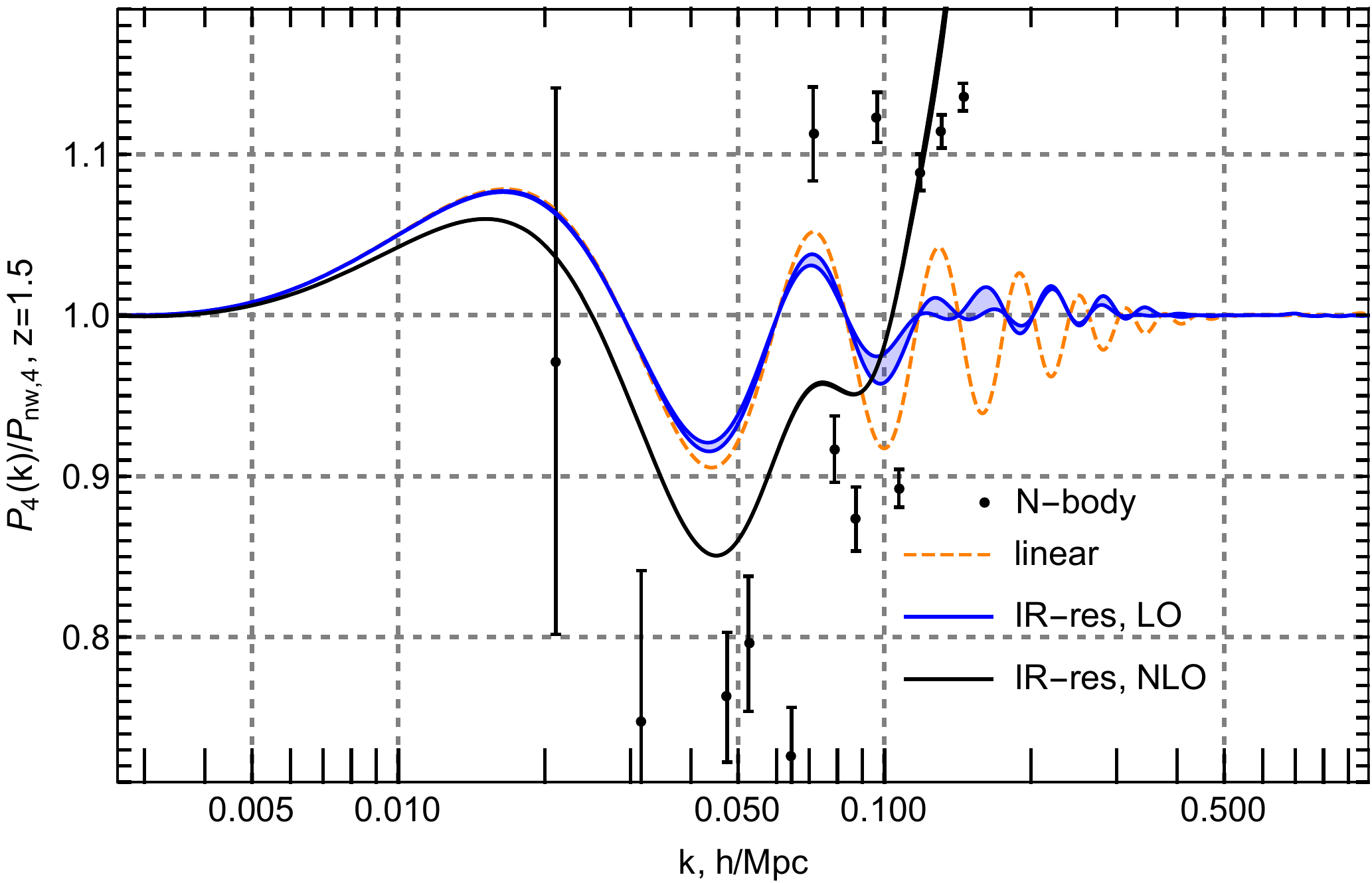}
\caption{\label{fig:pm0z1p5} 
Matter power spectrum in real space (upper left panel) 
and power spectrum multipoles in redshift space:
monopole (upper right panel), quadrupole (lower left panel), and hexadecapole 
(lower right panel), normalized to the corresponding 
linear non-wiggly power spectra. 
Bands show variation of the IR-resummed results when $k_S$
changes between $0.05$ and $0.2\,h$/Mpc. For NLO results the bands are
barely visible. 
All results are shown for $z=1.5$, $f=0.916$. 
}
\end{center}
\end{figure}

In the left panel of Fig.~\ref{fig:ximonopole} 
we show the leading-order IR resummed 
monopole correlation function for three different values\footnote{
Alternatively, one could
consider a $k$-dependent 
separation scale \cite{Baldauf:2015xfa}
to account for the fact that the
enhancement only takes place for modes with $k\gg q$.
In order to avoid the uncertainty related to the precise form 
of $k$-dependence, we prefer to keep $k_S$ as a free parameter that allows us to 
control
the theoretical error of our method.
} of $k_S$.
For comparison we also show the prediction of linear theory.
As expected, the damping of the BAO described by a simple 
exponential suppression of the wiggly component 
translates into a suppression of the BAO peak.
On the other hand, the scatter induced by the choice of $k_S$ 
is quite sizable at leading order. To reduce this uncertainty 
one has to go to next-to-leading order. 
The corresponding correlation function is displayed in the right panel 
of Fig.~\ref{fig:ximonopole}. 
For comparison we also show the LO result for $k_S=0.2\,h$/Mpc. 
We observe that the 1-loop contribution slightly lifts the correlation function
at short scales. The NLO predictions have a very mild (sub-percent) 
dependence 
on the separation scale $k_S$ which indicates the  
convergence of our resummation scheme.

Fig.~\ref{fig:xiquad} shows the result for the quadrupole, $\ell=2$.
In the left panel we plot the correlation function in linear theory 
and at the leading order of IR resummation for three
choices of $k_S$. Since the quadrupole contribution is proportional to the 
derivatives of the real-space correlation function \cite{Hamilton:1992zz}, 
instead of a single peak we observe an oscillating pattern at the BAO scales in linear theory.  
After IR resummation this pattern becomes almost invisible.
This happens because the broadband part of the quadrupole 
has a significant amplitude at the BAO scale, which makes it difficult 
to distinguish a much smaller BAO contribution. 
We note that the dependence on the separation scale is quite mild
both at leading (left panel) and next-to-leading order (right panel). 
We also observe the relative 
impact of the one-loop contribution becomes more sizable
as compared to the monopole.

Fig.~\ref{fig:xihexa} shows the result for the hexadecapole, $\ell=4$.
Similarly to the previous case, we observe that the oscillating
pattern corresponding  
to the BAO is strongly suppressed after the IR resummation. The dependence on 
the separation scale is mild both at LO and NLO. We observe that the
broadband part  
is significantly altered by the one-loop correction to the smooth power spectrum.

\subsection{Matter power spectrum: comparison with N-body data}

In this section we compare our predictions for the power spectrum and its 
multipoles
at LO and NLO with the N-body simulations performed in \cite{GilMarin:2012nb}.
The results of this section should be taken with a grain of salt 
as they do not take into account UV counterterms
whose inclusion is necessary for a consistent description 
of the short-scale dynamics. 
The analysis including 
UV counterterms will be reported elsewhere. 

Fig.~\ref{fig:pm0z0} shows the results for the power spectrum 
in real space at $z=0$ (upper left panel) and the power spectrum multipoles:
monopole (upper right panel), quadrupole (lower left panel), and hexadecapole 
(lower right panel) divided by the corresponding linear non-wiggly
power spectra. 
We show the predictions of linear theory (orange, dashed), 
leading order
(blue) and next-to-leading order (black) 
IR~resummation models.
For IR resummed power spectra we show bands
corresponding to the theoretical uncertainty caused by 
variation of $k_S$ in the range $(0.05\div 0.2)\,h$/Mpc.
Note that for the NLO spectra this band is barely visible.
The error bars correspond to sample variance from an overall simulation volume 
$160\times (2.4\text{Gpc}/h)^3$. 
They do not take into account systematic errors due to discreteness effects, which 
become big for higher-order power spectrum multipoles at large scales.
In particular, we observe that the fluctuations in the measured 
hexadecapole power spectrum 
are very large even at mildly non-linear scales.

Qualitatively, we observe that the LO result does not improve much over
linear theory as it misses the correct broadband information, while upon
including the 1-loop corrections at NLO the agreement between the data and 
the theory improves. The monopole and quadrupole moments clearly exhibit the finger-of-God
suppression at short scales. 
It reduces the range of agreement between the data and the 
theory as compared  
to real space and implies that UV counterterms should
play a significant role is redshift space \cite{Perko:2016puo,delaBella:2017qjy}.

In Fig.~\ref{fig:pm0z1p5} we demonstrate the results for $z=1.5$,
which display an improvement in the agreement between the data and the theory 
over a wider range of scales in line 
with the suppression of non-linearities at large
redshifts.

\section{Summary and outlook}
\label{sec:summary}

In this paper we embedded redshift space distortions and bias in 
time-sliced perturbation theory.
We developed a manifestly IR-safe framework which allows us 
to perturbatively compute
non-linear equal-time correlation
functions of biased tracers in redshift space. 
The key observation is that the coordinate transformation from real to redshift 
space can be viewed as a fictitious 1D fluid 
flow, which maps real space correlation functions 
to the redshift space ones.
Once this mapping is done, one can systematically resum
the enhanced IR corrections affecting the BAO feature.
The IR resummation of cosmological correlators
proceeds in a straightforward manner along the lines of \cite{Blas:2016sfa}.
IR resummation in TSPT is based on physically motivated power counting rules 
and has a clear diagrammatic representation,
which allows to compute the relevant corrections in a systematic and
controllable way. 

Our analysis gives a simple prescription for the numerical evaluation of the 
IR-resummed cosmological correlation functions. 
First, one has to isolate the oscillating part of the power spectrum 
as only this contribution is susceptible to non-linear damping due to
bulk flows. 
IR resummation at leading order
amounts to
replacing the usual linear power spectrum by the ``improved'' one,
\be
\label{eq:irresfinal}
P(k)\to P_{nw}(k)+e^{-k^2\Sigma^2_{\text{tot}}(\mu;k_S)} 
P_{w}(k)\,,
\ee 
where $\mu$ is the cosine of the angle between the wavevector $\k$ and
the line-of-sight, 
$k_S$ is the separation scale defining the range of modes which are resummed,
and the damping factor $\Sigma^2_{\text{tot}}$ is given in
\eqref{Sigmadefk}. Applying this prescription we
obtained the explicit expressions 
for the IR-resummed power spectrum \eqref{eq:finaltree}
and bispectrum \eqref{BLO}
of biased tracers in redshift space. 

At first order in hard loops IR resummation amounts 
to computing loop diagrams using the IR-resummed 
power spectrum \eqref{eq:irresfinal} as an input.
This must be accompanied by modification of the input power spectrum
in the tree-level part to avoid double counting. The general formula
for $n$-point IR-resummed redshift-space correlator reads,
\be
\label{Cnfinal}
\begin{split}
\mathfrak{C}_n^{(s)\,\text{IR res,LO+NLO}}(\k_1,...\k_n)=
&\mathfrak{C}_n^{(s)\,tree}\big[P_{nw}+(1+k^2\Sigma_{\rm tot}^2)
e^{-k^2\Sigma_{\rm tot}^2}P_w\big](\k_1,...\k_n)\\
&+\mathfrak{C}_n^{(s)\,1-loop}\big[P_{nw}+e^{-k^2\Sigma_{\rm
    tot}^2}P_w\big](\k_1,...\k_n)\;,
\end{split}
\ee
where $\mathfrak{C}_n^{(s)\,tree}$ and $\mathfrak{C}_n^{(s)\,1-loop}$
are the tree-level and 1-loop contributions understood as functionals
of the input power spectrum. Equation (\ref{Cnfinal})
applies both to the density and
velocity correlators, as well as to biased tracers. It also admits a
straightforward generalization to higher orders in hard loops.

The angular dependence of the damping factor $\Sigma_{tot}^2$ in
(\ref{Cnfinal}) reduces the 
symmetry of one-loop integrands
and complicates the numerical evaluation. 
We have shown that for the one-loop power spectrum 
it is possible to rearrange our result in an equivalent form,
Eq.~(\ref{eq:rsd1loopps2}),  suitable 
for numerical implementation using standard algorithms. We also
derived a simplified expression for the redshift-space 1-loop
bispectrum, see Eq.~(\ref{Bs1loop}).

The separation scale $k_S$ is \textit{a priori} arbitrary and any dependence 
on it should be considered as part of the
theoretical uncertainty. 
We show that the scatter of our results w.r.t variations of this scale 
in the reasonable range $(0.05\div 0.2)\,h$/Mpc is not negligible at LO,
but
substantially reduces when including NLO corrections.
This testifies the 
convergence of our resummation procedure.

We compared our results with available N-body data on the power spectrum 
of matter in redshift space and found that IR resummation 
at NLO significantly improves the range of agreement between theory and
data compared to linear  
theory and LO IR resummation. 
The results of our comparison are preliminary at the moment
as we have not included into calculation the UV-counterterms. Taking
into account these counterterms is expected to further improve the agreement.  
We leave this task for future study.

Our results suggest several
directions for future research.
On one hand, one can accurately assess the shift of the BAO
peak in redshift space  
and for biased tracers. Although the expression 
\eqref{eq:rsd1loopps2} already contains some contributions into the shift,
its precise value must be validated by computing full
NLO soft corrections.
On the other hand, our results may be useful for 
elucidating systematic uncertainties of the
reconstruction algorithms, see Ref.~\cite{Ding:2017gad} for a recent
work in this direction.  
Finally, our theoretical template may be used
for analyzing the data from full-shape 
measurements of galaxy clustering without reconstruction.

\section*{Acknowledgments}

We are grateful to V.~Assassi, D.~Blas, H.~Gil-Mar\'in, 
F.~Schmidt,  M.~Schmittfull, G.~Trevisan and M.~Zaldarriaga
for illuminating discussions. 
We are indebted to Z.~Vlah for sharing with us his numerical results and
for valuable comments.
We thank M.~Simonovi\'c for help with the FFTLog algorithm and for
encouraging interest.
We are grateful to D.~Blas and M.~Garny for useful comments on the draft.
This work is supported by the Swiss National Science Foundation. 
M.I. also acknowledges a partial support by the RFBR grant No. 17-02-01008.
S.S. is supported by the RFBR grant No. 17-02-00651.

\appendix 

\section{Review of time-sliced perturbation theory for matter in real
  space} 
\label{app:A}

In this section we give a brief review of TSPT 
for matter in real space
with emphasis on IR resummation \cite{Blas:2015qsi,Blas:2016sfa}.
For clarity we will omit the superscripts $(r)$ in the notations for
cosmological fields. 
We are interested in the correlation functions of
the overdensity field $\delta=(\rho-\bar \rho)/\bar\rho$
and the velocity divergence field $\Theta \propto \nabla \cdot {\bm v}$, whose time-evolution is
governed by the continuity and Euler equations for
the peculiar flow velocity ${\bm v}$,
\bseq
\label{eq:hydro}
\begin{align}
 & \frac{\d\delta}{\d \tau} + \nabla \cdot [(1+\delta){\bm v}] = 0 \,, \\
 & \frac{\d{\bm v}}{\d \tau} + {\cal H}{\bm v} + ({\bm v}\cdot \nabla) {\bm v} =  -\nabla\phi \,,
\end{align}
\eseq
where $\nabla^2\phi=\frac{3}{2}{\cal H}^2\Omega_m\delta$ and ${\cal
  H}=aH$. 
Here $\tau$ is conformal time and $\Omega_m$ is the matter
density fraction. It is well-known \cite{Bernardeau:2001qr} that in
the case of an Einstein--de Sitter universe these equations can be cast
in a form free from any explicit time dependence by introducing  
the time parameter $\eta=\ln D$, where $D$ is the linear growth
factor, and appropriately rescaling the velocity divergence
\be
\label{eq:thetadef}
\Theta =-\frac{\nabla \cdot {\bm v}}{{\cal H}f}
\ee
with $f=d\ln D/d\ln a$. For the realistic $\Lambda$CDM cosmology,
the substitution \eqref{eq:thetadef} into \eqref{eq:hydro} 
leaves a mild residual time dependence which,
however, has little effect on the dynamics. Following conventional
practice we will neglect this explicit time dependence in the
equations of motion, but keep the factor $f$ 
when it appears in redshift space quantities.

In Fourier space Eqs.~\eqref{eq:hydro} can be rewritten as
\be
\label{eq:eomsreal}
\begin{split}
& \d_\e\delta_\k -\T_\k=\int [dq]^2\delta^{(3)}(\k-\q_{12})
\alpha(\q_1,\q_2)\T_{\q_1}\delta_{\q_2}\,,\\
& \d_\e \Theta_\k +\frac{1}{2}\Theta_\k - \frac{3}{2}\delta_\k=  \int [dq]^2\delta^{(3)}(\k-\q_{12})\beta(\q_1,\q_2)\T_{\q_1}\T_{\q_2}\,,
\end{split}
\ee
with non-linear kernels
\be
\label{alphabetareal}
\alpha(\k_1,\k_2)\equiv\frac{(\k_1+\k_2)\cdot \k_1}{k_1^2}\,, \quad \quad 
\b(\k_1,\k_2)\equiv\frac{(\k_1+\k_2)^2(\k_1\cdot \k_2)}{2k_1^2k_2^2}\,. 
\ee
The main idea of the TSPT approach
is to substitute the time evolution of the overdensity and velocity
divergence fields, $\delta$ and $\Theta$, by that of the their time
dependent probability distribution functional.
This idea is particularly useful when one 
is only interested in equal time correlation functions.
For adiabatic initial conditions only one of the two fields is
statistically independent. 
We choose it to be the velocity divergence field $\T$ and denote its
probability 
distribution functional by 
$\mathcal{P}[\T;\e]$.
At any moment in time, the field $\delta$ can be expressed in terms of  $\T$
as 
\be
\label{eq:psi1}
\delta_\k=\delta[\T;\e,\k]\equiv \sum_{n=1}^\infty\frac{1}{n!}
\int [dq]^n K_n^{(r)}(\q_1,...,\q_n)\,\delta^{(3)}(\k-\q_{1...n})
\prod_{j=1}^n\T(\e,\q_j)  \,,
\ee
with $K_1^{(r)}=1$.
Equation \eqref{eq:psi1} can be used to eliminate
the density field from Eq.~\eqref{eq:hydro}
and obtain a closed equation for the velocity divergence,
\be
 \label{eq:In}
\d_\e\T(\e,\k)= \mathcal{I}[\T]\equiv \sum_{n=1}^\infty\frac{1}{n!}
\int [dq]^n I^{(r)}_n(\q_1,...,\q_n)\,\delta^{(3)}(\k-\q_{1...n})
\prod_{j=1}^n\T(\e,\q_j)  \,,
\ee
with $I^{(r)}_1\equiv 1$ corresponding to the growing mode
in the perfect fluid approximation. 
The kernels $K^{(r)}_n$ and $I^{(r)}_n$  
are found recursively using the relations,
\bseq
\label{eq:recGED1}
\begin{align}
&K_2^{(r)}(\k_1,\k_2)=\frac{4}{7}
\bigg(1-\frac{(\k_1\cdot\k_2)^2}{k_1^2k_2^2}\bigg)\;,\\
& I^{(r)}_2(\k_1,\k_2)=2\beta(\k_1,\k_2)+\frac{3}{2}K^{(r)}_2(\k_1,\k_2)\,,\\
&K^{(r)}_n(\k_1,...,\k_n)=
\frac{2}{2n+3}\bigg[ \sum_{i=1}^{n}\a\Big(\k_i,\!\sum_{1\leq j\leq n,j\neq i} \!\!\k_j\Big)\,
K^{(r)}_{n-1}(\k_1,...,\check{\k}_i,...,\k_{n})\notag\\
&-\sum_{1\leq i<j \leq n}I^{(r)}_2(\k_i,\k_j)\, 
K^{(r)}_{n-1}(\k_i+\k_j,\k_1,...,\check\k_i,...,\check\k_j,...,\k_n)\notag\\
&-\frac{3}{2} \sum_{p=3}^{n-1}\frac{1}{p!(n-p)!} \sum_{\sigma}
K^{(r)}_{p}\big(\k_{\sigma(1)},...,\k_{\sigma(p)}\big)\,K^{(r)}_{n-p+1}
\Big(\sum_{l=1}^p\k_{\sigma(l)},\k_{\sigma(p+1)},...,\k_{\sigma(n)}\Big)
\Bigg] \,,\label{Kn}\\
& I^{(r)}_n(\k_1,...,\k_n)=\frac{3}{2}K^{(r)}_n(\k_1,...,\k_n)~,\qquad
n\geq 3\,.
\end{align}
\eseq

Equal-time correlation functions for $\T$ and $\delta$ can be obtained by
taking functional derivatives with respect to the external sources $J$ and
$J_\delta$, respectively, of the following partition function,
\be
\label{eq:ztfp}
Z[J,J_\delta;\e]=\int [\mathcal{D}\T]\;{\mathcal P}[\T;\e]\;
\exp\bigg\{\int [dk] \T_{\k} J(-\k)+\int[dk] \delta[\T;\e,\k]J_{\delta}(-\k)\bigg\}\,.
\ee
The probability density functional satisfies the Liouiville equation which
reflects the conservation of probability,
\be
\label{Liouville}
\frac{\d}{\d \e} \mathcal{P}[\T;\e] + \int [dk]\frac{\delta}{\delta
  \Theta(\k)}(\mathcal{I}[\T;\e]\mathcal{P}[\T;\e])=0\,. 
\ee
In perturbation theory one can represent (logarithm of)
$\mathcal{P}[\T;\e]$ as a power series 
in $\T$,
\be
\label{eq:statweight}
\mathcal{P}[\T;\e]=\mathcal{N}^{-1}\exp\Bigg\{-\sum_{n=1}^{\infty}\frac{1}{n!}\int
[dk]^n\; 
\G_n^{(r)\,tot}(\e;\k_{1},...,\k_{n})\; \prod^n_{j=1} \T_{\k_j}\Bigg\} \,,
\ee
where ${\cal N}$ is a normalization factor. Substituting this
representation into (\ref{Liouville}) and 
using
Eq.~\eqref{eq:In} we obtain the following chain of equations on  
the vertices,
\begin{equation}
\begin{split}
  & \d_{\e}\Gamma_n^{(r)\,tot}(\eta;\k_1,...,\k_n) + \sum_{m=1}^n \frac{1}{m!(n-m)!} \sum_{\sigma} I^{(r)}_m(\eta;\k_{1},...,\k_{m}) \\
  &\times \Gamma_{n-m+1}^{(r)\,tot}(\eta; \sum_{l=1}^m \k_{\sigma(l)},\k_{\sigma(m+1)},...,\k_{\sigma(n)}) 
   =  \delta^{(3)}\left(\k_{1...n}\right) \int [dp] I^{(r)}_{n+1}(\eta; \p, \k_1,...,\k_n)\;.
  \label{eq:liouvillegamma}
\end{split}
\end{equation}
It is convenient to decompose the solution of these equations into two pieces:
\be 
\G_n^{(r)\,tot}=\G^{(r)}_n+C^{(r)}_n\,,
\ee
where $\G^{(r)}_n$ is the solution of the homogeneous equations 
\eqref{eq:liouvillegamma} with the initial conditions reflecting 
the initial statistical distribution,
and $C^{(r)}_n$ is the solution of the inhomogeneous equations 
with vanishing initial conditions.
The $\G^{(r)}_n$ vertices
have the physical meaning of 1-particle irreducible tree-level correlators 
with amputated external propagators, 
and $C^{(r)}_n$ are {\it counterterms}, whose role is to cancel divergences
in the loop corrections \cite{Blas:2015qsi}. 

For the Gaussian initial conditions the time-dependence of  the vertices $\G^{(r)}_n$ factorizes,
\be
\Gamma^{(r)}_n= \delta^{(3)}(\k_{1...n})\frac{\bar \Gamma'^{(r)}_n}{g^2(\e)}\,,
\ee
where the time-independent kernels $\bar \Gamma'^{(r)}_n$ are given by
\bseq
\label{eq:recGED}
\begin{align}
&\bar\G'^{(r)}_2(\k_1,\k_2) = \frac{1}{\bar P(k_1)}\,,
\label{recGED1}\\
&\bar\G'^{(r)}_n(\k_1,...,\k_n)=-\frac{1}{n-2} 
\!\sum_{1\leq i<j\leq n}\!\!
I^{(r)}_2(\k_i,\k_j)
\bar\G^{(r)}_{n-1}(\k_i\!+\!\k_j,\k_1,...,\check\k_i,...,\check\k_j,...,\k_n)
\notag\\
&\qquad\qquad\qquad\quad~~-\frac{3}{2(n-2)} \sum_{p=3}^{n-1}
\frac{1}{p!(n-p)!}
\sum_{\sigma}
K^{(r)}_{p}\big(\k_{\sigma(1)},...,\k_{\sigma(p)}\big)\notag\\
&\qquad\qquad\qquad\qquad\quad\times\bar\G^{(r)}_{n-p+1}\Big(\sum_{l=1}^p
\k_{\sigma(l)},\k_{\sigma(p+1)},...,\k_{\sigma(n)}\Big)\,,~~~~~~n\geq 3\;.
\label{Gned}
\end{align}
\eseq
The counterterms $C^{(r)}_n$ do not depend on time and are given by:
\bseq
\label{eq:recCounter}
\begin{align}
&C^{(r)}_n(\k_1,...,\k_n)=\frac{1}{n} \Bigg[\delta^{(3)}(\k_{1...n})\int [dp]I_{n+1}^{(r)}(\p,\k_1,...,\k_n)
\notag\\
&-\sum_{p=2}^{n}
\frac{1}{p!(n-p)!}
\sum_{\sigma}
I^{(r)}_{p}\big(\k_{\sigma(1)},...,\k_{\sigma(p)}\big)
C^{(r)}_{n-p+1}\Big(\sum_{l=1}^p
\k_{\sigma(l)},\k_{\sigma(p+1)},...,\k_{\sigma(n)}\Big)\Bigg]\,. 
\end{align}
\eseq

\subsection{IR resummation in real space}

IR resummation in real-space TSPT proceeds in three steps \cite{Blas:2016sfa}:

1. One notices that the 
vertices $\bar \G^{(r)}_n$ 
are functionals of the linear power spectrum $\bar P(k)$.
Hence, the decomposition of $\bar P(k)$ into the smooth and wiggly parts
induces a similar decomposition of the vertices,
\be 
\bar\G^{(r)}_n=\bar\G^{nw\,(r)}_n+\bar\G^{w\,(r)}_n\,,
\ee
where the terms $O(\bar P_w^2/\bar P_{nw}^2)$ are neglected.

2. One identifies IR-enhanced contributions. These are characterized 
by inverse powers of the small parameter $\varepsilon \sim q/k \ll 1$, 
where $q$ is a soft (loop) momentum and $k$ is an external momentum.
The enhancement takes place only for the wiggly vertices which have the 
following asymptotic behavior in the limit $q/k\to 0$:
\be 
\label{eq:recreal}
\begin{split}
\bar{\G}'^{w\,(r)}_n\Big(\k_1,...,&\k_m-\sum_{i=1}^{n-m}\q_i,\q_1,...,\q_{n-m}\Big)\\
&=(-1)^{n-m}\bigg(\prod_{i=1}^{n-m}\D^{(r)}_{\q_i} \bigg)
\big[\bar{\G}'^{w\,(r)}_n(\k_1,...,\k_m)\big]\,(1+\mathcal{O}(\ve))\,,
\end{split}
\ee  
where the operator $\D^{(r)}_\q$ acts on the wiggly power spectrum as follows, 
\be 
\label{eq:drealspace}
\D^{(r)}_\q [\bar P_w(k)]=\frac{(\k\cdot \q)}{q^2}(e^{\q\cdot \nabla_{\k'}}-1)\bar P_w(k')\Big|_{k'=k}\,.
\ee
From \eqref{eq:drealspace} we see that the wiggly vertex is enhanced as
\be 
\bar{\G}'^{w\,(r)}_n\Big(\k_1,...,\k_m-\sum_{i=1}^{n-m}\q_i,
\q_1,...,\q_{n-m}\Big)\sim O(\ve^{-n+m})\,.
\ee

3. One introduces power counting rules and resum all the contributions
at a desired 
order. 
The power counting rules in real and redshift spaces are the 
same, see Sec.~\ref{sec:pc}.
The procedure of IR resummation in TSPT 
has a simple diagrammatic interpretation. 
The leading soft corrections correspond to {\em daisy} diagrams with
multiple soft loops dressing a single wiggly vertex. This is true
irrespective of the number of hard loops in the diagram. 
The soft loop contributions factorize, which allows to easily 
resum them to all orders in perturbation theory.
For a generic wiggly vertex with $n$ hard wavenumbers 
dressed by $L$ soft loops one has
\be
\begin{split}
V_{w,n}^{L-loop}=&~~
\begin{gathered}
\begin{fmffile}{dressed_n}
\begin{fmfgraph*}(110,100)
\fmfpen{thick}
\fmfleft{l0,l1,l2,l3,l4,l5,l6,l7}
\fmfright{r0,r1,r2,r3,r4,r5,r6,r7}
\fmfv{d.sh=circle,label=${\bar \G}^{w (r)}_{n+2L}$,l.a=90,d.filled=shaded,d.si=5thick,l.d=17}{b1}
\fmf{phantom,tension=5,label={\footnotesize $\bullet$},l.s=left}{r7,b2}
\fmf{phantom,tension=5,label={\footnotesize $\bullet$},l.s=right}{l7,b2}
\fmf{phantom}{b1,b2}
\fmfv{d.sh=circle,d.filled=full,label={\footnotesize $\bullet$},d.si=0.05thick,l.d=0}{b2}
\fmf{phantom}{b1,vv3}
\fmf{phantom,label=$ $,l.s=left,l.d=4}{r3,vv3}
\fmf{plain}{b1,vv6}
\fmf{plain,label=$\k_{n-1}$,l.s=left,l.d=2}{r6,vv6}
\fmf{phantom}{b1,vv5}
\fmf{phantom,label=$ $,l.s=right,l.d=0}{r5,vv5}
\fmf{plain}{b1,vv4}
\fmf{plain,label=$\k_n$,l.s=left,l.d=4}{r4,vv4}
\fmf{plain}{b1,v6}
\fmf{plain,label=$\k_2$,l.s=left,l.d=2}{v6,l6}
\fmf{phantom}{b1,v5}
\fmf{phantom,label=$ $,l.s=left,l.d=1.5}{v5,l5}
\fmf{plain}{b1,v4}
\fmf{plain,label=$\k_1$,l.s=left,l.d=2}{v4,l4}
\fmf{phantom}{b1,v3}
\fmf{phantom,label=$ $,l.s=left,l.d=4}{v3,l3}
\fmf{phantom}{b1,v2}
\fmf{phantom,label=$ $,l.s=right,l.d=0}{v2,l2}
\fmf{phantom}{b1,vv2}
\fmf{phantom,label=$ $,l.s=right,l.d=0}{vv2,r2}
\fmf{plain,tension=1,right=0.5,label=$\q_{L}$,l.s=left,l.d=8}{r1,b1}
\fmf{plain,tension=1,right=0.5}{b1,r1}
\fmf{plain,tension=1,right=0.5,label=$\q_1$,l.s=left,l.d=11}{l1,b1}
\fmf{plain,tension=1,right=0.5}{b1,l1}
\fmfv{d.sh=circle,d.filled=full,label={\footnotesize $\bullet$},d.si=0.05thick,l.d=0}{b3}
\fmf{phantom,tension=2,label={\footnotesize $\bullet$},l.s=right}{r0,b3}
\fmf{phantom,tension=2,label={\footnotesize $\bullet$},l.s=left}{l0,b3}
\fmf{phantom}{b1,b2}
\end{fmfgraph*}
\end{fmffile}
\end{gathered}\\
=&~ \frac{1}{(n+2L)!}\cdot(2L+n)...(2L+1)\cdot (2L-1)!! \\
&\times \prod_{i=1}^{L}\left[\int_{q_i< k_S}[dq_i]\,
g^2\bar P_{nw}(q_i)\right]g^{-2}\bar
\G'^{w\,(r)}_{n+2L}(\k_1,...,\k_n,\q_1,-\q_1,...,\q_L,-\q_L)\,.
\end{split}
\label{vertdress0}
\ee  
Using Eq.~\eqref{eq:recreal} we obtain 
\be 
V_{w,n}^{L-loop}=\frac{1}{L!}\left[\frac{g^2}{2}\int_{q< k_S}[dq]\,
\bar P_{nw}(q)\D^{(r)}_{\q}
\D^{(r)}_{-\q}\right]^L g^{-2}\bar{\G}'^w_{n}(\k_1,...,\k_n)\,. 
\ee
Clearly, the summation over the number of soft loops leads to an exponentiation 
of the differential operator 
\be
-g^2\S^{(r)}\equiv \frac{g^2}{2}\int_{q< k_S}[dq]\,
\bar P_{nw}(q)\D^{(r)}_{\q} \D^{(r)}_{-\q}\,.
\ee
For $n=2$ we obtain 
\be
\G'^{w,(r)\,\text{IR res,LO}}_2= e^{-g^2\S^{(r)}}{\G}'^w_2\,.
\ee
Adding the smooth part 
and inverting the whole vertex we obtain the LO result for the power
spectrum 
\be 
\label{eq:LOreal}
P^{(r)\,\text{IR res,LO}}(\e;k)	=
g^2(\e)\Big(\bar P_{nw}(k)+e^{-g^2(\e)\S^{(r)}}\bar P_w(k)\Big)\,.
\ee
Approximating the differential operator as \eqref{eq:diffoper} leads to
\be 
P^{(r)\,\text{IR res,LO}}(\e;k)= P_{nw}(k)+e^{-k^2\Sigma^2(\e)}P_w(k)\,,
\ee
where the damping factor $\Sigma^2$ is given in Eq.~\eqref{Sigmanorm}
and we have switched to the compact notation for the time-dependent linear power spectrum \eqref{eq:notcomp}.
IR resummation at first order in hard loops 
amounts to evaluating the loop integrals using  
\eqref{eq:LOreal} and accounting for double counting in the tree-level
result, e.g. 
for the power spectrum we have
\be
\begin{split} 
P^{(r)\,\text{IR-res,LO+NLO}}=&
 P_{nw}+(1+k^2\Sigma^2(\eta))
e^{-k^2\Sigma^2(\eta)} P_{w}\\
&+P^{(r)\,1-loop}[P_{nw}+e^{-k^2\Sigma^2}P_{w}]\,.
\end{split}
\ee   
This procedure generalizes to higher point functions $\mathfrak{C}^{(r)}_n$, i.e.
\be
\label{Cncomp2}
\begin{split}
\mathfrak{C}_n^{(r)\,\text{IR
    res,LO+NLO}}(\k_1,...,\k_n)=&\mathfrak{C}_n^{(r)\,tree}\big[ 
P_{nw}+(1+k^2\Sigma^2)e^{-k^2\Sigma^2}P_w\big](\k_1,...,\k_n)\\
&+\mathfrak{C}_n^{(r)\,1-loop}\big[
P_{nw}+e^{-k^2\Sigma^2}P_w\big](\k_1,...,\k_n)\,.
\end{split}
\ee
These expressions are valid up to next-to-leading order soft corrections which are numerically small.
	    
\section{Asymptotic behavior of RSD vertices in the soft limit}
\label{app:B}

In this Appendix we prove the asymptotic formula (\ref{eq:GnmindMain}) 
for the redshift space vertices.
For clarity we will omit the superscript $(s)$ in the notation for the wiggly
vertices
and keep in mind that these quantities 
are evaluated in redshift space.
In what follows it is useful to define the operator
\be
\begin{split}
\D^z_{\q}[\bar P_w(k)]=\frac{k_zq_{z}}{q^2}(e^{\q\cdot
  \nabla_{\k'}}-1)
\bar P_w(k')\Big|_{\k'=\k}\,.
\end{split} 
\ee

To prove (\ref{eq:GnmindMain}) we proceed by induction. In
Eq.~(\ref{eq:G3w2}) we have verified this formula for $n=3$. Now, suppose
it is valid for $n-1$ and any $m$. Our aim is to prove it for
$n$. Using the recursion relation (\ref{eq:Gammaseq}) we write
\be
\begin{split}
\G'^{w}_{n}(f;\k_1,...,&\k_m-{\bf Q}, \q_1,...,\q_{n-m})=
\G'^{w\,(r)}_{n}(f;\k_1,...,\k_m-{\bf Q}, \q_1,...,\q_{n-m})\\
-\int_0^f d\F\bigg[&
\sum_{1\leq i<j<m}I^{(s)}_2(\k_i,\k_j)\;
\G'^{w}_{n-1}(\F;\k_i+\k_j,...,\check{\k}_i,...,\check{\k}_j,...)\\
&+\sum_{i=1}^{m-1}I^{(s)}_2(\k_i,\k_m-{\bf Q})\;
\G'^w_{n-1}(\F;...,\check{\k}_i,...,\k_m+\k_i-{\bf Q},...)\\
&+\sum_{i=1}^{m-1}\sum_{j=1}^{n-m}I^{(s)}_2(\k_i,\q_j)\;
\G'^{w}_{n-1}(\F;...,\k_i+\q_j,...,\check{\q}_j,...)\\
&+\sum_{j=1}^{n-m}I^{(s)}_2(\k_m-{\bf Q},\q_j)\;
\G'^{w}_{n-1}\Big(\F;...,\k_m-\sum_{l\neq j}\q_l,...,\check{\q}_j,...\Big)\\
&+\sum_{1\leq i<j\neq n-m}I^{(s)}_2(\q_i,\q_j)\;
\G'^{w}_{n-1}(\F;...,\q_i+\q_j,...,\check{\q}_i,...,\check{\q}_j,...)\bigg]\,,
\end{split} 
\ee
where we have introduced a shorthand notation ${\bf
  Q}\equiv\sum_{i=1}^{n-m}\q_i$. 
Let us estimate the enhancement of various terms in this expression. 
The real-space vertex $\G_n'^{w\,(r)}$ is of order $O(\ve^{-n+m})$,
as seen from Eq.~(\ref{eq:recreal}).  
The vertices in the second and third 
lines have $n-m$ soft arguments and thus are also
enhanced as $O(\varepsilon^{-n+m})$ according to our 
induction hypothesis. 
The vertices in the last three lines have one soft argument less
and thus are only $O(\varepsilon^{-n+m+1})$. 
In the fourth and fifth lines, 
however, this is compensated by the poles in the $I^{(s)}_2$ kernels,
\be 
I^{(s)}_2(\k_i,\q_j)\approx \frac{k_{i,z}q_{j,z}}{q_j^2}=O(1/\varepsilon)\,.
\ee
Keeping only the terms $O(\varepsilon^{-n+m})$ we arrive at
\be
\label{allterms}
\begin{split}
\G'^{w}_{n}=\G'^{w\,(r)}_{n}
-\int_0^f d\F\bigg[&
\sum_{1\leq i<j\leq m}I^{(s)}_2(\k_i,\k_j)\;
\G'^{w}_{n-1}(\F;\k_i+\k_j,...,\check{\k}_i,...,\check{\k}_j,...)\\
&+\sum_{j=1}^{n-m}\sum_{i=1}^{m-1}\frac{k_{i,z}q_{j,z}}{q_j^2}\;
\G'^{w}_{n-1}(\F;...,\k_i+\q_j,...,\check{\q}_j,...)\\
&+\sum_{j=1}^{n-m}\frac{k_{m,z}q_{j,z}}{q_j^2}\;
\G'^{w}_{n-1}\Big(\F;...,\k_m-\sum_{l\neq
  j}\q_l,...,\check{\q}_j,...\Big)
\bigg]\,.
\end{split} 
\ee
Next, we use that $\k_m=-\sum_{i=1}^{m-1}\k_i$ due to momentum
conservation and rewrite the last two terms as 
\be 
\label{lasttwo}
\begin{split}
&\sum_{j=1}^{n-m}\sum_{i=1}^{m-1}
\frac{k_{i,z}q_{j,z}}{q_j^2}\Big[
\G'^{w}_{n-1}(\F;...,\k_i+\q_j,...,\check{\q}_j,...)
-\G'^{w}_{n-1}(\F;...,\k_m-\sum_{l\neq j}\q_l,...,\check{\q}_j,...)\Big]\\
&=\sum_{j=1}^{n-m}\D^z_{\q_j}
\G'^{w}_{n-1}(...,\k_m-\sum_{l\neq j}\q_l,...,\check{\q}_j,...)\\
&\approx (-1)^{n-m-1}\sum_{j=1}^{n-m}\D^z_{\q_j}
\prod_{l\neq j}\big(\D_{\q_l}^{(r)}+\F \D^z_{\q_l}\big)\;
\G'^w_m(\F;\k_1,...,\k_m)\\
&=(-1)^{n-m-1}\frac{\d}{\d\F}\Bigg(
\prod_{l=1}^{n-m}\big(\D_{\q_l}^{(r)}+\F \D^z_{\q_l}\big)\bigg)\;
\G'^w_m(\F;\k_1,...,\k_m)\,.
\end{split}
\ee
On the other hand, the first term in square brackets in
(\ref{allterms}) reads,
\be
\label{firstterm}
\begin{split}
&\sum_{1\leq i<j\leq m}I^{(s)}_2(\k_i,\k_j)\;
\G'^{w}_{n-1}(\F;\k_i+\k_j,...,\check{\k}_i,...,\check{\k}_j,...)\\
&\approx (-1)^{n-m}\prod_{l=1}^{n-m}\big(\D_{\q_l}^{(r)}+\F
\D^z_{\q_l}\big)
\sum_{1\leq i<j\leq m}
I^{(s)}_2(\k_i,\k_j)\;
\G'^{w}_{m-1}(\F;\k_i+\k_j,...,\check{\k}_i,...,\check{\k}_j,...)\\
&=(-1)^{n-m-1}\prod_{l=1}^{n-m}\big(\D_{\q_l}^{(r)}+\F
\D^z_{\q_l}\big)\;
\frac{\d}{\d\F}\G_m'^w(\F;\k_1,...\k_m)\;,
\end{split} 
\ee
where in the last equality we again used the relation
(\ref{eq:Gammaseq}). Combining Eqs.~(\ref{lasttwo}) and
(\ref{firstterm}) we obtain,
\be
\G_n'^w=\G_n'^{w\,(r)}+(-1)^{n-m}\int_0^f d\F\,
\frac{\d}{\d\F}\bigg(
\prod_{l=1}^{n-m}\big(\D_{\q_l}^{(r)}+\F
\D^z_{\q_l}\big)\;
\G_m'^w(\F;\k_1,...\k_m)\bigg)\;.
\ee
Integration and use of Eq.~(\ref{eq:recreal}) for the real space vertex 
$\G_n'^{w\,(r)}$
yields the formula
(\ref{eq:GnmindMain}). QED

\section{Bias expansion at one loop}
\label{app:C}

In order to obtain the bias kernels $M_n$ at one loop we have to go to the 
second order in $\Pi$, i.e. we need only
\be
\begin{split}
& \Pi^{[1]}_{ij}=\frac{\d_i\d_j \delta}{\Delta}\,,\\
& \Pi^{[2]}_{ij}=\frac{\d_i\d_j}{\Delta}(\T(1+\delta)-\delta)+\frac{\d_i\d_j}{\Delta}\left(
\d_l \delta\frac{\d_l\Theta}{\Delta}
\right) - \frac{\d_i\d_j\d_l \delta}{\Delta}\frac{\d_l\Theta}{\Delta}
\,.
\end{split} 
\ee  
At first order there is a single 
operator in the bias expansion, $\text{tr} \Pi^{[1]}=\delta$.
At second order there are two operators,
\be 
\begin{split}
&\mathcal{O}^{[2]}_1\equiv \frac{1}{2} (\text{tr}[\Pi^{[1]}])^2
=\frac{\delta^2}{2}\,,\\
& \mathcal{O}^{[2]}_2\equiv \frac{1}{2}\text{tr}[(\Pi^{[1]})^2]
=\frac{1}{2}\frac{\d_i\d_j \delta}{\Delta}\frac{\d_i\d_j \delta}{\Delta}\,.
\end{split} 
\ee
At third order we have:
\be 
\begin{split}
 \mathcal{O}^{[3]}_1\equiv &\frac{1}{6}(\text{tr}[\Pi^{[1]}])^3
=\frac{\delta^3}{6}\,,\\
 \mathcal{O}^{[3]}_2\equiv& \frac{1}{2}
\text{tr}[(\Pi^{[1]})^2]\text{tr}[\Pi^{[1]}]
=\frac{1}{2}\frac{\d_i\d_j \delta}{\Delta}\frac{\d_i\d_j
  \delta}{\Delta}\delta \,,\\ 
 \mathcal{O}^{[3]}_3\equiv& \frac{1}{6}\text{tr}[(\Pi^{[1]})^3]
=\frac{1}{6}\frac{\d_i\d_j \delta}{\Delta}\frac{\d_i\d_l \delta}{\Delta}\frac{\d_l\d_i \delta}{\Delta} \,,\\
 \mathcal{O}^{[3]}_4\equiv&\frac{1}{2}\text{tr}[\Pi^{[2]}\Pi^{[1]}]
=\frac{1}{2}\bigg\{\frac{\d_i\d_j}{\Delta}(\T\delta +\T-\delta
)\frac{\d_i\d_j \delta}{\Delta} \\
&\qquad\qquad\qquad~+\left[\frac{\d_i\d_j}{\Delta}\left(
\d_l \delta\frac{\d_l\Theta}{\Delta}
\right) - \frac{\d_i\d_j\d_l \delta}{\Delta}\frac{\d_l\Theta}{\Delta}\right]
\frac{\d_i\d_j \delta}{\Delta} \bigg\}
\,.
\end{split} 
\ee
Going into Fourier space and using the decomposition \eqref{eq:psi1} 
we obtain \eqref{eq:dhbnkern} with the following kernels:
\be 
\begin{split}
 & M_1^{(r)}(\k)=~b_1\,, \\
& M_2^{(r)}(\k_1,\k_2)= b_1 K_2(\k_1,\k_2)+b_2
+b_{\mathcal{O}^{[2]}_2}\frac{(\k_1\cdot \k_2)^2}{k_1^2 k_2^2}\,,\\
& M_3^{(r)}(\k_1,\k_2,\k_3)= b_1K_3(\k_1,\k_2,\k_3)+
b_2 [K_2(\k_1,\k_2)+{\rm perm.}]\\
&+b_{\mathcal{O}^{[2]}_2}\left[\frac{(\k_1\cdot \k_{23})^2}{k_1^2 k_{23}^2}K_2(\k_2,\k_3)+{\rm perm.}\right]
+ b_3
+b_{\mathcal{O}^{[3]}_2}
\left[\frac{(\k_1\cdot \k_2)^2}{k_1^2 k_2^2}+{\rm perm.}\right]\\
&+b_{\mathcal{O}^{[3]}_3}
\frac{(\k_1\cdot \k_2)(\k_2\cdot \k_3)(\k_3\cdot \k_1)}{k_1^2 k_2^2 k_3^2}
+b_{\mathcal{O}^{[3]}_4}
\Bigg[\frac{(\k_1\cdot \k_{23})^2}{k_1^2 k_{23}^2}\left(1-\frac{1}{2}K_2(\k_2,\k_3)\right)\\
& +
\frac{(\k_3\cdot \k_2)}{k_{23}^2k_1^2k_2^2k_3^2}
\Big[(\k_1\cdot \k_2)(\k_1\cdot\k_3)(k_2^2+k_3^2)
-(\k_2\cdot \k_3)\big((\k_1\cdot\k_3)^2+(\k_1\cdot \k_2)^2\big)
\Big]\!\!+\!{\rm perm.}\!
\Bigg],
\end{split} 
\ee
where `perm.' means terms obtained by cyclic permutations of the
momenta $\k_1,\k_2,\k_3$. 
Note that the $M_n^{(r)}$ kernels written above are manifestly IR
safe.
The kernels in redshift space are obtained using the recursion
relations similar to (\ref{eq:Meqs}). In particular, we have
\be
\begin{split}
&M_1^{(s)}(\k)=b_1+f\mu^2\;,\\
&M_2^{(s)}(\k_1,\k_2)=M_2^{(r)}(\k_1,\k_2)+
\bigg\{\mu_1^2+\mu_2^2-2\frac{(\k_1\cdot\k_2)}{k_1k_2}\mu_1\mu_2\bigg\}b_1f\;,
\end{split}
\ee
where $\m_i\equiv (\hat{\k}_i\cdot \hat{\bf z})$. The expression for
$M_3^{(s)}$ is rather cumbersome and we do not present it here.

For reference we also write down the SPT kernels for
biased tracers in redshift space. Compared to
\cite{Bernardeau:2001qr} we add the tidal bias. 
It appears convenient to change the bias basis and consider
\be
\delta_{h}=b_1\delta+\frac{b_2}{2}\delta^2+b_{\mathcal{G}_2} \mathcal{G}_2
+\frac{b_3}{6}\delta^3 + b_{\mathcal{G}_3}\mathcal{G}_3
+b_{(\mathcal{G}_2\delta)}\mathcal{G}_2\delta
+b_{\G_3}\G_3\,,
\ee
where 
\be
\begin{split}
& \mathcal{G}_2(\Phi)=(\d_i\d_j \Phi)^2-(\d^2\Phi)^2\,,\\
& \mathcal{G}_3(\Phi)=- \d_i\d_j \Phi \d_j\d_k \Phi \d_k\d_i\Phi
-\frac{1}{2}(\d^2\Phi)^3 +\frac{3}{2}(\d_i\d_j \Phi)^2\d^2\Phi \,,\\
& \G_3 = \mathcal{G}_2(\Phi)-\mathcal{G}_2(\Phi_v)\,,
\end{split}
\ee
and we introduced we velocity potential defined via $\Delta \Phi_v=\T$.
Acting along the lines of Sec. 7.4.1 of \cite{Bernardeau:2001qr} we obtain
\bseq 
\begin{align}
&Z_1(\k)  = b_1+f\mu^2\,,\\
&Z_2(\k_1,\k_2)  =\frac{b_2}{2}+b_{\mathcal{G}_2}\left(\frac{(\k_1\cdot \k_2)^2}{k_1^2k_2^2}-1\right)
+b_1 F_2(\k_1,\k_2)+f\mu^2 G_2(\k_1,\k_2)\notag\\
&\qquad\qquad\quad~~+\frac{f\mu k}{2}\left(\frac{\mu_1}{k_1}(b_1+f\mu_2^2)+
\frac{\mu_2}{k_2}(b_1+f\mu_1^2)
\right)
\,,\\
&Z_3(\k_1,\k_2,\k_3)  =\frac{b_3}{6}
+b_{\mathcal{G}_3}\left[-\frac{(\k_1\cdot \k_2)(\k_2\cdot
    \k_3)(\k_3\cdot
    \k_1)}{k_1^2k_2^2k_3^2}-\frac{1}{2}+\frac{3}{2}\frac{(\k_1\cdot
    \k_2)^2}{k_1^2k_2^2}\right]\notag\\
&\quad+b_{(\mathcal{G}_2\delta)}\left[\frac{(\k_1\cdot
    \k_2)^2}{k_1^2k_2^2}-1\right]
 +2b_{\G_3}\left[\frac{(\k_1\cdot
     (\k_2+\k_3))^2}{k_1^2(\k_2+\k_3)^2}-1\right]
\big[F_2(\k_2,\k_3)-G_2(\k_2,\k_3)\big]\notag\\  
&\quad
+b_1 F_3(\k_1,\k_2,\k_3)+f\mu^2 G_3(\k_1,\k_2,\k_3)+\frac{(f\m k)^2}{2}(b_1+f \mu_1^2)\frac{\m_2}{k_2}\frac{\m_3}{k_3}\notag\\
&\quad
+f\mu k\frac{\mu_3}{k_3}\left[b_1 F_2(\k_1,\k_2) + f \mu^2_{12} G_2(\k_1,\k_2)\right]
+f\m k (b_1+f \mu^2_1)\frac{\m_{23}}{k_{23}}G_2(\k_2,\k_3)\notag\\
&\quad+b_2 F_2(\k_1,\k_2)+2b_{\mathcal{G}_2}\left[\frac{(\k_1\cdot (\k_2+\k_3))^2}{k_1^2(\k_2+\k_3)^2}-1\right]F_2(\k_2,\k_3)
+\frac{b_2f\mu k}{2}\frac{\mu_1}{k_1}\notag\\
&\quad+b_{\mathcal{G}_2}f\mu k\frac{\m_1}{k_1}\left[\frac{(\k_2\cdot \k_3)^2}{k_2^2k_3^2}-1\right]
\,,
\end{align} 
\eseq
where $\k=\k_1+\k_2+\k_3$ and
the kernel $Z_3$ must be symmetrized in its arguments.

\section{Simplification of NLO IR resummed integrands: example of 
bispectrum in redshift space}
\label{app:D}

At face value, IR resummation requires using the dressed anisotropic
power spectrum 
\be
\label{eq:dressedL1}
P_{nw}(p)+e^{-p^2\Sigma_{tot}^2(\mu_\p)}P_w(p)
\ee
as an input in loop calculations. 
This prescription causes technical complications that one would like to
minimize. We have seen in Sec.~\ref{sec:NLOeval} that in the case of
one-loop power spectrum one can pull the 
anisotropic
damping factor outside the momentum integral without changing the
order of approximation of the final result.
Here we discuss the general situation for an 1-loop $n$-point
correlator and illustrate it on the example of bispectrum. We will
work in terms of SPT diagrams.

Consider an SPT one-loop 
diagram for some $n$-point function and substitute the linear power
spectrum in it by its IR-resummed counterpart
(\ref{eq:dressedL1}). 
Depending on the topology of the diagram the wiggly power spectrum can be: 
\begin{itemize}
\item[(a)]
 outside the loop and depend only on an external momentum. In this
 case the loop integral contains only the isotropic smooth power
 spectrum $P_{nw}$ and its evaluation is straightforward.

\item[(b)] 
inside the loop and be multiplied by a combination of kernels
without IR singularities at finite values of the loop momentum 
$\p$ (we choose $\p$ to coincide with the argument of $P_w$). An
example of such contribution is given by 
the second term in (\ref{eq:deltaP}) which is part of $P_{13}$
diagram in SPT language. 
As we discussed in Sec.~\ref{sec:NLOeval}, such contributions
are exponentially suppressed and can be safely neglected.

\item[(c)]
 inside the loop and be multiplied by a combination of kernels with a
 single 
IR singularity at $p=p_0$, where $p_0$ is a linear combinations
of external momenta. This is the case of the third term in
(\ref{eq:deltaP}) which comes from the $P_{22}$ diagram. In such
diagram the argument of the damping factor can be substituted by
$\p_0$,  
\be 
e^{-p^2\Sigma_{tot}^2(\mu_\p)}
P_w(p)\to e^{-p_0^2\Sigma^2_{\text{tot}}(\mu_{\p_0})} P_w(p)\,.
\ee
Then the damping factor can be taken out of the loop integral.

\item[(d)] 
inside the loop and be multiplied by a combination of kernels with
more than one  
IR singularity. We are not aware of any convenient method to simplify
these contributions, so in this case the anisotropic damping factor
must be kept inside the integral.  

\end{itemize}

Applying the above algorithm to the one-loop correction to the bispectrum
we obtain the following expression,
\be
\begin{split}
B^{(s)\,\text{IR res,NLO}}(\k_1,\k_2,\k_3)=
B^{(s)\,1-loop}[P_{nw}]
+\tilde B_{w}^{(s)\,1-loop}(\k_1,\k_2,\k_3)\,,
\end{split} 
\ee
where $B^{(s)\,1-loop}[P_{nw}]$ is evaluated using only the smooth power 
spectrum and 
\be
\begin{split}
\tilde B_{ w}^{(s)\,1-loop}=
\tilde B_{411, w}^{(s)}
+\tilde B_{321-II, w}^{(s)}
+\tilde B_{321-I, w}^{(s)}
+\tilde B_{222, w}^{(s)}\;.
\end{split} 
\ee
Here the individual terms read,
\bseq
\label{tildeBsw}
\be 
\label{B411}
\begin{split}
\tilde B_{411, w}^{(s)}&= 12Z_1(\k_2)Z_1(\k_3)
  \left[
e^{-k_2^2\Sigma^2_{\rm tot}(\mu_2)}P_{w}(k_2)P_{nw}(k_3)
+ e^{-k_3^2\Sigma^2_{\rm tot}(\mu_3)}P_{nw}(k_2)P_{w}(k_3)
\right]\\
  &~~~~~~\times \int [dp] Z_4(-\p,\p,-\k_2,-\k_3)P_{nw}(p)\quad
  +\text{2 cyclic perm.}\,, 
\end{split}
\ee
\be 
\label{B321II}
\begin{split}
\tilde  B_{321-II, w}^{(s)}&= 6Z_1(\k_2)Z_2(\k_2,\k_3)
  \!\left[
e^{-k_2^2\Sigma^2_{\rm tot}(\mu_2)}P_{w}(k_2)P_{nw}(k_3)
\!+\! e^{-k_3^2\Sigma^2_{\rm tot}(\mu_3)}P_{nw}(k_2)P_{w}(k_3)
\right]\\
  &~~~~~~\times \int [dp] Z_3(-\p,\p,\k_3)P_{nw}(p)\quad +\text{5 permutations}\,,
\end{split}
\ee
\be 
\label{B321I}
\begin{split}
 \tilde B_{321-I, w}^{(s)}=& 6Z_1(\k_3)e^{-k_3^2\Sigma^2_{\rm tot}(\mu_3)}
P_{w}(k_3)\\
&~\times\int [dp] Z_3(-\p,\p-\k_2,-\k_3)
Z_2(\p,\k_2-\p)P_{nw}(p)P_{nw}(|\k_2-\p|)\\
&+12 Z_1(\k_3) P_{nw}(k_3)e^{-k_2^2\Sigma^2_{\rm tot}(\mu_2)}\\
&~\times\int [dp] Z_3(-\p,\p-\k_2,-\k_3)
Z_2(\p,\k_2-\p)P_{w}(p)P_{nw}(|\k_2-\p|)\\
&+\text{5 permutations},
\end{split}
\ee

\be 
\label{B222}
\begin{split}
\tilde B_{222, w}^{(s)}=8\int& [dp]
Z_2(-\p,\p+\k_1)
Z_2(-\p-\k_1,\p-\k_2)
Z_2(\p+\k_2,\p) \\
&\times e^{-p^2\Sigma^2_{\rm tot}(\mu_\p)}P_w(p)
P_{nw}(|\p+\k_1|) P_{nw}(|\p-\k_2|)
\quad +\text{2 cyclic perm.}
\end{split}
\ee
\eseq
We observe that
the terms (\ref{B411}), (\ref{B321II}) are of type (a), the term
(\ref{B321I}) is of type (c), whereas (\ref{B222}) is of type (d).

\newpage{\pagestyle{empty}\cleardoublepage}
\end{document}